\documentclass[11pt]{article}

\renewcommand{\baselinestretch}{1.1}

\usepackage{a4wide}

\usepackage{epsfig}

\usepackage{fancyheadings}
\usepackage{cite}
\usepackage{amsmath}
\usepackage{amssymb}

\begin{document}
%

%
%

\def\lap{\nabla^2}

\def\gtt{G^{t}_{~t}}
\def\gaa{G^{\theta}_{~\theta}}
\def\grrzz{(G^{r}_{~r}+G^{z}_{~z})}
\def\crz{G^{r}_{~z}}
\def\crrzz{(G^{r}_{~r}-G^{z}_{~z})}

\def\figuremode{\small}

%
%

\title{\bf Static Axisymmetric Vacuum Solutions and Non-Uniform Black
  Strings }

\author{{\bf Toby Wiseman}\thanks{e-mail: {\tt
      T.A.J.Wiseman@damtp.cam.ac.uk}} \\ \\
  Department of Applied Mathematics and Theoretical Physics,\\
  Centre for Mathematical Sciences,\\
  Wilberforce Road,\\
  Cambridge CB3 0WA, UK}

\date{September 2002}

\maketitle

%
\begin{abstract}
%
  
  We describe new numerical methods to solve the static axisymmetric
  vacuum Einstein equations in more than four dimensions. As an
  illustration, we study the compactified non-uniform black string
  phase connected to the uniform strings at the Gregory-Laflamme
  critical point.  We compute solutions with a ratio of maximum to
  minimum horizon radius up to nine.  For a fixed compactification
  radius, the mass of these solutions is larger than the mass of the
  classically unstable uniform strings. Thus they cannot be the end
  state of the instability.

%
\end{abstract}
%

\vspace{7.0cm}
\begin{flushright}
DAMTP-2002-116 \\
hep-th/0209051
\end{flushright}

\newpage

\tableofcontents

\newpage

%
\section{Introduction}
\label{sec:intro}
%

Static axisymmetric vacuum gravity in four dimensions is generally
solved by the elegant Weyl solutions \cite{Weyl} which reduce the 3
independent Einstein equations to one elliptic Laplace equation.
However, solutions other than Schwarzschild and flat space are nakedly
singular, or contain conical deficits. In $d$ dimensions higher than
four, one can consider static axisymmetric solutions with a rotational
$O(d-2)$ spatial isometry group. Then there are a far richer variety
of solutions, notably including the black strings, which evade such
problems.  However no general analytic solution is known for more than
four dimensions \cite{Emparan_Reall1,Emparan_Reall2}.

Interest in black strings was aroused when Gregory and Laflamme
\cite{Gregory_Laflamme1,Gregory_Laflamme2,Gregory_Laflamme3} showed
that the uniform black string, a simple product of a Schwarzschild
solution with a line, is unstable to modes with wavelength larger than
a critical value. Using entropy arguments they concluded that the end
state of the classical instability would be a sequence of black holes.
The critical deformation mode is static, and led Gregory and Laflamme
to point out the existence of a new family of black string solutions
without translational invariance, the non-uniform strings.

Compactifying the line direction to a circle allows uniform black
strings to be stabilised for values of the circle radius smaller than
the critical instability length. In the following, when we refer to a
string we implicitly mean a compactified one unless otherwise stated.
Such string solutions are important in understanding the process of
black hole formation in extra dimensions, both in Kaluza-Klein theory
and more generally in the presence of branes \cite{Kol1, Kol2}, and
their classical stability has also been considered in AdS
\cite{Reall2, Gregory, Kang, Gibbons_Hartnoll1}.

Gubser and Mitra used AdS-CFT arguments to conjecture that the onset
of the classical instability of \emph{uniform} strings coincides with
local thermodynamic instability \cite{Gubser_Mitra1, Gubser_Mitra2},
and their work has subsequently been generalised
\cite{Hubeny_Rangamani}.  This was later proved using Euclidean
methods by Reall \cite{Reall}, the translational invariance playing a
crucial role, and further considered in \cite{Ross}.  The dynamics of
the uniform black string classical instability, and its subsequent
decay to black holes was questioned by Horowitz and Maeda
\cite{Horowitz_Maeda1} who showed that the horizon could not pinch off
in a finite affine time. They then argued that the end state might
instead be a non-uniform string. Gregory and Laflamme had constructed
the non-uniform strings to linear order in perturbation theory. In
order to understand the properties of these solutions, Gubser
\cite{Gubser} went to higher order in the marginal deformation, and
showed that the mass of these non-uniform solutions increases from
that of the critical string, for the same circle length.

In light of Horowitz and Maeda's result, Gubser suggested that
at \emph{finite} deformation the mass of the non-uniform strings would
decrease below that of the critical string mass, and their entropy
would be larger than a uniform string of the same mass, and circle
length, therefore allowing a non-uniform string to be the end state of
the critical string instability. This indicated the transition from
uniform to non-uniform string would be first order, and thus
non-adiabatic.  It is important to note that the Horowitz-Maeda result
does not remove the possibility of a black hole end state.  This could
involve infinite affine time, with possibly complicated dynamics
associated with the horizon pinching, although they give arguments why
this would be unlikely.  Alternatively the dynamics could be more
complicated, and evade the assumptions of the Horowitz and Maeda
proof. For example, naked singularities might be present in the
intermediate stages, signalling a dynamic violation of cosmic
censorship, as originally postulated by Gregory and Laflamme
\cite{Gregory_Laflamme1}. Indeed the end state may even be nakedly
singular itself.

Further evidence supporting the first order nature of the transition
was reported in \cite{Horowitz} where unpublished numerical work by
Choptuik et al \cite{Choptuik} is said not to have found smooth
evolution to a non-uniform phase. However they cannot yet determine
the end state. Note that there is no known generally stable algorithm
for numerical evolution of general relativity \cite{Lehner} and
therefore studying an unstable system for long dynamical times is an
extremely difficult problem.

Analytically finding these solutions for finite deformation would
allow the issue of their classical stability and thermodynamic
properties to be settled. A very interesting analytic attempt was made
by Harmark and Obers \cite{Harmark_Obers} where the usual 3 degrees of
freedom in the axisymmetric metric were reduced to only one using a
conjectured ansatz. It seems difficult to show whether this is
indeed a consistent ansatz, and at first sight it appears very
optimistic.  What lends considerable weight to their conjecture is
that they have shown that it is consistent to second order in an
asymptotic expansion, which is a highly non-trivial result.
Unfortunately, the resulting equation for the unknown is very
complicated, and indicates, should the ansatz be consistent, that
there is little hope in having a convenient integrable form.  Using
some assumptions, they claim their ansatz indicates new neutral
non-uniform solutions exist, which are distinct from those emerging
from the uniform strings via the Gregory-Laflamme marginal modes. This
remains to be shown, but would naturally be extremely interesting if
it were confirmed.  \footnote{They claim the mass of these new
  non-uniform strings is likely to be greater than that of the
  critical uniform string, and thus they do not consider them a
  possible end state of the Gregory-Laflamme (GL) instability.}  The
existence of charged near extremal solutions was also considered by
Horowitz and Maeda who have shown that non-uniform solutions exist
\cite{Horowitz_Maeda2} by considering initial data.  The
classification of the compactified black hole metric was considered in
\cite{DeSmet}.

In a recent paper Kol considered the black string/hole phase diagram.
He argues that a classically stable non-uniform phase is too difficult
to include in the phase diagram and therefore probably does not exist.
However it must be noted that, amongst other assumptions, the work
uses thermodynamics to understand the classical stability. The
relation between the two has been shown in the case of strings with a
non-compact translational symmetry \cite{Gubser_Mitra1, Gubser_Mitra2,
  Reall}. To date, there is no known relation between classical and
local thermodynamic stability for cases without this translation
symmetry, such as the non-uniform strings.  Kol conjectures that the
mass of the non-uniform phase is always greater than that of the
critical string, and furthermore the family of solutions is continuous
through a topology changing point where non-uniform strings become
black holes.

In this paper we hope to resolve the thermodynamic properties of the
non-uniform string solutions. The methods we use will be numerical,
and follow from techniques developed in \cite{Wiseman} to study the
geometry of stars on a UV Randall-Sundrum brane
\cite{Randall_Sundrum1, Randall_Sundrum2}. There we observed that a
particular gauge made the second derivatives appearing in a subset of
the Einstein equations appear elliptic, in line with our expectation
that the static equations should indeed be elliptic. We showed how to
employ elementary numerical methods to solve these equations, and in
addition how to implement the constraints on the boundary data, and to
ensure they are globally satisfied. 

In section \ref{sec:elliptic}, we begin by briefly reviewing this
method in a more general context, and then in section
\ref{sec:blackstring}, show how to apply it to the black string in 6
dimensions.  Note that we study the 6 dimensional string for various
simplifying technical reasons, although the previous results in the
literature all apply equivalently in the 5 and 6 dimensional cases. In
particular, repetition of Gubser's work in the 6 dimensional case
again yields the same conclusion, that a non-uniform string emerging
from the critical point has larger mass than the critical string, and
lower entropy than a uniform string of the same mass.  Thus the
transition from the unstable uniform string to whatever end state it
eventually reaches is again non-adiabatic, as there is no other family
of connected solutions.

The compactified black strings provide a clean application for our
numerical method, where it is very simple to implement. This is
compared to the previous work of the star on the UV brane
\cite{Wiseman}, where various technical issues associated with the
coordinate system at the symmetry axis complicated matters. The
method, based on elementary relaxation techniques, yields solutions
with large deformation parameter. The implementation of the method
will be made available at \cite{Website}. The performance of the
method is evaluated in section \ref{sec:performance}, and
self-consistency is demonstrated.  Using moderate resolution and
computing time we present solutions with a maximum/minimum horizon
radius of around ten. This can no doubt be considerably improved by
further development of the algorithm and relaxation methods.  Due to
the technical nature of this section, the reader might wish to pass
over this directly to the following section \ref{sec:solutions}, where
the solutions and their properties are discussed.  These solutions
allow us to see the prevailing asymptotic behaviour for fixed circle
length, namely that the mass, which increases with increasing
non-uniformity near the critical solution as shown by Gubser,
continues to increase until it appears to reach an asymptotic value of
approximately twice that of the critical string, for maximally
non-uniform string solutions. Several independent consistency checks
are performed to estimate the systematic errors present in the method,
and we find that they are small, and thus this mass result is expected
to be robust. In addition we compute the difference of the entropy of
the non-uniform strings to uniform ones of the same mass, finding that
the non-uniform solutions always have the lower entropy. However, this
entropy difference is very small, and thus is `difficult' to measure
numerically using the current implementation, and so we consider this
result less concrete.

Although we are using relaxation methods, we do not have an energy
functional to minimise. Therefore whilst we find the non-uniform
solutions, we cannot infer their classical stability. It is, however,
intriguing that when applying the method to find uniform string
solutions, only the classically stable solutions can be found by our
method.

\begin{figure}[htb]
\centerline{\psfig{file=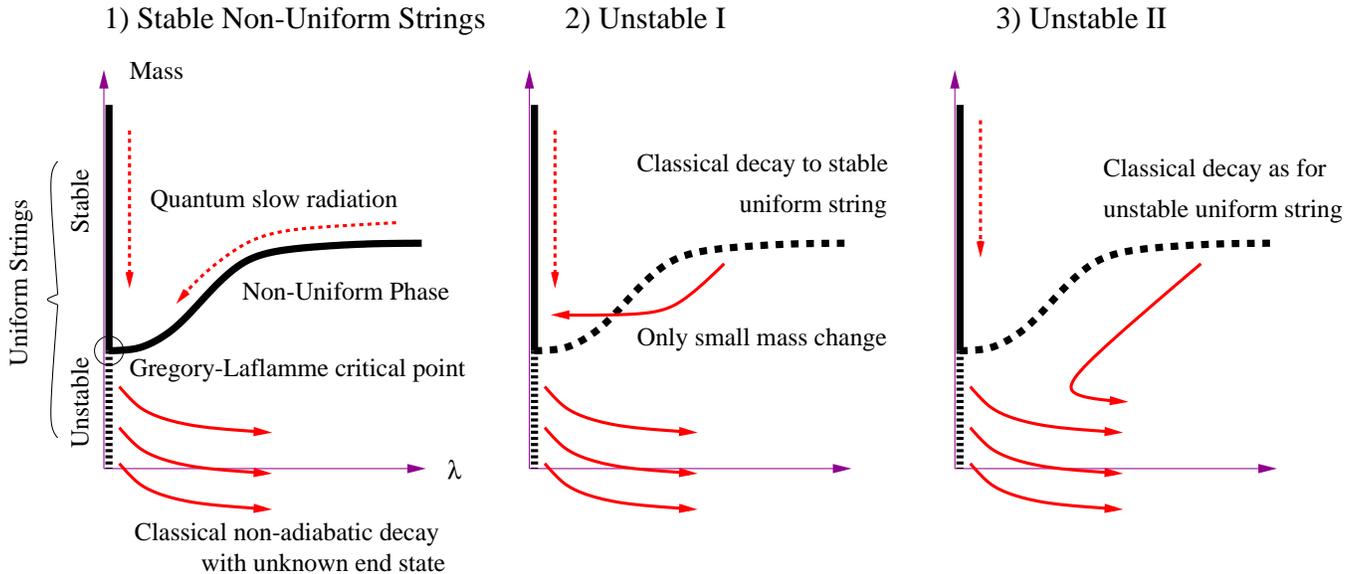,width=7in}}
\caption{ \figuremode
  Summary of results: We have shown by numerical methods that the
  non-uniform branch of solutions always has larger mass, for a fixed
  compactification radius, than the critical uniform string. This
  implies that the non-uniform strings cannot be the end state of the
  Gregory-Laflamme instability for the unstable uniform strings. Not
  knowing the classical stability of these non-uniform solutions there
  are 3 plausible scenarios. 1) The non-uniform strings are
  classically \emph{stable}, and may quantum mechanically decay by
  Hawking radiation to the critical point, like uniform stable
  strings. 2) They are classically \emph{unstable} and decay to stable
  uniform strings. We have shown the horizon volume, or entropy, of
  non-uniform strings is always less than that of a uniform string of
  the same mass, so this is allowed by the second law, although the
  mass lost in the decay must be small.  3) They are classically
  \emph{unstable} and decay in the same way as the unstable uniform
  strings.  If stable, it is likely these solutions may play an
  important role in black hole formation in compactified theories.
\label{fig:results} 
}
\end{figure}

In section \ref{sec:discussion}, we discuss the implications, that the
non-uniform static solutions do indeed exist non-perturbatively, but
are not accessible as the decay product of a classically unstable
uniform black string due to their higher mass. If they are classically
stable, their behaviour is presumably similar to uniform strings with
greater than critical mass, namely that they quantum mechanically
radiate until the critical point is reached, and then classically
decay. They would possibly play an important role in higher
dimensional dynamics, particularly in black hole formation with
compactified extra dimensions, where the black hole mass is in the
range where both the uniform and non-uniform solutions exist. If they
are classically unstable, as conjectured by Kol, then the solutions
would classically decay, and are likely to play little role in higher
dimensional dynamics.  Starting with a solution, we have shown that
the decay to the uniform strings with the same, or slightly lower mass
is allowed by the second law, although the mass difference cannot be
very large, as the difference in horizon volume, or entropy, between
the non-uniform and uniform strings of the same mass is always small.
This possibility is then quite the opposite of the Horowitz and Maeda
picture. Here the non-uniform solutions might decay to the uniform
stable solutions.  Alternatively, if much radiation is given off in
the decay, it is likely that the non-uniform strings will behave in
much the same way as the unstable uniform strings. These different
behaviours are summarised in figure \ref{fig:results}.  Whichever case
is true, the fascinating question of the end state of the GL
instability remains open, as members of this branch of non-uniform
solutions cannot be the end state.  Following the Horowitz-Maeda
result, this may signal the existence of new non-uniform solutions
unconnected to the GL critical point, or alternatively, novel decay
dynamics, possibly involving cosmic censorship violation.

%
\section{Static Axisymmetric Gravity with Elliptic Boundary Conditions}
\label{sec:elliptic}
%

The problem that concerns us is solving static vacuum gravity. This is
thought to be elliptic (ie. a boundary value problem) in the sense
that it is consistent with imposing data, say the induced metric, on
all coordinate patch boundaries. In referring to boundaries we mean to
include asymptotic boundaries.  The simplest example where the
Einstein equations are manifestly elliptic is in the Newtonian limit
of gravity. There we see that the Newtonian potential obeys a Laplace
equation, or Poisson equation in the presence of matter. Of course
this is only a perturbative description, but is strongly suggestive
that for small deformations of the boundary data away from a
non-linear solution, we might use elliptic methods.

Conventional numerical gravity problems are dynamical, where initial
data on some Cauchy surface is given, and the equations are then
evolved hyperbolically off that surface \cite{Lehner}. It is true that
elliptic equations may be solved in a `hyperbolic' fashion, if data
can be consistently imposed on a subset of the boundaries. However, if
the problem specifies data on all the boundaries, as is often
physically the case, then such methods will result in shooting
problems. Whilst shooting is acceptable for an ordinary differential
equation, it is no longer a good method for solving a 2 variable
partial differential equation. In such a case, solutions are best
obtained by relaxation, or in linear cases, by spectral methods. The
key question we begin to answer is how to implement a relaxation
scheme for gravity. The methods we outline in this paper are tailored
to the axisymmetric case, as we are primarily interested in exploring
the properties of the black string in higher dimensions. However,
generalisations of these methods may well apply in situations with
different, or even less symmetry.

In 4 dimensions the elegant Weyl solutions \cite{Weyl} provide a
non-linear general solution to the axisymmetric problem.  The solution
reduces to solving a scalar Laplace equation in flat 3 dimensional
space for axisymmetric solutions, and thus we see the elliptic nature
of the static problem.  Most vacuum Weyl solutions have some form of
naked curvature or angular deficit singularity, and there is no
asymptotically flat black string solution.  In higher dimensions,
where axisymmetric solutions may be asymptotically flat in the radial
direction, with no naked singularities, such as for the 5 dimensional
black string, there is no known general analytic solution.  Whilst the
Weyl solution does generalise to higher dimensions \cite{Myers,
  Emparan_Reall1, Emparan_Reall2}, it does not describe axisymmetric
geometries.  Finding this general axisymmetric solution is an
important open problem. In 4 dimensions the radial sub-manifold is an
$S^1$ and is therefore flat.  However this is no longer true in higher
dimensions, which considerably complicates the form of the Einstein
equations.  Indeed there is no reason to expect these to be
integrable, and there to exist a closed form general solution.

The methods used in this paper were first applied in \cite{Wiseman} to
solve the geometry of a star on a Randall-Sundrum brane near its upper
mass limit. This enabled the geometry of stars with radius smaller
than the AdS length to be found, and in addition, we confirmed that
effective 4 dimensional gravity is reproduced in the non-linear regime
for large stars. These are the only calculations for strongly
gravitating small sources on branes where the bulk is consistently
solved for, except special cases \cite{Gregory2, Horowitz2}, where the
Weyl, or generalised Weyl solutions \cite{Emparan_Reall1} can be
applied.  For numerical work on brane black holes, based on shooting
methods, see \cite{Shinkai}, where the difficulty of shooting with
partial differential equations is apparent.  However, generic
behaviour of the bulk geometry near the brane can be studied for some
initial guess on the brane, although it appears to be pathological far
from the brane, as one would expect from shooting.  A more recent
similar work is \cite{Casadio}.  Note that for strongly gravitating
large objects, analytic progress can be made when the extra dimension
is compact \cite{Wiseman2, Soda}.  However these derivative expansion
methods cannot be applied for the black strings considered here, as
the change in curvature radius due to the non-uniformity is large
compared to the compactification radius.

In some respects the case of the black string is technically simpler
to implement than the Randall-Sundrum star, and thus provides an
archetypal example of our method.  In the rest of the section we
review the general features of the axisymmetric relaxation method, and
in the following section we explicitly show how to apply it to the
black string.

%
\subsection{A Relaxation Method for Solving Static Gravity}
\label{sec:axisym}
%

Having resorted to a numerical solution we must ask why solving an
elliptic axisymmetric problem is difficult. In a scalar field theory
context, finding numerical solutions to the elliptic static equations
of motion is essentially trivial, simply a matter of using standard
relaxation techniques on the Hamiltonian. One difficulty for gravity
is that there is no local energy functional that is positive definite
in the metric and its derivatives. Therefore standard relaxation
techniques cannot be applied. More difficult still is the presence of
constraints in the Einstein equations. Let us consider axisymmetry and
use coordinate freedom to locally parametrise our metric with the
minimal number of functions generally compatible with this symmetry.
There are then 3 metric functions required, but 5 Einstein equations.
One crucial feature of our technique will be to choose a coordinate
system such that 3 of these 5 Einstein equations appear to be
elliptic, with their second derivative terms having a Laplace form.
We call these equations the `interior equations'. Let us then simply
assume that, although there is no energy function to minimise, we may
successfully solve these equations for the boundary data given.
However, how can we guarantee this solution also satisfies the 2
remaining Einstein equations, which we term the `constraints'?

In a hyperbolic ADM evolution \cite{ADM} the constraint equations must
only be imposed on the initial surface and then, in an ideal
evolution, can be ignored as the Bianchi identities ensure that they
remain satisfied. The constraint equations involve the induced metric
on the Cauchy surface and its normal derivative, the extrinsic
curvature. For a hyperbolic evolution this is exactly what must be
specified, the data and its normal time derivative, and thus the
constraints may be evaluated.

In our elliptic problem we envisage giving one piece of data on all
boundaries, such as the induced metric, rather than all the data, the
metric and normal derivative. Now we cannot evaluate the constraints
at the boundaries simply based on the data we specify. Knowing the
induced metric, we will not know the normal derivatives entering the
constraints until we have solved the interior equations.  Thus in a
hyperbolic evolution, ensuring the constraints are satisfied and
evolving the remaining interior equations are separable processes. For
the elliptic case, we cannot evaluate the constraints until we have
solved the interior equations. Thus we must impose the constraints in
an iterative manner, initially picking some boundary data, solving the
interior equations, evaluating the constraints and then modifying the
boundary data to hopefully improve the constraints on the boundaries,
and therefore in the interior. This is repeated until the desired
result is obtained. In addition, we must ensure that satisfying the
constraints on the boundaries does indeed imply they are satisfied in
the interior.

Thus naively we update 2 constraints on the boundaries, specifying 3
metric functions there. Unlike the case of field theory however, the
position of the boundaries may well be additional data, as the metric
solution defines the geometry of the space itself.  If a general
coordinate transformation that moves the boundaries does not preserve
the form of the metric, then extra degrees of freedom must be used to
parameterise the position of these boundaries. In addition to the 2
constraint equations, one must also iteratively update this boundary
position. In this general case we may count local data; (1 function
for the boundary position) + (3 metric functions) - (2 constraints) =
(2 remaining functions), providing the physical data.

Let us now be more explicit and outline a gauge choice that simplifies
the above procedure considerably and ensures that the interior
equations do have an elliptic form.

%
\subsection{The `Conformal Gauge'}
\label{sec:general_method}
%

Let us take the static axisymmetric metric to depend on a radial
variable $r$, and a cylindrical variable $z$.  Following
\cite{Wiseman}, we choose a metric that has
invariance under conformal transformations in the $r, z$ plane,
namely,
\begin{equation}
ds^2 = g_{\mu\nu}(r,z) dx^\mu dx^\nu = - e^{2 A} dt^2 + e^{2 B} ( dr^2 + dz^2 ) + e^{2 C}
d\Omega^2_{d-3}
\label{eq:general_metric}
\end{equation}
where $A, B, C$ are functions of $r, z$, and the line element
$d\Omega^2_{d-3}$ is that of a unit $(d-3)$-sphere.  We may always use
the $r, z$ coordinate degrees of freedom to locally choose a metric of
this form, therefore reducing the 5 possible metric functions of the
most general metric to only 3.  We note this is reminiscent of the
diagonal Weyl form of the metric in 4 dimensions.

We find 3 `interior' Einstein equations from $\gtt$, $\gaa$, and
$\grrzz$, where $\theta$ is an angle on the $(d-3)$-sphere, which yield
equations for $X_i = \{ A, B, C \}$ of the form;
\begin{equation}
\lap X_i = \mathrm{src}_{X_i}
\label{eq:general_equations}
\end{equation}
where $\lap = \partial_r^2 + \partial_z^2$, and the sources
$\mathrm{src}_{X_i}$ depend non-linearly on all the $X_j$, $\partial_r
X_j$ and $\partial_z X_j$. The diagonal form of the metric ensures
that no mixed second derivatives appear in these equations. The
conformal invariance in $r, z$ results in the simple Laplace form for
the second derivatives.

As in \cite{Wiseman}, we suppose that in general, an iterative scheme
may be implemented to solve these interior equations for $A, B, C$ for
some boundary data `near' to that of a known reference solution
$\tilde{A}, \tilde{B}, \tilde{C}$. Whilst there is no energy
functional, we may implement a simple Gauss-Seidel scheme
\cite{numrecp}, treating the sources as fixed. Given a starting
`guess' for $A, B, C$, usually chosen to be the known solution, so
that initially $X_i = \tilde{X}_i$, the source terms are calculated
and the resulting Poisson equations are then solved for the boundary
data, keeping these sources fixed. This gives rise to a new $A, B, C$.
The sources are updated with these new values and the process is
iterated.  Whilst there is certainly no guarantee of convergence, and
much freedom in implementation of the iterative scheme, we have found
in \cite{Wiseman} and in case of the black string presented here,
convergence is achieved using the simplest implementations. Thus
whilst these interior equations are extremely difficult to study
analytically, numerically they are in fact rather easy.

Now we consider the two remaining equations, the `constraints', $\crz$
and $\crrzz$, whose second derivative structure is not of Laplace
form. Instead $\crz$ contains only mixed second derivatives of the
form $\partial_r \partial_z X_i$ and $\crrzz$ contains only hyperbolic
second derivatives as $(\partial_r^2 - \partial_z^2) X_i$. These
equations are related to the interior ones by the Bianchi identities.

The `conformal gauge' has ensured we have interior equations with
Laplace second derivatives as we had hoped for. The next crucial
feature of this gauge is that we have residual coordinate freedom to
move the boundaries in the $r, z$ plane. These may then be placed
anywhere, and choosing the boundary locations completely fixes the
residual coordinate freedom. We might contrast this with a metric
choice of the form $ds^2 = - e^{2 A} dt^2 + e^{2 B} dr^2 + e^{2 C} r^2
d\Omega^2 + dz^2$, where such a coordinate transformation does not
preserve the form of the metric.

We might be confused that losing the one function parameterising the
coordinate position of the boundary would ruin the counting of degrees
of freedom given above. Now (3 boundary metric functions) - (2
constraints) seems to only yield one physical degree of freedom? The
second feature of the gauge is that only one of the constraint
equations must actually be satisfied on all boundaries, and then the
second automatically is too, provided it is enforced at just one
point. The key is the Bianchi identities. Assuming the interior
equations are satisfied, having been relaxed as discussed above, the
remaining terms in the Bianchi identities give simple Cauchy-Riemann
relations,
\begin{eqnarray}
\partial_r \left( g \, \crz \right) +
    \partial_z \left( \frac{g}{2} \, \crrzz \right) & = 0 \nonumber \\
\partial_z \left( g \, \crz \right) -
    \partial_r \left( \frac{g}{2} \, \crrzz \right) & = 0
\label{eq:bianchi}
\end{eqnarray}
where $g = \det{g_{\mu\nu}}$. This elegant result implies both
constraints, multiplied by the volume element, separately satisfy
Laplace equations. Thus if one of them is zero on all boundaries, it
must be zero over the whole interior. Furthermore, the Bianchi
identities then imply the other is determined to be a constant.
Therefore, consider that we implement a scheme which ensures that the
$\crz$ constraint is satisfied on all boundaries, so $\crz$ is
\emph{uniquely} determined to be zero in the interior. Then $\crrzz$
must only be imposed at a single point to ensure that it is also true
in the interior. Again, this is the unique solution.

We immediately see the power of the conformal gauge over other
choices.  Not only do we not have to include and relax extra degrees
of freedom to parametrise the boundary positions, we also only have to
ensure one constraint is satisfied on all the boundaries. We only need
satisfy the other constraint at a single point. In addition, it
guarantees Laplace like second derivatives for the interior equations.
Taken together, these features hugely simplify the task of
implementing an algorithm.

It is numerically sensible to redefine $A, B, C$, by subtracting off
$\tilde{A}, \tilde{B}, \tilde{C}$ so that when these functions are
zero, the metric is then the reference non-linear solution. Thus in
\cite{Wiseman}, the redefinition introduced a warp and radial
scale factor to give AdS in axisymmetric coordinates when $A, B, C$
vanished. The philosophy is to then consider deformations about this
non-linear solution. The linear theory is manifestly elliptic, and
gives much information regarding the correct boundary conditions to
impose. Thus we expect convergence for small perturbations. However,
the method will generically allow one to go beyond small deformations.

%
\section{A Prototype Example: Vacuum Black Strings}
\label{sec:blackstring}
%

We now discuss the application of the ideas outlined above, to the
construction of compactified non-uniform neutral black strings, in the
branch of solutions connected to the GL critical point. The
implementation of the numerical method developed here will be made
available at \cite{Website}. The problem of finding these solutions on
an $S^1$ is a prototype elliptic one.  Asymptotically the geometry is
a product of flat space with the $S^1$, horizon boundary conditions
with some degree of `wiggliness' must be imposed in the interior, and
periodicity must be imposed along the $S^1$ direction.  The scale
invariance of the vacuum Einstein equations means that finding the
solutions for a fixed $S^1$ size allows all other solutions to be
generated simply by a scaling. In a dynamical context we can take the
length of the asymptotic $S^1$ to be fixed, and thus we will be
imposing boundary conditions so that string solutions with different
uniformity are generated having the same asymptotic $S^1$ radius.

In fact we will consider the 6 dimensional black string solution,
rather than the 5 dimensional one examined by Gubser. It must be
stressed that the Horowitz-Maeda result is dimension independent, and
we repeat Gubser's analysis in Appendix \ref{app:gubser_PT}, finding
exactly the same thermodynamic character. Thus the physical behaviour
of the GL instability appears to be the same in both 5 and 6
dimensions. The reasons for considering the 6 dimensional string are
twofold. Firstly, the black string metric takes a particularly simple
form in the conformal gauge in 6 dimensions,
\begin{equation}
ds^2 = - \frac{r^2}{m + r^2} e^{2 A} dt^2 + e^{2 B} ( dr^2 + dz^2 ) + e^{2 C}
( m + r^2 ) d\Omega^2_{3}
\label{eq:bs_metric}
\end{equation}
where $z$ is now an interval as the string is wrapping an $S^1$. In
contrast to this elegant form, the 5-dimensional conformal gauge
metric is far less convenient. The second reason is that the metric
perturbation dies away faster, the higher the dimension. In 5
dimensions $C \sim \ln r / r$ whereas in 6 dimensions $C \sim 1 / r$.
As the lattice must be cut off at a finite $r$ for practical
calculations, we expect to get better accuracy for the faster fall
off.

When we deform the geometry from the uniform black string we will wish
to characterise the geometric deformation. Following Gubser we will
use the quantity,
\begin{equation}
\lambda = \frac{1}{2} \left( \frac{\mathcal{R}_{max}}{\mathcal{R}_{min}} - 1 \right)
\label{eq:def_lam}
\end{equation}
where $\mathcal{R}_{max}$ is the maximum radius of the 3-sphere at the
horizon and $\mathcal{R}_{min}$ is the minimum. Thus $\lambda$ is zero
for the homogeneous black string. We will consider other geometric
quantities to be functions of $\lambda$, and we take $\lambda$ to
parameterise the path of non-uniform solutions. Thermodynamic
quantities of interest will be the horizon temperature, the horizon
volume and thus entropy of the string, and the mass. In Appendix
\ref{app:gubser_PT} we describe Gubser's perturbation method applied
to the 6 dimensional string in conformal gauge. Using this
construction we gain valuable information about the asymptotic
behaviour of the metric. It also allows us to test how well our
non-linear method performs by directly comparing solutions for small
$\lambda$. In addition we can see the range of validity of the
perturbation results when $\lambda$ becomes of order unity.

Gubser's method generates a finite set of ordinary differential
equations at each order in the expansion, and these are solved using
shooting methods. This is achieved by decomposing the metric functions
into Fourier components as,
\begin{align}
X_i(r,z) & = \Sigma_{n=0}^{\infty} \bar{\lambda}^n X^{n}_i(r) \cos( n K z )  
\nonumber \\ \nonumber \\
X^{n}_i(r) & = X^{n (0)}_i(r) + \bar{\lambda}^2  X^{n (1)}_i(r) + \ldots 
\nonumber \\ \nonumber \\
K & = k^{(0)} + \bar{\lambda}^2  k^{(1)} + \ldots
\label{eq:fourier_decomp}
\end{align}
for $X_i = \{ A, B, C \}$. In fact $X^{0 (0)}_i = 0$, and we choose to
normalise the solution such that $C^{1 (0)} = 1$.  Note that the
perturbation series expansion parameter, $\bar{\lambda}$, only agrees
with the definition of $\lambda$ above in \eqref{eq:def_lam} to
leading order.  We will consider the horizon to be fixed at $r = 0$,
and therefore the perturbations to $A, B, C$ to be finite there. The
form of the expansion implies the lines $z = 0, \pi / K$ demark a half
periodic domain. Thus the horizon and periodic boundaries have been
specified in position. The only remaining residual coordinate freedom
in the conformal gauge is the periodicity, given by $K$, which
determines the proper length of the asymptotic $S^1$.

The accuracy of Gubser's method is high for calculating quantities of
interest up to third order in $\bar{\lambda}$. However it is not easy
to extend the method to higher orders. The task of extracting the
Einstein equations becomes exponentially more difficult, plus the
shooting problems become harder and errors will inevitably build up
\cite{Gubser}. Thus in order to even approach the fully non-linear
regime where higher order corrections become important, appears a
difficult task. Again this is important motivation for developing a
fully non-linear method. Furthermore, whilst one may perform a high
order perturbation expansion in $\bar{\lambda}$, which is
approximately equal to $\lambda$ for small perturbations, this tells
us little about the range of $\lambda$ where the low order
perturbation series approximates the non-linear solution.  Assuming
the solution exists for infinitesimal $\bar{\lambda}$ a crucial
question is what is the radius of convergence in $\bar{\lambda}$.
Does the solution have non analytic behaviour in $\bar{\lambda}$?
Should we expect the solution to be good up to $\lambda = 1$? Or
$\lambda = 10$?  Or $\lambda = 0.1$? All these numbers are
approximately order unity but would lead to very different physical
behaviour.  These are all questions which the non-linear method we
employ later will answer.

Let us now consider applying our non-linear method to the black string
case. Firstly we have the translationally invariant black string
solution which we take as a background to deform about, using the
metric \eqref{eq:bs_metric}. The Einstein equations take the form
discussed in \eqref{sec:general_method}. Without loss of generality we
may choose $m = 1$ for the uniform string background.  However, the
mass of this background is changed via the static perturbation mode,
$A \rightarrow A + \Delta$ and $C \rightarrow C - \Delta$ where,
\begin{equation}
\Delta = \frac{1}{2} \ln\left(\frac{m + r^2}{\tilde{m} + r^2}\right)
\label{eq:change_mass} 
\end{equation}
which scales the mass by $\tilde{m}/m$. This is compatible with our
boundary conditions, which fix the asymptotic length of the $S^1$.  Thus we
are free to choose any value for $m$. Fixing the length of the $S^1$,
the mass per unit length of the string can always adjust itself
through $A, C$. This is seen explicitly in Gubser's perturbation
theory as the freedom to add in the $z$ independent modes at second
order, by the choice of $C^{0 (1)}$.

In practice we choose the asymptotic $S^1$ length to be close to the
period of the critical uniform string with $m = 1$. Then, at least for
small non-uniformity we expect the asymptotic form of the metric to
involve a `minimal' perturbation of $A, B, C$. We see this explicitly
in the later section \ref{sec:different_L}, and in figure
\ref{fig:different_L}. Choosing a large disparity between the $S^1$
radius, and the critical period, would mean the $z$ independent
asymptotic modes of the solution would swamp the $z$ dependent modes.
The smaller $A, B, C$ are, and the smaller their gradients, the more
numerical accuracy we can expect.  Thus it is beneficial to choose the
$S^1$ length to be compatible with $m$ in the sense that the $z$
dependent modes and independent asymptotic modes will be similar in
magnitude.

We will use the conformal invariance to choose the horizon to be at $r
= 0$, compatible with having finite $A, B, C$ in our choice of metric
\eqref{eq:bs_metric}. Furthermore we will choose the periodic
boundaries to be at $z = 0$ and $z = L$, and take periodic boundary
conditions there. Numerically we will impose the `asymptotic' boundary
conditions at a finite, but large $r$, and of course check that the
solutions are insensitive to this value. We now discuss these boundary
conditions.

%
\subsection{Constraint Structure and the Asymptotic Boundary}
\label{sec:asym_bound}
%

Let us now consider the non-linear constraint structure
asymptotically.  The constraint structure in equation
\eqref{eq:bianchi} implies that the constraints multiplied by $g =
\det{g_{\mu\nu}}$ must obey Laplace equations. The form of the
constraint equations $\crz$ and $\crrzz$ guarantee that provided $A,
B, C \rightarrow 0$ as $r \rightarrow \infty$, the equations are
satisfied.  However, we should worry that although the constraints are
asymptotically satisfied, the measure blows up as $g \sim r^3$ at
large $r$ and might compensate this so that the product of the
constraint with the measure is finite.

We note that $\crz$ is satisfied if the metric has no $z$ dependence.
As the $z$ dependence asymptotically dies away exponentially in the
perturbation theory, we expect that the $\crz$ constraint equation
will be guaranteed exponentially well at large $r$. Our strategy will
be to enforce that the measure weighted $\crz$ is also true on the
remaining boundaries, so that the constraint structure implies that
$\crz$ is zero in the interior and that $\crrzz$ must be a constant.
Thus we must impose $\crz$ at the horizon boundary as we discuss in
the next section, and ensure that periodic boundary conditions are
imposed at $z = 0, L$. Note that we do not explicitly enforce $\crz$
at the periodic boundaries as the periodic solution to the Laplace
equation with zero data at $r = 0, \infty$ is uniquely zero.  This
still leaves the $\crrzz$ constant as a potential worry.  Unlike
$\crz$, there is no reason for $\crrzz$ to be asymptotically satisfied
even if the metric has no $z$ dependence.  This is why we choose to
explicitly satisfy $\crz$ on all other boundaries, rather than
$\crrzz$. We must still impose $\crrzz$ at a point on one of the
boundaries to ensure that $\crrzz$ is actually zero. We will do this
at the asymptotic boundary.  Let us discuss how to impose data to
ensure that $A, B, C$ go to zero and that $\crrzz$ is satisfied.

Boundary conditions asymptotically are indicated by the perturbation
theory in Appendix \ref{app:gubser_PT}. We expect $A, B$ to be
independent of $z$, going as $A \sim \mathrm{const} + a_0 / r^2$ and
$B \sim \mathrm{const} + b_0 / r^2$.  We must impose a condition on
each as we are solving a boundary value problem for $A, B$ using the
interior equations. We choose to set the constants in the asymptotic
form of $A, B$ to zero. As $B \rightarrow 0$ this choice ensures that
we set the asymptotic length of the $S^1$ simply using the period of
the $z$ coordinate, $L$.  Thus a crude boundary condition is simply to
impose that $A, B = 0$ on the asymptotic boundary. As the boundary
must actually be placed at finite $r$, we can do better by requiring
that $A, B$ decay as $1 / r^2$ giving a mixed Neumann-Dirichlet
condition.

The boundary condition for $C$ is considerably more subtle. In the
linear theory we find $C \sim c_0 \frac{1}{r^2} + c_1 \frac{1}{r}$.
Again there is no $z$ dependence, but there do remain the two
constants $c_0, c_1$ from the $z$ independent mode. So while imposing
$C = 0$ asymptotically does specify data for the $z$ dependent modes,
selecting only the exponentially decaying ones, the $z$ independent
mode is not constrained. We now understand that the previously
discussed constant value that $\crrzz$ will take is exactly determined
by the relation between the constants $c_0$ and $c_1$. Thus the
$\crrzz$ constraint equation relates these constants.  In practise we
use the $G^{z}_{~z}$ equation to determine the $z$ independent $C$
component in terms of $A, B$ on the asymptotic boundary, as described
in Appendix \ref{app:details}.  Assuming the interior equation
$\grrzz$ is satisfied, this is equivalent to using the constraint
$\crrzz$.

For the majority of the data presented in later sections we impose the
asymptotic boundary at $r_{max} = 6$. However, we also test the
sensitivity of this in Appendix \ref{app:boundary} and, as we would
have hoped, find that there is little sensitivity to the position of
the boundary providing it is approximately equal to or above this
value.

%
\subsection{Horizon Boundary}
\label{sec:horizon_bound}
%

A power expansion in $r$ of the interior equations results in boundary
conditions at the horizon for regularity, namely that $\partial_r A,
\partial_r C = 0$, although we note that they do
not require $\partial_r B = 0$. However, in addition to the interior
equations, finiteness of the constraints give conditions. For $\crz$
we find,
\begin{equation}
( \partial_z A - \partial_z B ) \mid_{r=0} = 0
\label{eq:horizon_const}
\end{equation}
as the constraint diverges as $1 / r$ multiplied by $(\partial_z A -
\partial_z B)$. Physically this is the condition that the horizon
temperature is a constant.  Similarly $\crrzz$ diverges as $1 / r$
multiplied by $\partial_r B$, so $\partial_r B = 0$ is indeed a
condition.

Thus it appears naively that specifying one condition on each metric
function on all the other boundaries, we have over constrained the
problem as we have 4 conditions for 3 metric functions at the horizon.
However we have seen that the two constraints do not need to be
imposed everywhere. One must be enforced at all points on the
boundary, and the other only at one point, due to the constraint
structure \eqref{eq:bianchi}. From the previous section we know that
we must just satisfy $\crz$ on the horizon, and then both constraints
will be true \emph{everywhere}, remembering the one point where
$\crrzz$ is enforced was chosen to be on the asymptotic boundary.

Thus we use \eqref{eq:horizon_const} to determine $B$ on the horizon.
Now consider the constraint structure at $r = 0$. The measure factor
in \eqref{eq:bianchi} goes as $g \sim r$.  Imposing
\eqref{eq:horizon_const}, $\crz$ will no longer diverge but tend to a
constant as $r \rightarrow 0$, and then the measure multiplied by
$\crz$ will indeed be zero as required. As discussed, the remaining
constraint $\crrzz$, multiplied by $g$, must then be zero everywhere
too, and this implies $\partial_r B = 0$ at $r = 0$.

Note that whilst $\partial_r A, \partial_r C = 0$ must be imposed at
$r = 0$ during the stages of relaxation, or else the source terms in
the interior equations will be singular, we do not need to impose this
for $B$. This is very important, as violation of the constraints are
inevitable during relaxation, and small violations will be present in
the final solution due to numerical error, so $\partial_r B$ will not
be exactly zero.

To summarise, imposing $A, C$ to be even at $r = 0$ allows us to solve
the interior equations. Data for the remaining metric function $B$ is
used to satisfy $\crz$. The fourth condition $\partial_r B = 0$ will
be true because $\crrzz$ is zero everywhere, due to $g \, \crz$ being
imposed on all boundaries and $g \, \crrzz$ being enforced
asymptotically.

We now discuss how to deform the solution away from the
translationally invariant black string solution. The corresponding
degree of freedom in the horizon data is visible as the integration
constant in solving \eqref{eq:horizon_const}. In order to implement
this condition we will fix the value of $B$ at some location on the
horizon. Imposing even periodic boundary conditions at $z = 0, L$
means the natural place to specify the deformation of $B$ is at $z =
0$ or $z = L$ on the horizon, and the value will give the value of the
local maximum or minimum of $B$. In practice we shall pick $z = L$,
and choose $B$ to be positive so that it is the maximum. We term this
value $B_{max}$.  Then $B$ is generated along the rest of the horizon
simply by integration of \eqref{eq:horizon_const}.

%
\subsection{Stability of the Relaxation Scheme}
\label{sec:stability}
%

In this section we discuss the stability of the relaxation scheme, ie.
whether we expect to find a solution by the relaxation method. 

The first point to note is that the relaxation procedure we employ
here is not based on an energy functional. If we were minimising such
a functional, we would not find a solution that is dynamically
unstable with respect to the imposed boundary conditions, assuming all
directions on the energy surface are probed, as is likely in practice.
However, we have no such functional to minimise. Instead our method is
equivalent to extremising the action functional that gives rise to the
interior equations. Since we are only extremising, we do not expect to
be able to make statements about the classical stability of solutions
found. A simple example illustrating this is static solutions to the
wave equation on a line interval, with Dirichlet boundary conditions
at each end of the interval.  To find these solutions reduces to
solving the Helmholtz equation. For both negative and positive `mass',
solutions can be found (for generic boundary data) using action
extremisation, or equivalently Gauss-Seidel relaxation on the field
equation, although the negative mass solutions have time dependent
exponential growing modes. Using an energy minimisation we could not
find the unstable solution, simply because the energy functional is
not bounded from below.

Of course our method will be sensitive to static perturbation modes
present in a solution that preserve the boundary conditions, and may
be seen by the solution not converging and `drifting', or dependence
of the final state on the initial `guess'.  There are two obvious
static modes for the strings when the asymptotic $S^1$ has fixed
radius.

There is the critical GL mode that moves one away from the
critical black string along the line of non-uniform solutions. However
this mode involves changing the value of $B_{max}$. Therefore fixing
$B_{max}$ does indeed constrain this deformation, choosing how much of
this mode to include in our solution. This is exactly what allows us
to control the non-uniformity of the solution.

However the generic uniform string does have a mode of deformation,
given earlier in equation \eqref{eq:change_mass}, that preserves the
boundary conditions we will impose, namely that we fix $B_{max}$ and
the asymptotic size of the $S^1$.  We will refer to this as the
`static mass mode'. Ironically this static mass mode means that the
translationally invariant black string is not a convergent solution in
our relaxation scheme.  Setting the deformation $B_{max}$ to zero and
starting with the an initial guess metric with $A, B, C \simeq 0$, and
taking the $S^1$ size so that the $A, B, C = 0$ string is classically
stable, we find that $A, B, C$ go very `quickly' (in iteration time)
to zero, a correct uniform string solution, but then drift `slowly'
via this mode and do not converge to an particular uniform string.
Whilst we have fixed the $S^1$ length, we have no way to fix the mass,
due to the presence of this mode. We discuss the interesting initial
unstable $S^1$ length shortly.

However, this scale invariance is broken once the string becomes
non-uniform. Then the period, and thus the length of the asymptotic
$S^1$, becomes related to the mass per unit length. For example, a
small perturbation of a string with mass per unit length corresponding
to $m = 1$, has proper wavenumber $k^{(0)} = 1.269$. Alternatively,
fixing the asymptotic $S^1$ length should select a specific mass per
unit length for the non-uniform strings. In fact the method is
extremely stable. If we do give a non-zero value for $B_{max}$ to
create a non-uniform solution, the algorithm adjusts the mass per unit
length, via what is asymptotically the mass mode, so that the solution
fits inside the fixed asymptotic length of the $S^1$.  Furthermore, it
means that we can run the method with different $S^1$ lengths, which
is analogous to a change of `scheme' in Gubser's perturbative method,
and allows consistency checks of the solutions as detailed in sections
\ref{sec:different_L} and \ref{sec:solutions}.

Whilst we have no reason to believe that finding a solution using our
method ensures it is dynamically stable, the behaviour of the method
on uniform strings is intriguing.  If we try to relax a uniform black
string solution, $B_{max} = 0$, from a slightly perturbed initial
guess, so that $A, B, C$ are non-zero but small, the behaviour of the
method depends on whether the string compactification period is above
or below the critical length. As mentioned above, if it is below, the
string very quickly (in algorithm iteration time) settles down to the
uniform solution, but then slowly drifts, due to the mass mode.
However, if we start with a period greater than the critical value,
the initial motion of the metric under iteration is for the mass per
unit length of the string to `suddenly' (again, in iteration time)
change to a greater value, via the mass mode so that now $B = 0$, but
$A, C$ are far from zero, making the period less than the new critical
length. Thus the stable black strings are found, ignoring the mass
mode drifting, but the unstable strings cannot be. We might be tempted
to say this shows the method `knows' about the dynamical stability,
and it would be very interesting future work to see if we could infer
information about classical stability from solutions found by the
method.

%
\section{Performance of the Method}
\label{sec:performance}
%

The method, as described above, functions very well. Details of the
implementation are given in Appendix \ref{app:details}. In this
section we discuss the qualitative behaviour of the method. Due to the
rather technical nature of this section, the reader may prefer to skip
straight to the following section \ref{sec:solutions}, which details
the properties of the solutions. Essentially we discuss 3 checks of
the method; direct examination of the constraints, calculating the
same quantities for different $L$, and direct comparison with
perturbation theory for small $\lambda$. There is one further
consistency check we perform later in section \ref{sec:solutions},
where the asymptotic mass is compared for direct calculation and
integration from the first law. Before discussing the checks, we now
make some general comments.

Firstly we consider small perturbations. As discussed in the previous
section, setting $B_{max} = 0$, we cannot relax stable uniform
strings due to the presence of the static mass mode. However, provided
a small but non-zero (the value being dependent on the resolution and
exact details of the algorithm) $B_{max}$ is taken the method
converges very well.

\begin{figure}[htb]
\centerline{\psfig{file=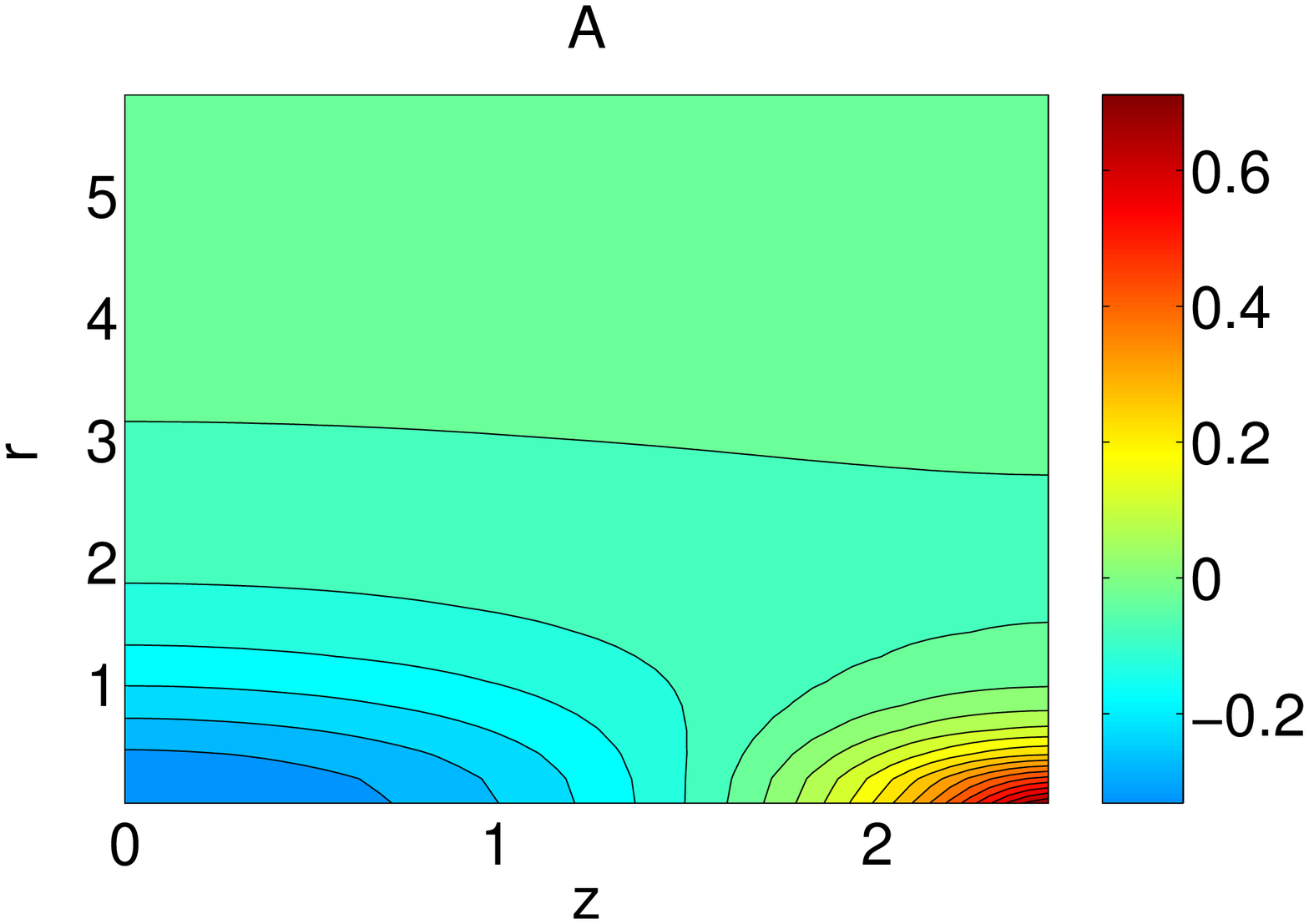,width=3in}\hspace{0.5cm}\psfig{file=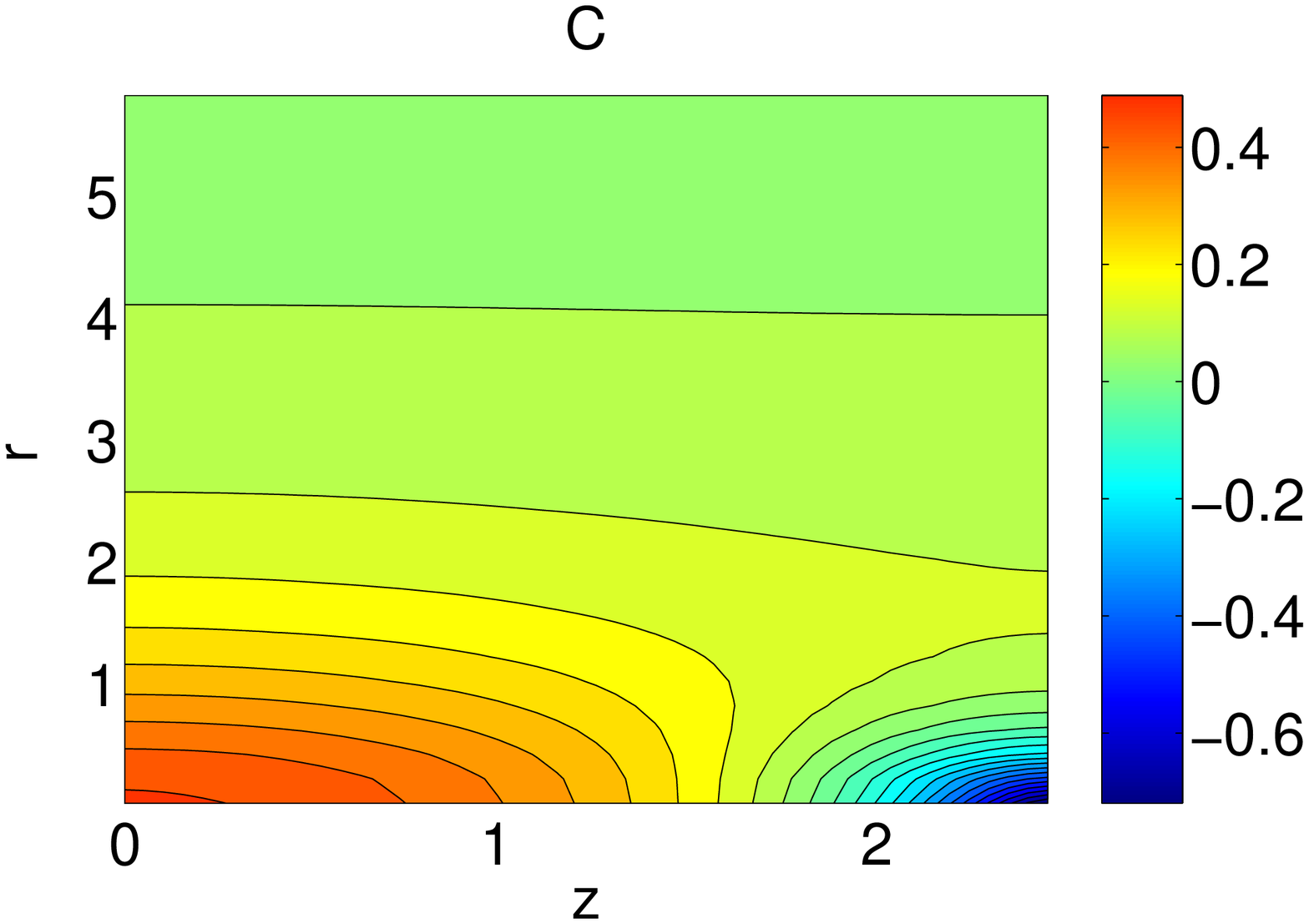,width=3in}}
\centerline{\psfig{file=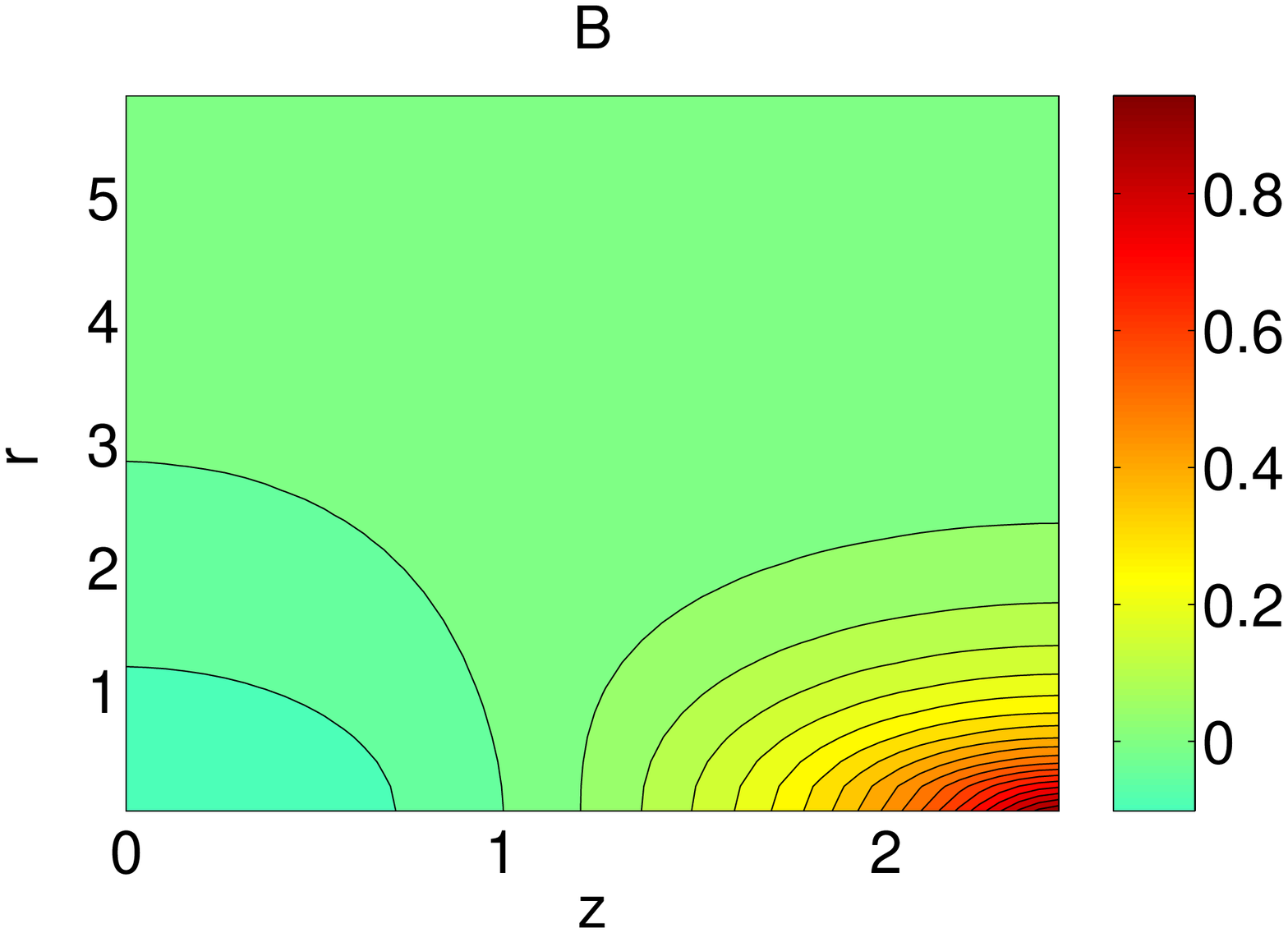,width=3in}}
\caption{ \figuremode
  Contour plots of the metric functions $A, B, C$ for a perturbation
  with intermediate non-uniformity, $\lambda \simeq 1.2$. $B_{max}$,
  the value of $B$ on the horizon at $r = 0$, $z = L$ is in the bottom
  right hand corner in the $B$ plot. Note that the gradients are
  largest in this corner. (Generated using $m = 1$, $L = 2.4758$, with grid
  resolution $240*100$, $dr, dz \simeq 0.025$)
\label{fig:example1} 
}
\end{figure}

To orient the reader, we display the output metric functions, $A, B,
C$, for a moderate value of $B_{max} = 1.0$, giving $\lambda = 1.2$,
in figure \ref{fig:example1}. In the bottom right hand corner, at $r =
0$, $z = L$, $B$ is fixed to $B_{max}$. Having set $m = 1$ in the
background metric \eqref{eq:bs_metric}, the value $L$ is chosen to be
the critical value for $A, B, C = 0$, given by the (half) period of
the Gregory Laflamme zero mode, so $L = 2.4758$. This will be our
`standard' $S^1$ length.  We generate data using this value, but also
check later that using different `schemes' or values of $L$ give
consistent results.

\begin{figure}[phtb]
  \centerline{\psfig{file=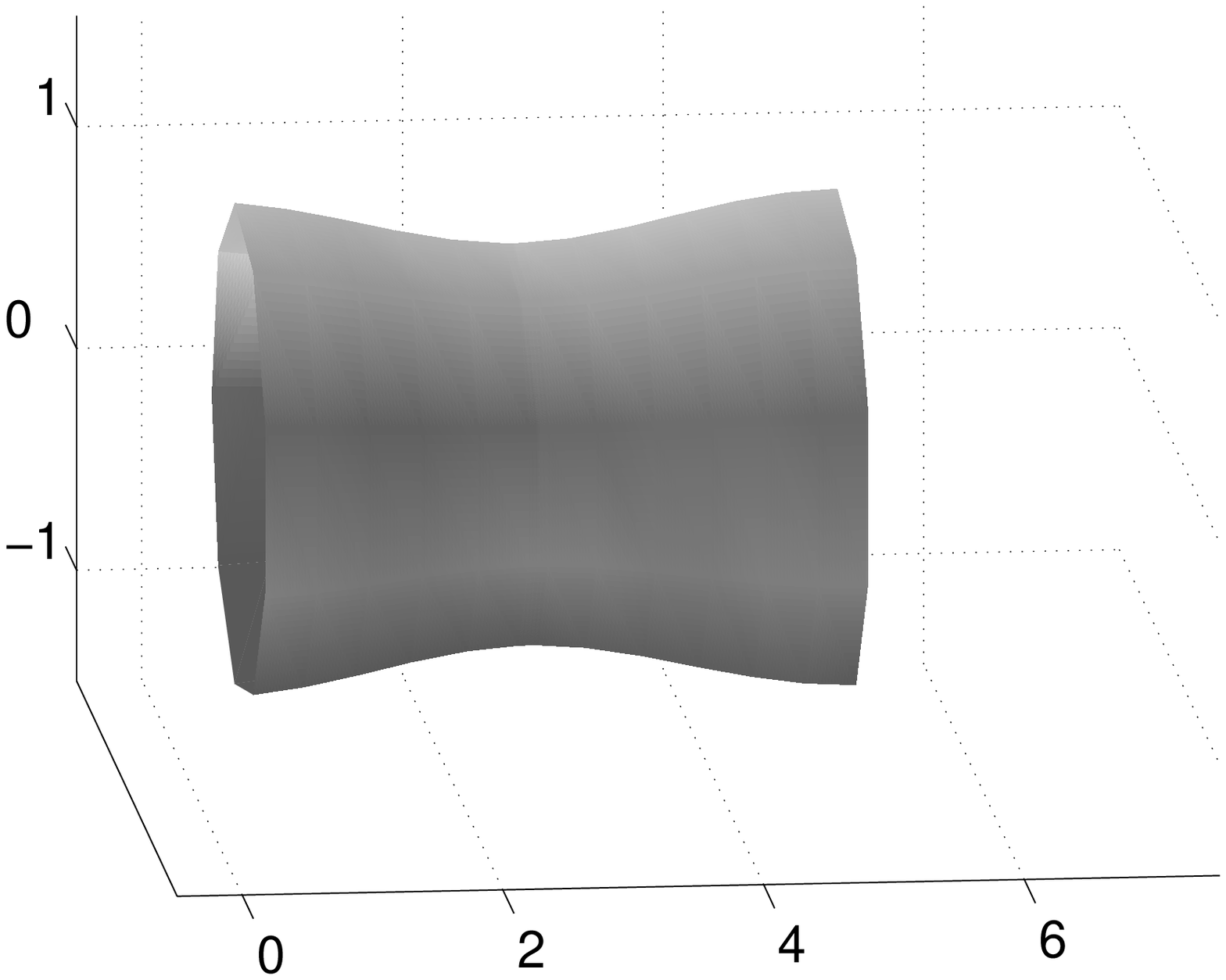,width=3.5in}}
  \vspace{1cm}
  \centerline{\psfig{file=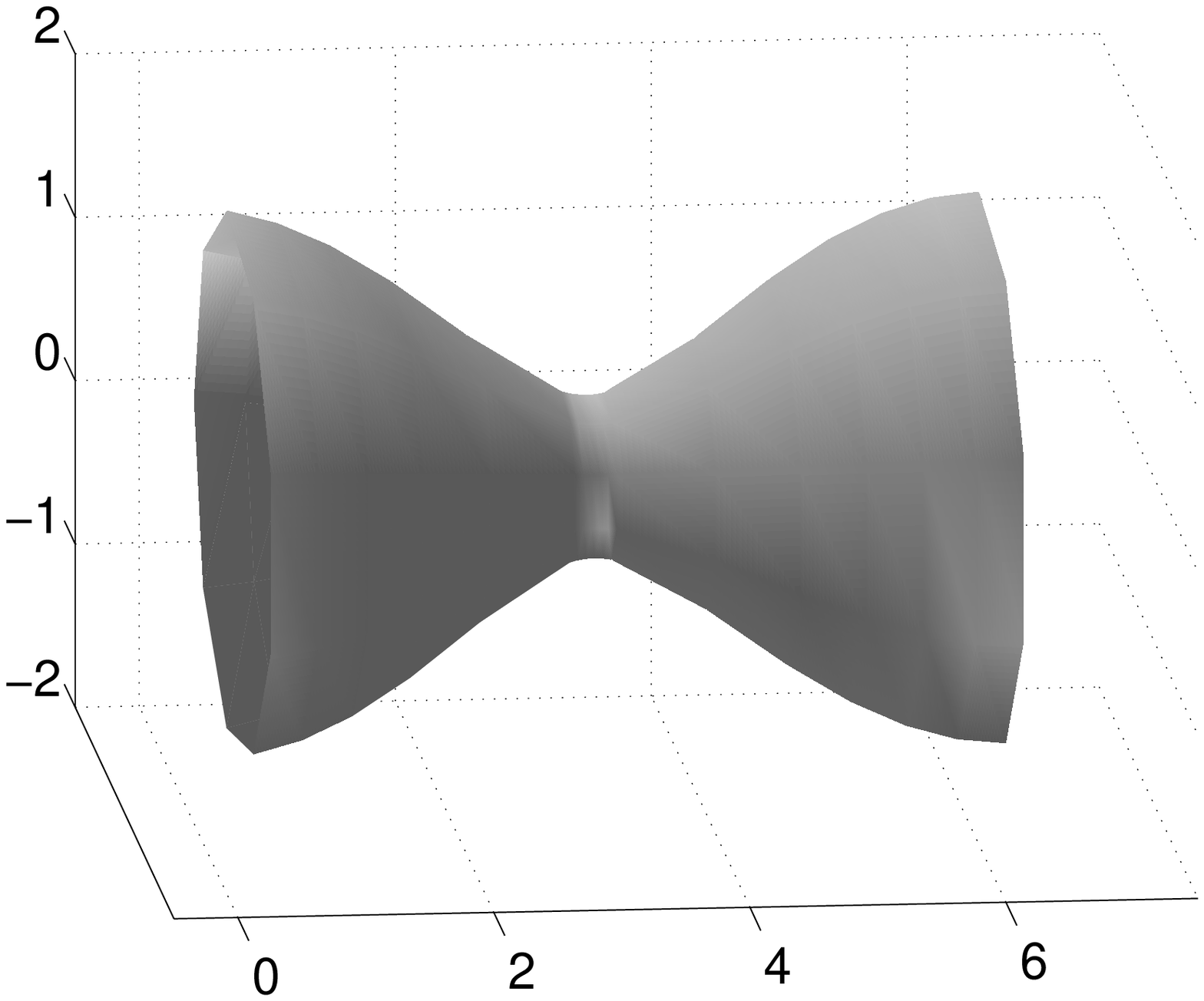,width=3.5in}
    \hspace{0.5cm} \psfig{file=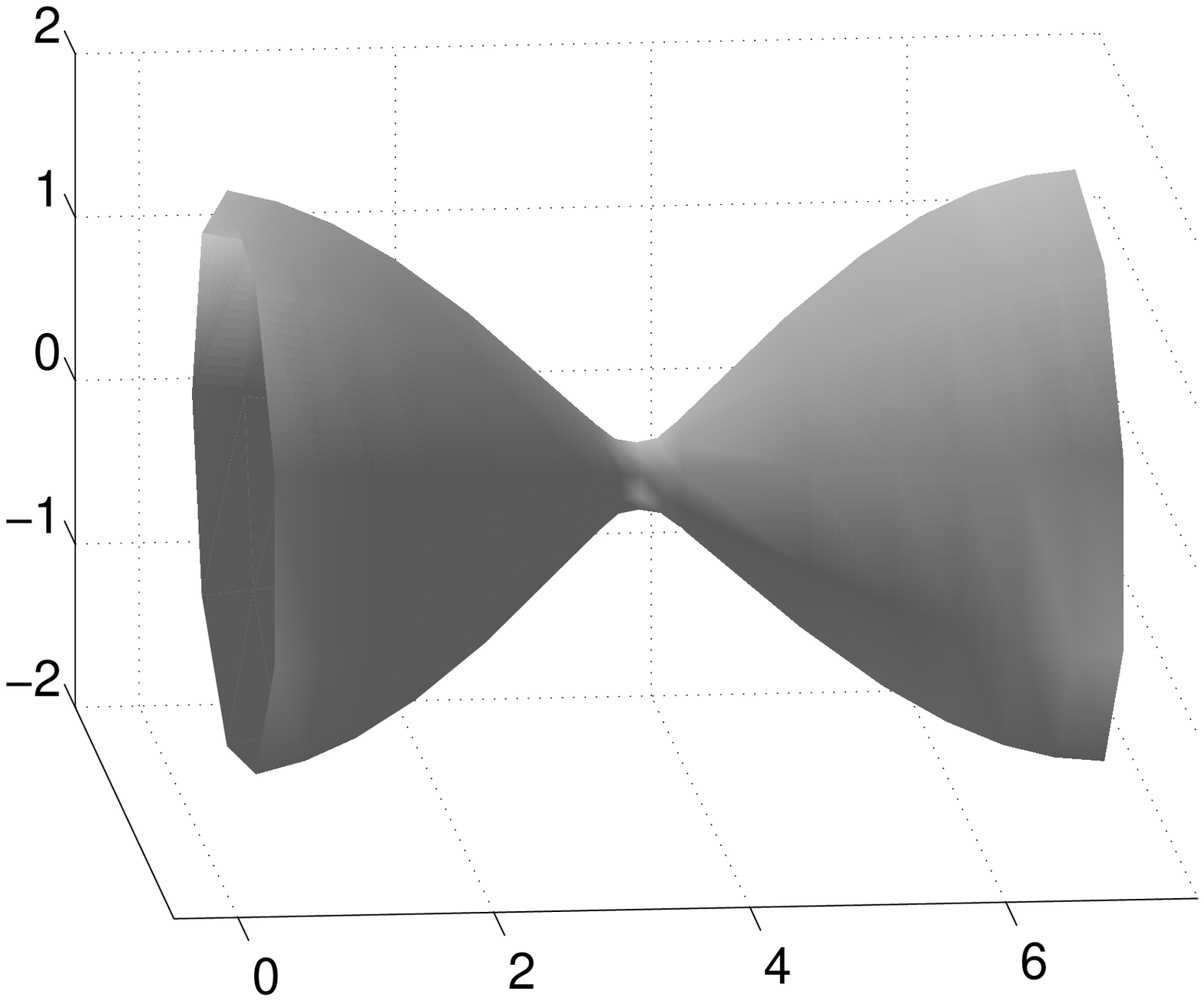,width=3.5in}}
  \vspace{1cm}
\caption{ \figuremode
  Embeddings of the spatial geometry of the event horizon into 3
  dimensional space, projecting out the 2 trivial sphere directions.
  The top plot is for $\lambda = 0.1$, where the leading order
  perturbation theory gives good results. Already for $\lambda = 1.2$,
  the bottom left plot, the deformation is large and in the
  non-perturbative regime. The bottom right plot is of the most
  non-uniform string relaxed in this paper, having $\lambda = 3.9$.
\label{fig:for_fun} 
}
\end{figure}

In figure \ref{fig:for_fun} we show a spatial embedding of the horizon
for this solution. We embed the 5-dimensional spatial horizon geometry
into $\mathbb{R}^3$, having projected out 2 of the trivial sphere directions.
We also do the same for a nearly uniform string with $\lambda =
0.1$ and also the most non-uniform relaxed, with $\lambda = 3.9$.
Hopefully this allows the reader to have a more intuitive view of the
horizon geometry for different $\lambda$.

\begin{figure}[htb]
\centerline{\psfig{file=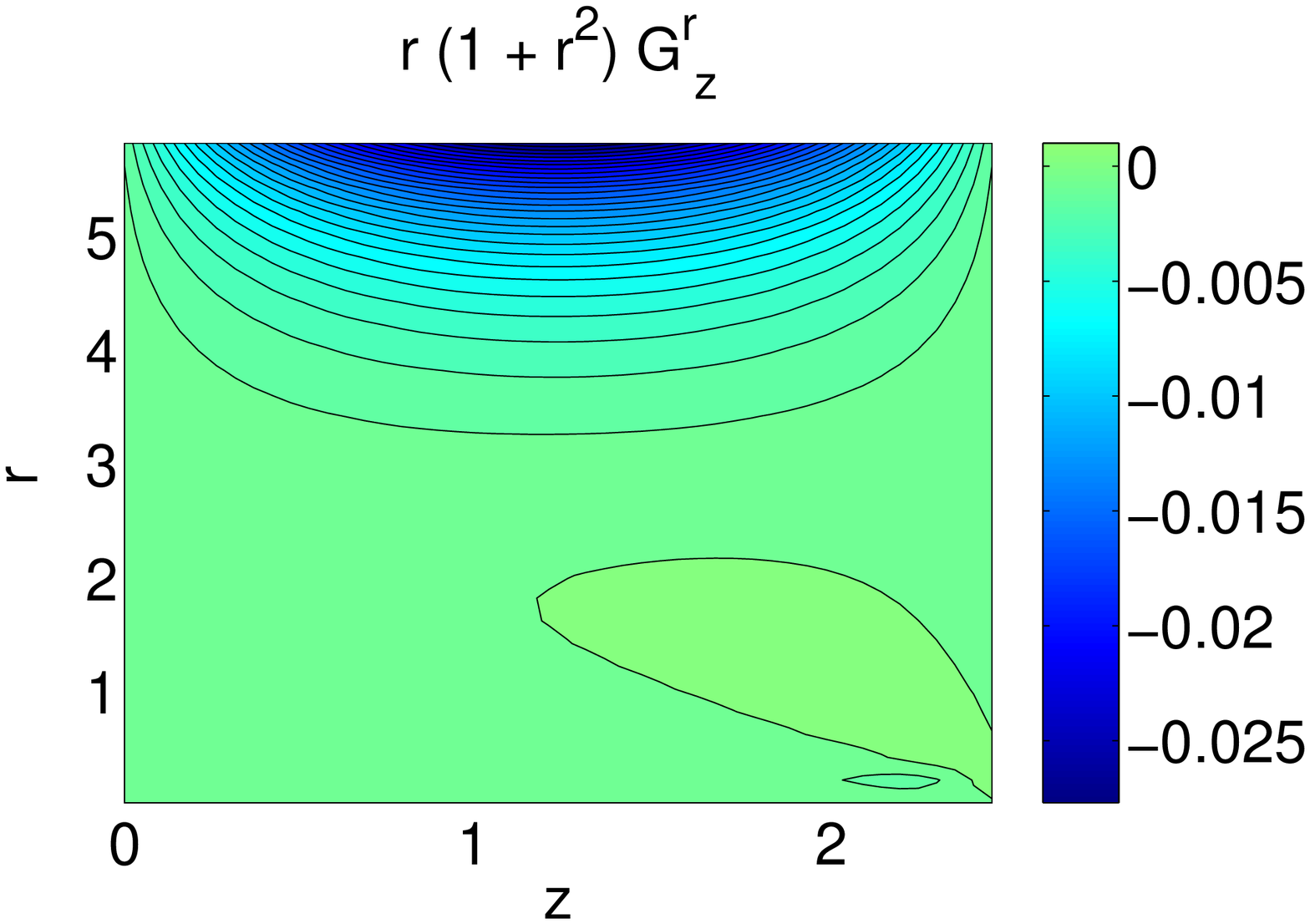,width=3in}\hspace{0.5cm}\psfig{file=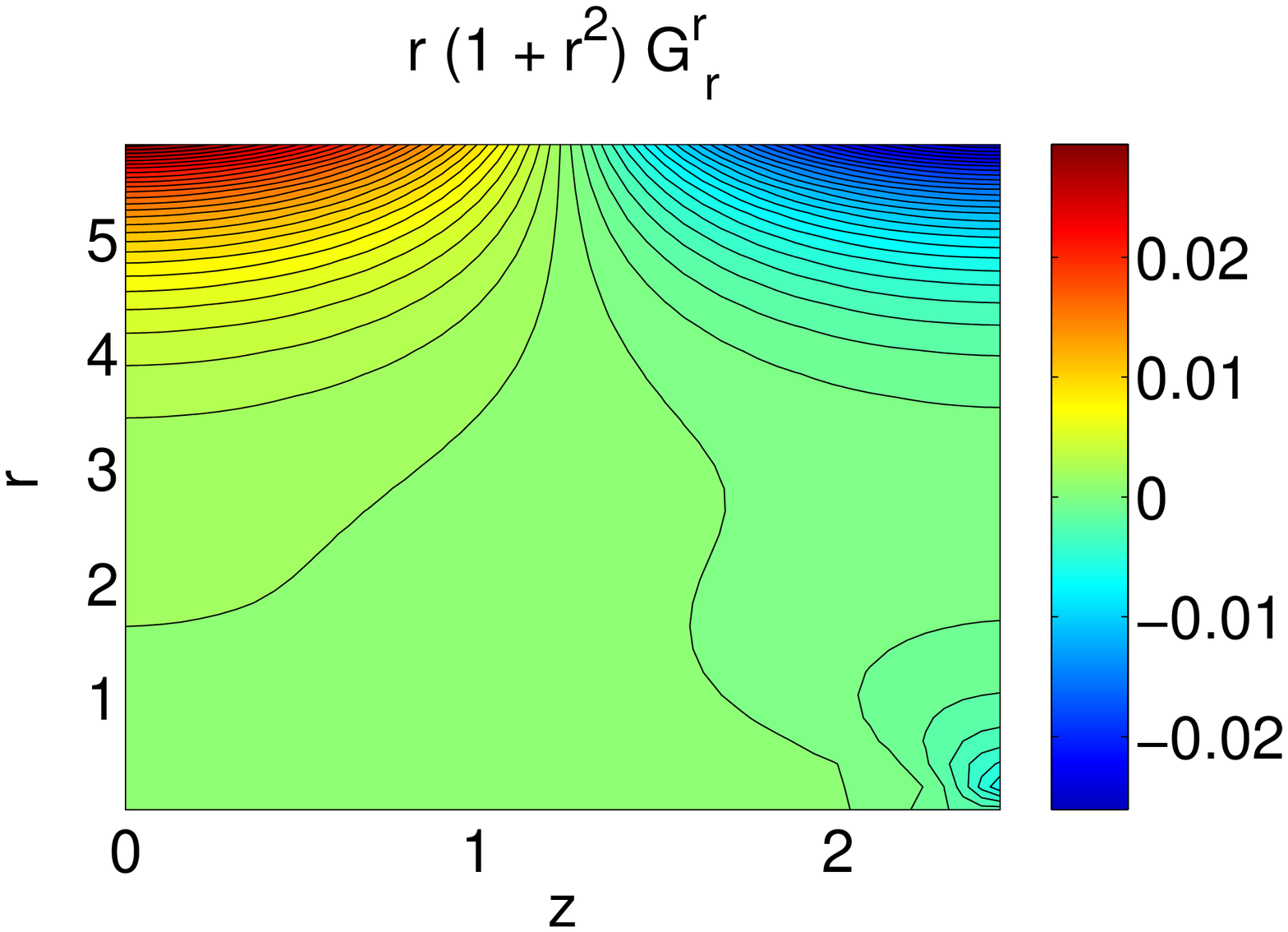,width=3in}}
\centerline{\psfig{file=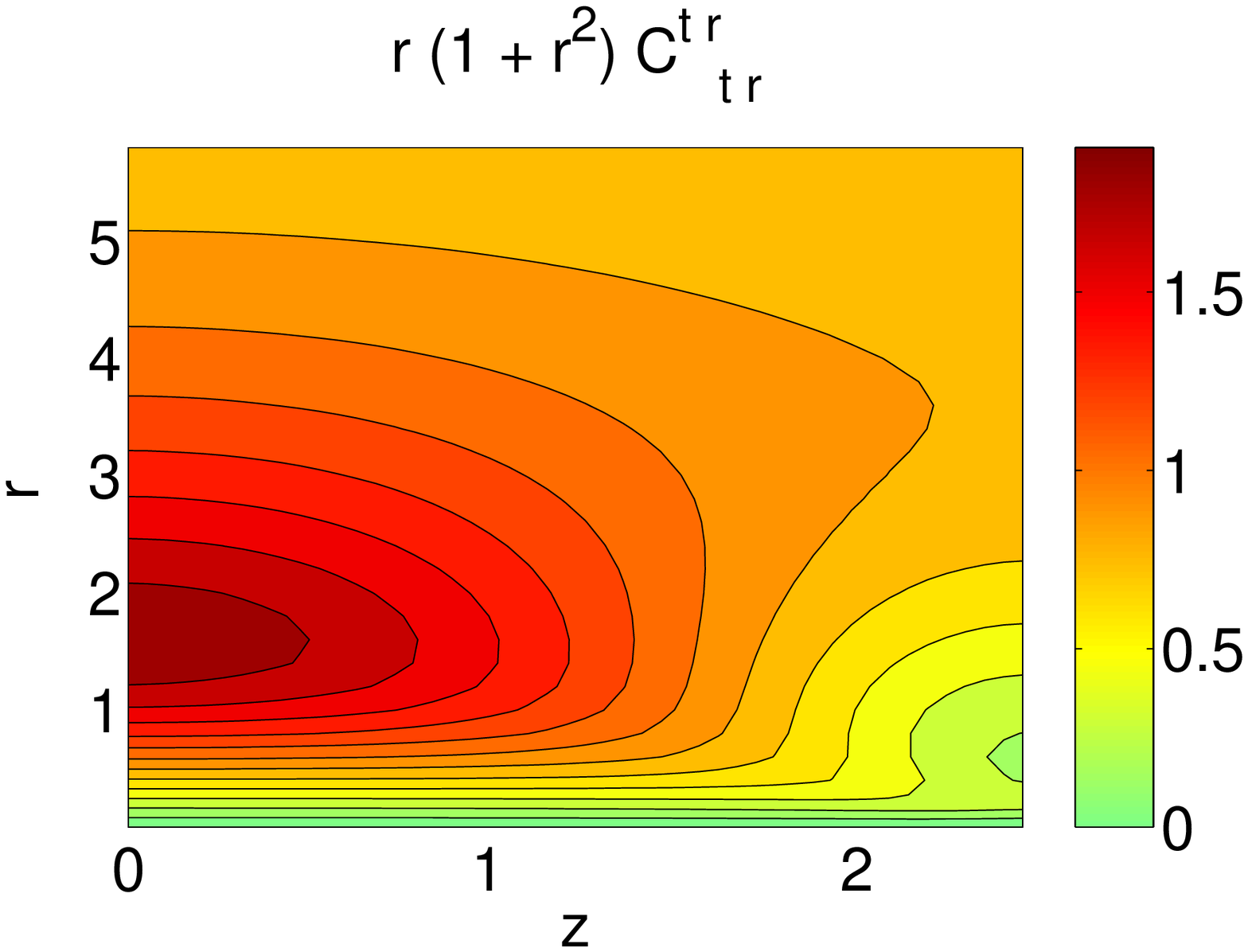,width=3in}\hspace{0.5cm}\psfig{file=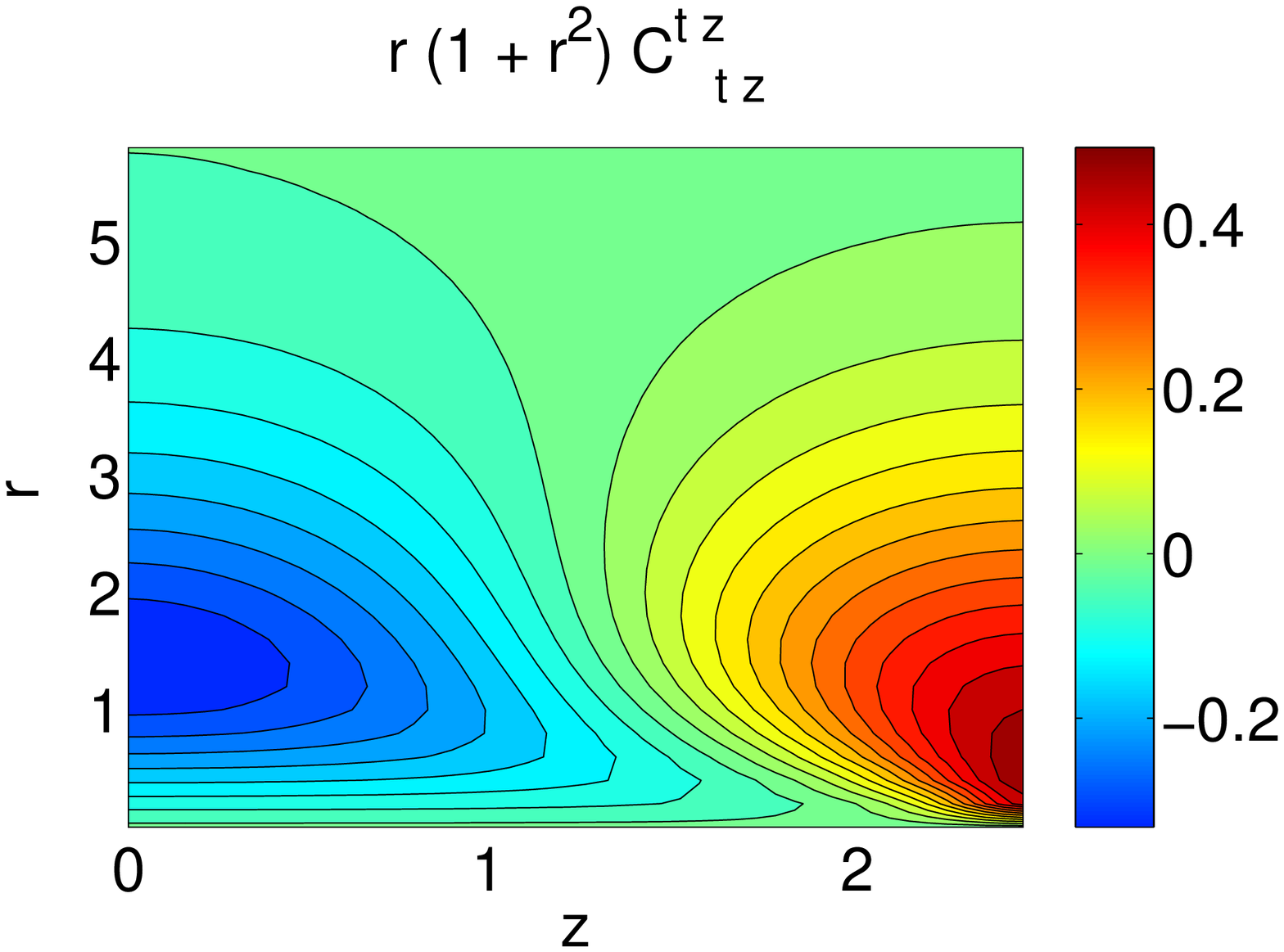,width=3in}}
\caption{ \figuremode
  Top left and right: Again for $\lambda = 1.2$, plots of the measure
  weighted $\crz$ and $G^{r}_{~r}$ respectively. Note $G^{r}_{~r}$ is
  equivalent to the constraint $\crrzz$ due to the interior equations
  being satisfied. These should be compared to two components of the
  measure weighted Weyl tensor, $C^{tr}_{~~tr}$ and $C^{tz}_{~~tz}$ on
  the bottom left and right, with indices as they would appear in
  curvature invariants. The weighted curvatures are of order one, but
  the weighted constraints are much smaller, showing the constraints
  are well satisfied, the peak constraint values being a few percent
  of the typical curvatures.
\label{fig:example2} 
}
\end{figure}

In figure \ref{fig:example2}, for the same $\lambda = 1.2$ solution we
plot the constraint equations, $\crz$ and $\crrzz$.  For comparison we
also plot, 2 of the 4 independent components of the Weyl tensor;
$C^{tr}_{~~tr}$ and $C^{tz}_{~~tz}$. In fact we weight both the
constraints and the Weyl curvatures by the explicit background $r$
dependence of the measure, $r (1 + r^2)$. We then term these
quantities the `measure weighted constraints/curvatures', although
technically we have not really weighted by the actual measure, but
only the relevant factor of it.  We weight the constraints for two
reasons; firstly, because this more closely resembles the quantity
that enters the Bianchi identities \eqref{eq:bianchi}, and secondly,
because even tiny violations of the constraints at the horizon (which
we always expect numerically) result in $1 / r$ divergences.  Thus
multiplying by this factor of the measure converts this unphysical
divergence to a finite value.  Note that the indices of the Weyl
components are raised and lowered as they would appear in curvature
invariants, and that the $C^{tr}_{~~tr}$ component receives a
contribution $3/(1+r^2)^2$ from the background metric. This allows one
to see the significant curvatures due to the non-uniformity, compared
to the homogeneous string background, for this intermediate value of
$\lambda$.

We see that the constraints are very well obeyed over the whole space.
Of course, there is no well defined measure of numerical error, but a
useful rule of thumb is to compare typical, and peak values of the
constraints with the Weyl curvature components, which have both been
multiplied by the measure factor in the same way. We see that the
typical value of the weighted Weyl curvatures is of order one, and the
weighted constraints are far less, their peak values being a few
percent of the typical curvatures.  In fact for this intermediate
value of $\lambda$ we see that the majority of constraint violation is
at large $r$, and results from imposing the asymptotic boundary at
finite $r$. We later show (Appendix \ref{app:boundary}) that this
violation is reduced when we increase this value of $r$, $r_{max}$,
and that for $r_{max} \gtrsim 6$, the metric functions are virtually
unchanged by increasing it. We choose $r_{max}$ to be as low as
possible, and still give good results, simply to decrease
computational time.  For more detailed discussion of constraint
violation with increasing resolution the reader is referred to
Appendix \ref{app:constraints}.  There it is shown that the numerical
constraint violation does indeed improve for a given $\lambda$ as the
resolution is increased, as we expect.

%
\subsection{Self-adjustment and Consistency for Different Choices of $L$}
\label{sec:different_L}
%

As discussed, non-uniform solutions break the scale invariance of the
uniform solutions that allows the mass per unit length and asymptotic
$S^1$ size to be independent parameters. Fortunately our method
automatically finds the correct solution without us having to tune
$m$, the background mass per unit length, for a given asymptotic $S^1$
length $L$.

\begin{figure}[htb]
  \centerline{\psfig{file=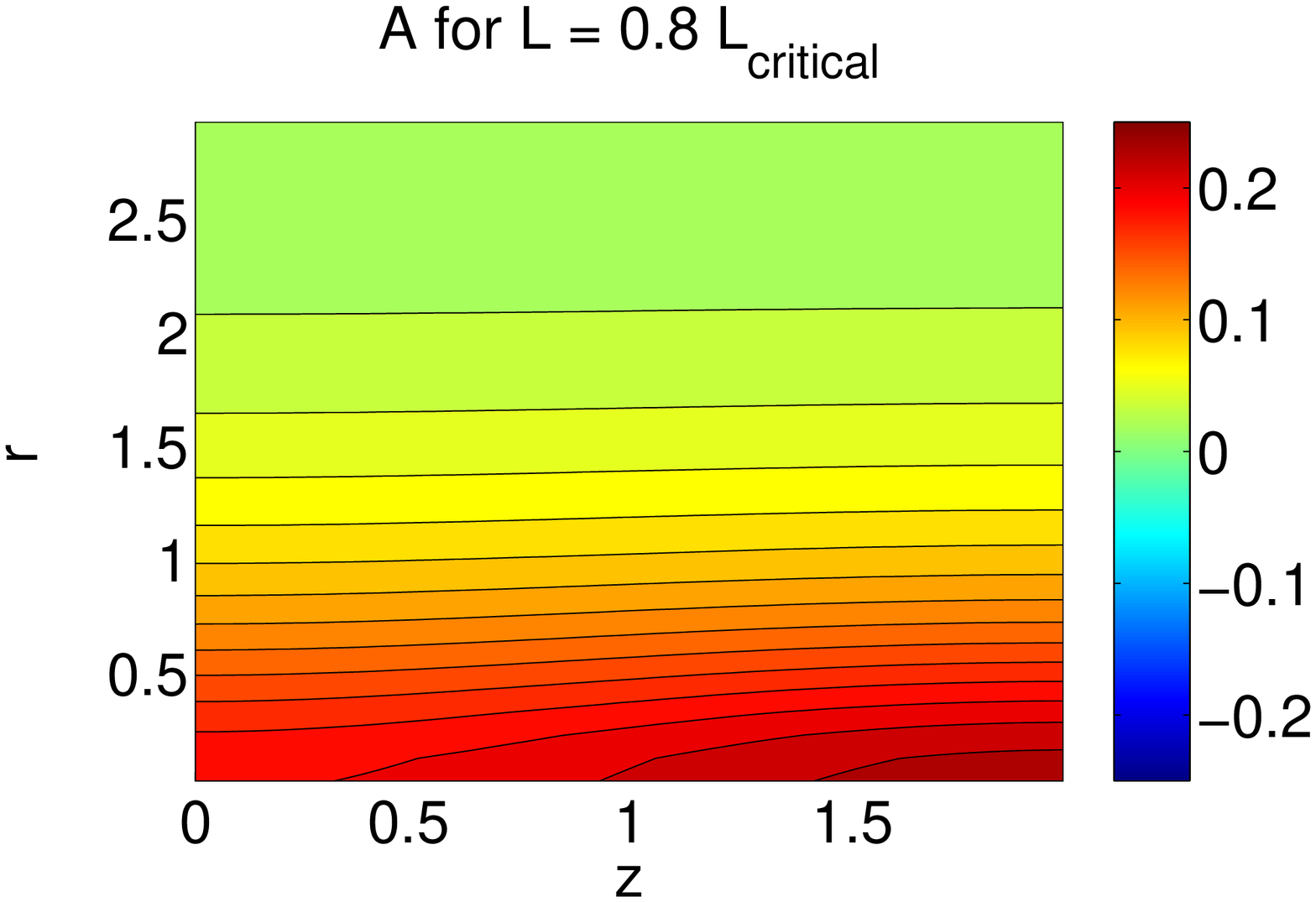,width=3in}\hspace{0.5cm}\psfig{file=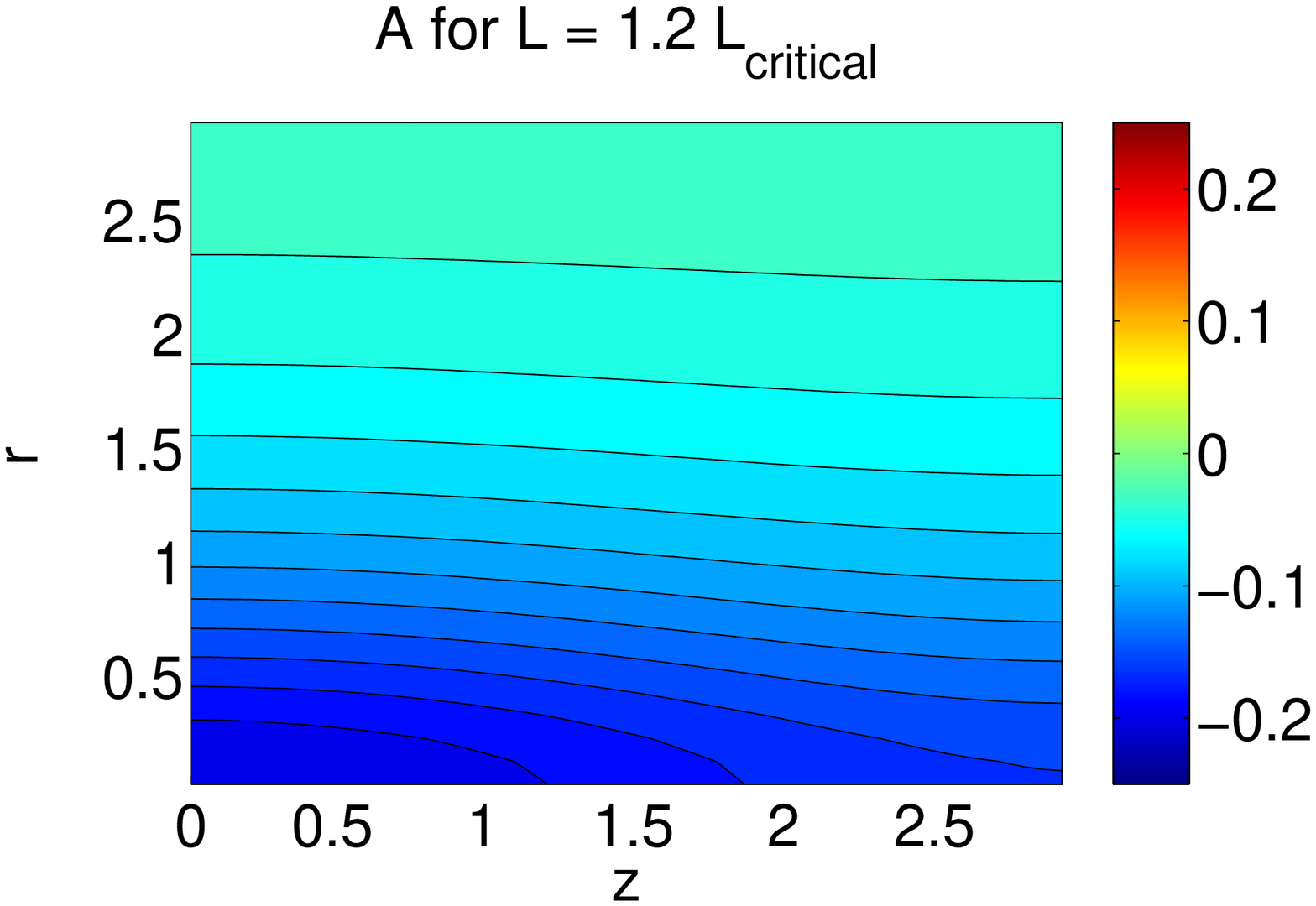,width=3in}}
  \centerline{\psfig{file=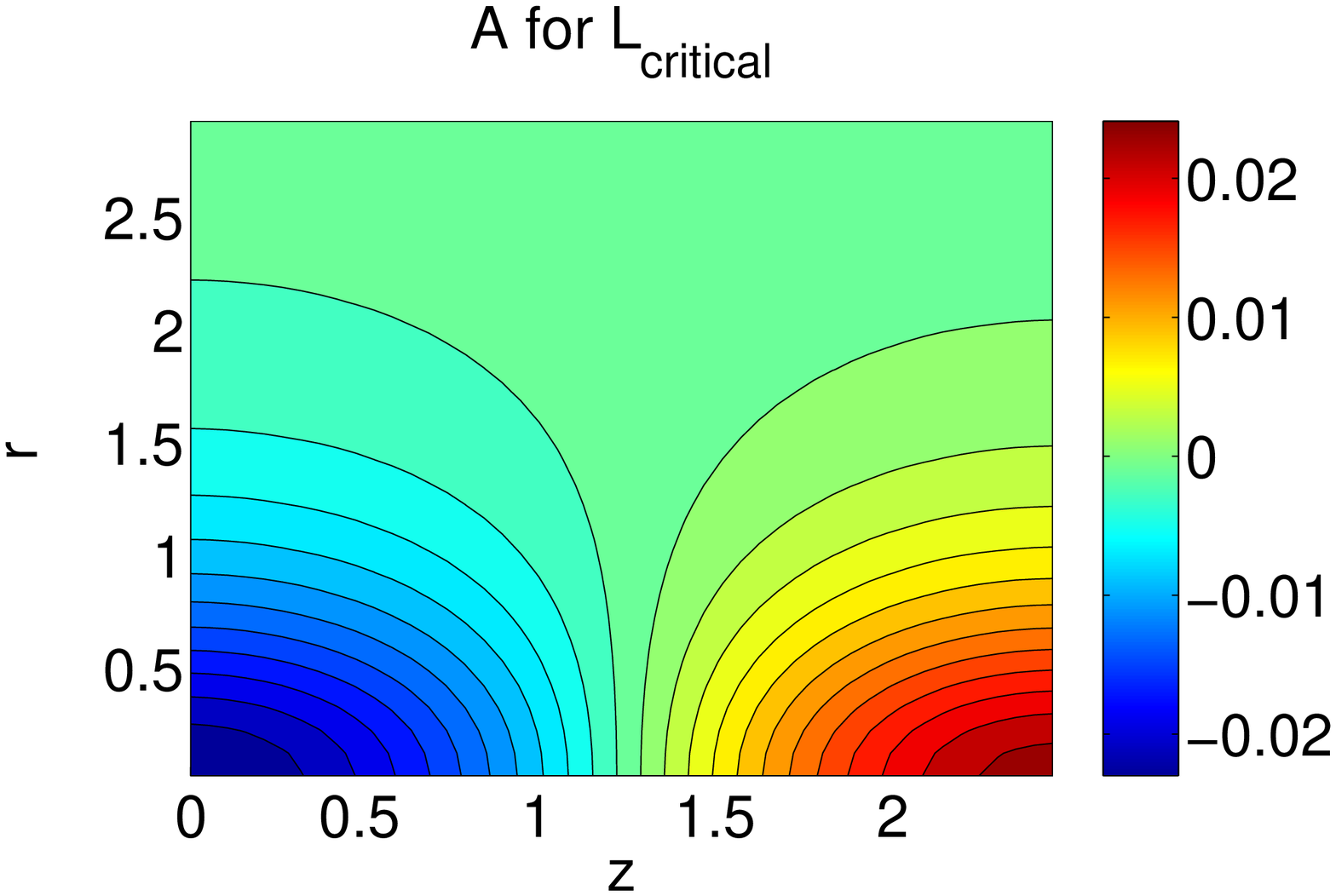,width=3in}}
\caption{ \figuremode
  Plots of the metric function $A$ for $B_{max} = 0.025$. We fix $m =
  1$, the mass per unit length of the background metric, and vary the
  asymptotic $S^1$ length, $L$.  The lower plot uses the critical
  value $L = 2.4758$. The top left uses a value of $L$ $20 \%$ smaller
  than the critical, and the top right uses $20 \%$ larger. One sees
  that the small perturbation in the critical case is simply
  superimposed on the static mass mode background for the two top
  plots. In these cases, the mass mode readjusts the mass per unit
  length so that the solution has the correct periodic length. We find
  we may start with any $L$ and have $m = 1$, and the method works.
  This allows consistency checks to be made, as in the perturbation
  approach, and is analogous to a non-perturbative `change of scheme'.
\label{fig:different_L} 
}
\end{figure}

We illustrate the configurations found fixing $m = 1$ and choosing
different values for $L$ in figure \ref{fig:different_L}.  We have
chosen $B_{max} = 0.025$, so a very small deformation, and have
chosen the critical $L = 2.4758$, the value corresponding to the
periodicity for small deformations for this $m = 1$, and 2 other
values of $L$, 20\% larger and smaller.  We see that the method
beautifully adjusts the mass per unit length, via the static mass
mode, deforming $A$ and $C$ as in \eqref{eq:change_mass}.  On top of
this homogeneous background there is the very small perturbation
imposed.

Thus we can compute all results using several values of $L$.  Indeed
we will use this method in the later section \ref{sec:solutions} to
show the very good consistency of the solutions generated. This is
similar to changing the `scheme' in Gubser's perturbative approach.

%
\subsection{Comparison with Perturbation Theory for Small $\lambda$}
\label{sec:compare_PT}
%

Another check of the method is a comparison with perturbation theory
for solutions near the critical point. By changing the perturbation
`scheme' to fit that singled out by our conformal gauge choice and
asymptotic boundary conditions, we may directly compare the two
methods. The scheme that the non-linear method chooses is that the
asymptotic $S^1$ size is fixed. In order to compare with the
perturbative method we pick this asymptotic size to be the critical
one for the $m = 1$ background solution with $A, B, C = 0$, ie. $L =
2.4758$. In this scheme we must choose $C^{0 (1)}$ so that $k^{(1)} =
0$. This is a shooting problem that we may perform by hand and the
results are given in Appendix \ref{app:gubser_PT}.

We empirically find that the absolute error in quantities calculated
by our non-linear method, inferred by comparing different resolutions,
is approximately constant for different $\lambda$. Therefore choosing
a very small $\lambda$ to compare the perturbation theory and the
non-linear method means that the quantities being measured are tiny,
and thus the fractional errors will be enormous. Conversely, choosing
to compare at a large value of $\lambda$ means that non-linear
corrections will be large, and we only compute fully up to second
order in the perturbation theory. So we choose to use $B_{max}
= 0.1$, corresponding to $\lambda \sim 0.1$, which is not too small,
or too large. For various quantities, the following table shows the
fractional differences between the value calculated in the
perturbation theory and the value given by our non-linear method for 3
different resolutions.
\begin{center}
\begin{tabular}{c|ccc}
  Quantity & 240*100 & 120*50 & 60*25 \\ \hline \\
  $\{A, B, C\}^{1 (0)}$ & * & * & *  \\ \\
  $B^{0 (1)}$ & * & * & *  \\
  $\{A, C\}^{0 (1)}$ & 0.06 & 0.16 & 0.64  \\ \\
  $\{A, B, C\}^{2 (0)}$ & * & * & *  \\ \\
  $\delta \mathcal{S} / \mathcal{S}$ & 0.08 & 0.40 & 1.87  \\
  $\delta \mathcal{T} / \mathcal{T}$ & 0.04 & 0.14 & 0.61  \\
  $\delta \mathcal{M} / \mathcal{M}$ & 0.07 & 0.27 & 1.23
\end{tabular}
\end{center}
Here $\mathcal{S}, \mathcal{T}$ and $\mathcal{M}$ are the entropy,
horizon temperature and mass of the string. The values for the metric
at the horizon were calculated by Fourier transforming the solution
generated by the non-linear method at the horizon, to extract the
various perturbation theory contributions.  We calculate
$\bar{\lambda}$ from the amplitude of the $\cos(k^{(0)} z)$ component
of $C$ at the horizon (as $C^{1 (0)} = 1$), and then use this to
normalise the other components in the expansion
\eqref{eq:fourier_decomp}.  As noted earlier, the $\bar{\lambda}$ in
the perturbation expansion agrees with $\lambda$, defined
geometrically \eqref{eq:def_lam}, to leading order in $\lambda$. The
comparison of values in the table above can only be meaningful to
leading order in $\lambda$, as only the leading order contributions
were calculated in the perturbation theory. A `*' indicates a small
difference, of order the expected size of next order corrections,
estimated by $\lambda^2 \sim 0.01$.

Again, we wish to stress that the fractional errors shown only apply
for this value of $\lambda$. It is the absolute errors that are
approximately constant with $\lambda$, the fractional errors
decreasing as the quantity of interest increases in magnitude. Thus
the fact that the fractional error in the lowest resolution in the
above table is very large for some quantities is merely a reflection
of taking a small $\lambda$ so that the quantity measured is very
small itself. The best way to assess errors due to finite resolution
is in the later figure \ref{fig:properties}, where the same
thermodynamic quantities are plotted for the 3 resolutions. We then
see that the absolute errors are very small, even for the lowest
resolution, $60*25$. For all quantities measured, over the whole range
of $\lambda$, the 3 resolutions converge consistently with second
order scaling.

It is easy to see from the table that the errors in the horizon metric
functions appear to be localised in $\{A, C\}^{0 (1)}$. These are the
$z$ independent modes, and are responsible for the asymptotic
behaviour of the solution at large $r$. The asymptotic boundary
conditions discussed in section \ref{sec:asym_bound} constrain exactly
these modes.  In addition, $\delta \mathcal{M} / \mathcal{M}$ shows
that the mass estimation carries some error too. It is technically
difficult to extract the mass from the metric solution, as firstly,
the $z$ independent component must be extracted by averaging at the
asymptotic boundary, and then this must be integrated to large $r$ to
find the mass (this is detailed in Appendix \ref{app:mass}).  Thus the
quantities with the greatest error are those associated with the large
$r$ asymptotics of the solution, as we might have expected.

What is crucial is that the scaling of the errors towards the
perturbation theory values with increasing resolution appears to be
compatible with a second order scaling. This is exactly what we
expected to find, and indicates that, at least for low $\lambda$ the
method performs very well with no obvious systematics.

%
\subsection{Large $\lambda$}
\label{sec:large_lambda}
%

Let us now turn to our primary interest, the behaviour of the
solutions at large $\lambda$.  The highest resolutions used in this
paper allow $\lambda$ to be as large as $\sim 3.9$. The first issue is
how to relax solutions with large $\lambda$.

The initial `guess' configuration we relax from is very crude, simply
taking the metric functions to be zero everywhere except on the
horizon where we put a crude cosine function for $B$ with the correct
value of $B_{max}$ at $z = L$. Note that we have also checked
that for a different initial guess we obtain the same final solution.
Unsurprisingly, if we start with a moderate $\lambda$, say $\lambda >
0.3$, the method diverges immediately. We must gently increase the
value of $B_{max}$, say in steps of $\delta B_{max} =
0.05$, building up from the small values where the initial convergence
does work. Thus we take the solution $B_{max} = 0.30$ and then
run the algorithm using this as a starting guess, but perturbing
$B_{max}$ to $0.40$.

The next question is then how far we can proceed with this method. We
see in figure \ref{fig:example1} that the conformal gauge allocates
more of the coordinate space to the region where $B$ is negative than
positive. For large $\lambda$ we find that large numerical gradients
accumulate in the $r = 0$, $z = L$ corner of the lattice.  For a fixed
resolution, there appears to be a critical value of $\lambda$ where
the constraints start to become severely violated in this corner. We
believe this is simply due to the lack of resolution for the large
gradients. Continuing to even larger $\lambda$ the method fails to
produce a convergent solution.  If we double the resolution in both
$r$ and $z$, the constraints are reduced and we can proceed to higher
$\lambda$ finding convergent solutions. Thus we see that this lack of
convergence is simply an artifact of the algorithm.  However, it does
mean that we must use higher and higher resolutions to find very
non-uniform solutions.

For the implementation outlined in Appendix \ref{app:details}, a
resolution of $60*25$ fails to give convergence for $B_{max}
\simeq 1.2$, and as this value is approached, the constraints become
increasingly violated in the $r = 0$, $z = L$ corner of the lattice.
Increasing the resolution to $120*50$ allows us to proceed up to
$B_{max} \simeq 1.4$ before similar constraint violations
occur, and convergence breaks down.  To explore larger $\lambda$ we
use the resolution $240*100$, the highest used in this paper, which
appears to give good results up to $B_{max} \simeq 1.9$, above which
convergence fails. The corresponding $\lambda \simeq 3.9$.  See
Appendix \ref{app:constraints} for detailed plots of these effects.

It is difficult to access the physical effect of the constraint
violation.  We might compare the weighted constraints with the
weighted curvatures, as in figure \ref{fig:example2}. However, we note
that the physical properties extracted from the solutions, and shown
in the following sections (for example, see figure
\ref{fig:properties}), agree well for measurements of different
resolutions, even near the maximum $\lambda$ for the lower
resolutions, where convergence breaks down and the constraints are
most violated.  In particular we can see this comparing $60*25$
solutions properties for $\lambda$ near the limit of convergence, with
the corresponding $240*100$ solutions of the same $\lambda$, for which
the constraints are very well satisfied.  While the constraints are
violated there, and the large gradients are not well resolved on the
$60*25$ lattice, the quality of the solutions still appears to be very
good.  This is why we trust the $240*100$ results presented later up
to the point where convergence is lost.

In principle we could proceed to higher resolution than $240*100$.
However the computation time obviously increases with resolution. The
algorithm was implemented using elementary numerical methods.  A
$240*100$ grid would take several days of relaxation time on a typical
desktop PC. We discuss areas of numerical improvement in section
\ref{sec:improvement} which would almost certainly make the next
resolution jump accessible, allowing much larger $\lambda$ to be
found. For the purposes of this paper, this is left for further
research.

In summary we appear to have a consistent non-linear scheme that
converges with increasing resolution, and apparently allows us to
access finite values of $\lambda$ deformation. Ideally we would have
hoped that we could access all available $\lambda$. Whilst this may be
so, it appears that in the implementation of the elliptic method
presented here, the resolution is required to increase for increasing
$\lambda$ to ensure the constraints are well satisfied and convergent
solutions are found. This being said, even with very modest processing
power, and elementary numerical methods, we have been able to probe
large finite values of $\lambda$.

%
\section{Thermodynamic Properties of the Non-Uniform Strings}
\label{sec:solutions}
%

We have demonstrated a numerical method which computes the non-uniform
black string geometries, and using limited resolution and computing
power, have found configurations up to $\lambda \simeq 3.9$. We now
discuss the properties of these solutions. In the first sub-section we
consider the basic thermodynamic quantities that we can compute
robustly, and show various consistency checks. In particular, we will
show that the mass of the non-uniform solutions, for fixed asymptotic
$S^1$ radius, is always greater than the mass of unstable uniform
strings.  In the second sub-section, we discuss quantities which are
harder to determine using the numerical results we have available, due
to requiring differencing of almost equal values, or division of small
quantities.

%
\subsection{Temperature, Entropy and Mass}
%

\begin{figure}[phtb]
  \centerline{\psfig{file=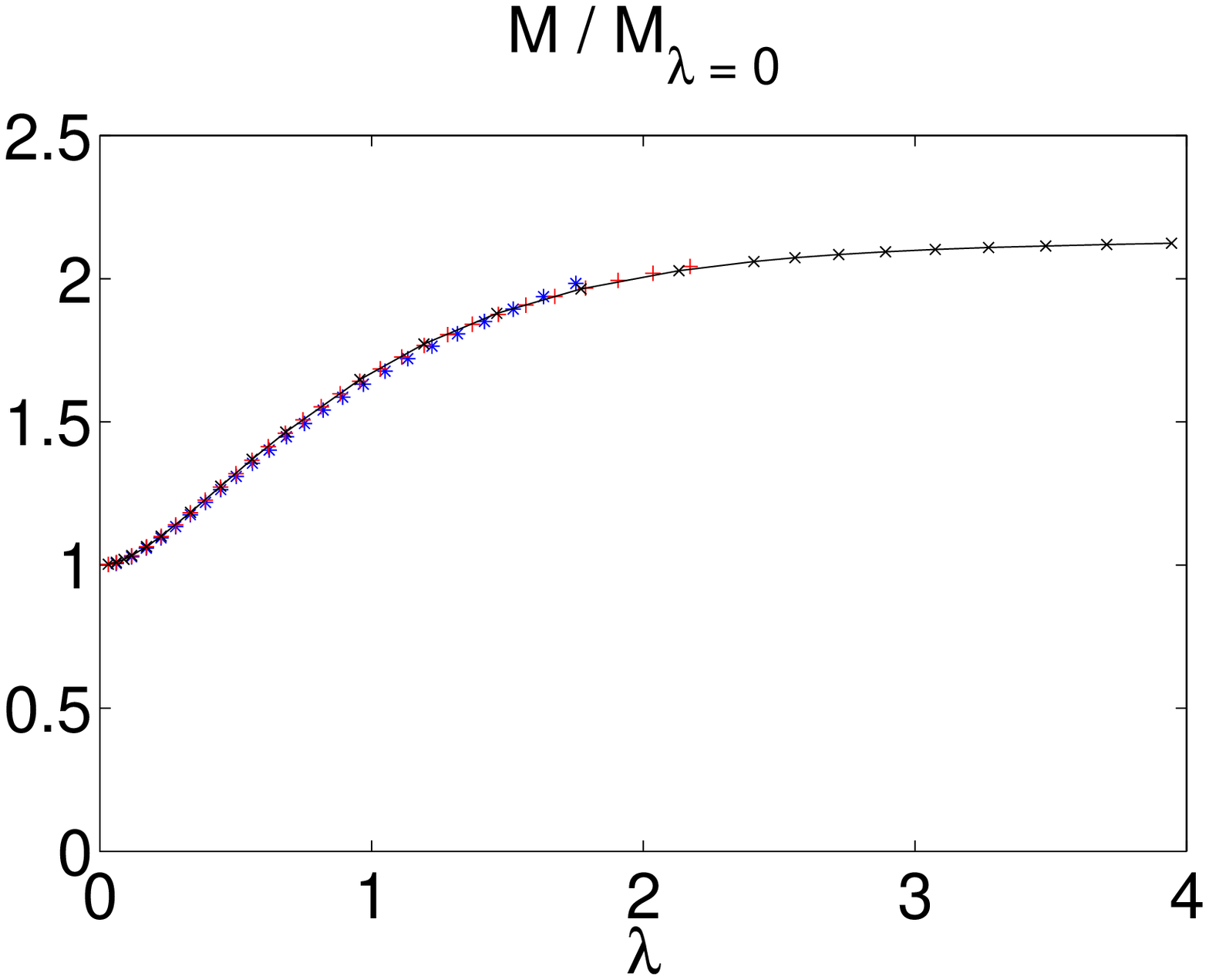,width=3in}\hspace{0.5cm}\psfig{file=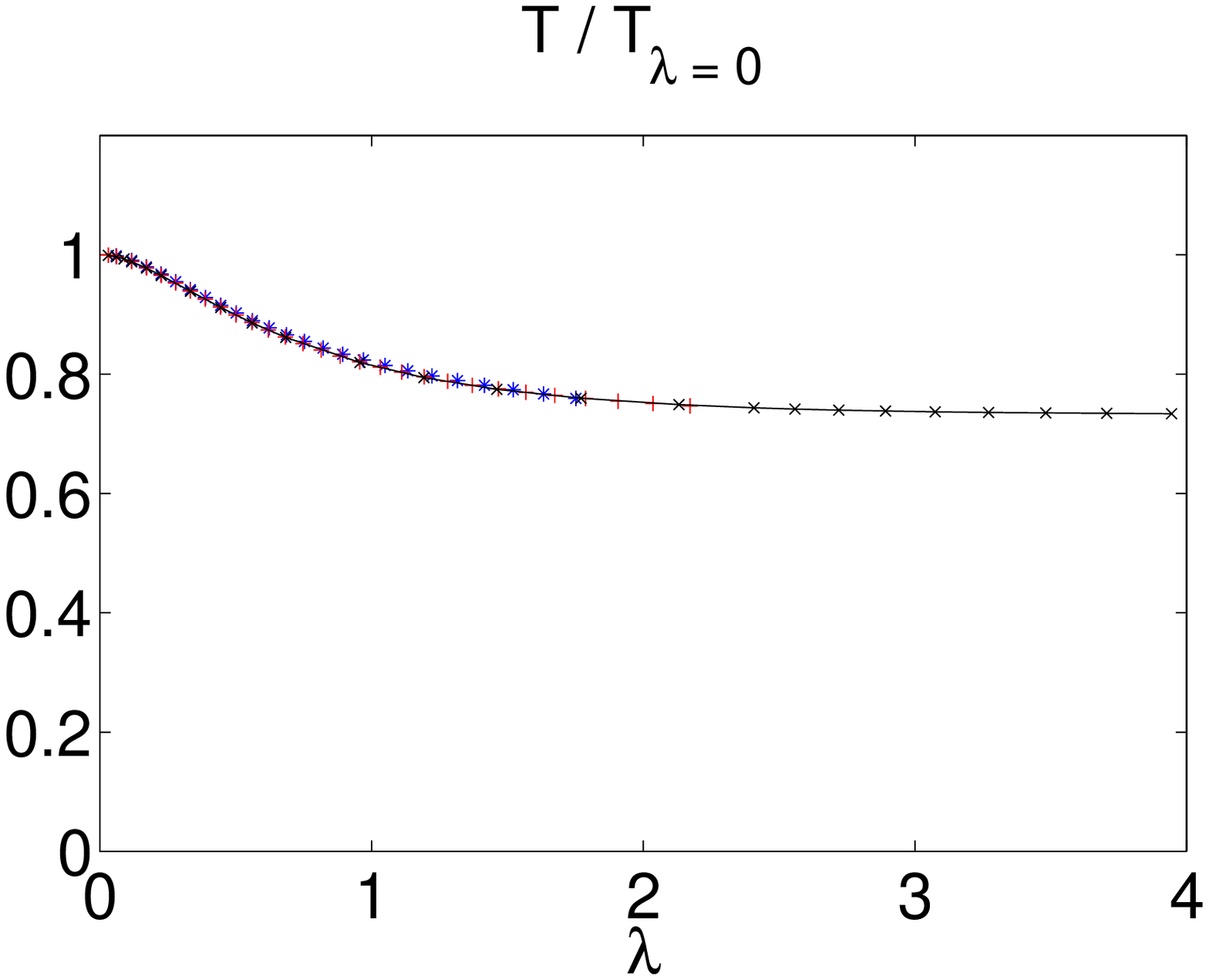,width=3in}}
  \centerline{\psfig{file=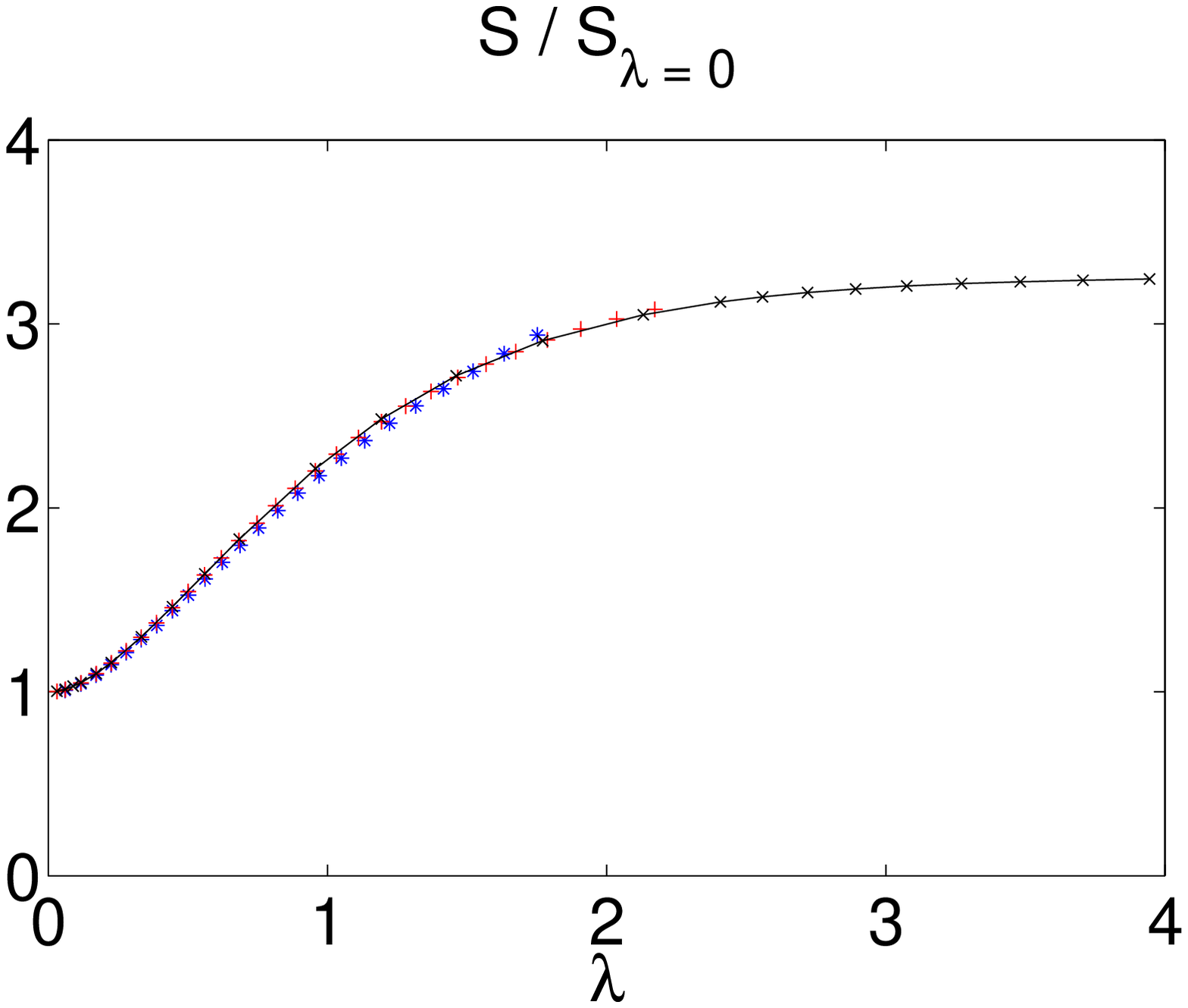,width=3in}}
\caption{ \figuremode
  The key results of this paper: for fixed asymptotic compactification
  length, the thermodynamic quantities for the non-uniform black
  strings plotted against $\lambda$, normalised by their value for the
  critical, $\lambda = 0$, string. We see that all these quantities
  appear to asymptote to a constant for large $\lambda$. In particular
  the mass is always greater than that of the critical uniform string,
  and therefore the non-uniform strings cannot be the end state of
  decay for the Gregory-Laflamme instability.  Three resolutions are
  used, $60*25$ in blue, $120*50$ in red and $240*100$ in black. Very
  good consistency is found, even near the $\lambda$ where convergence
  breaks down for the lower two resolutions, and we would expect the
  lower resolution to have errors due to constraint violation.
  (Results generated using $m = 1$ and the critical value of $L =
  2.4758$)
\label{fig:properties} 
}
\end{figure}

The natural geometric properties of interest are the horizon
temperature $\mathcal{T}$, the horizon volume or entropy
$\mathcal{S}$, and the total mass $\mathcal{M}$, and their variation
with $\lambda$. In figure \ref{fig:properties} we fix the asymptotic
$S^1$ length and plot these quantities normalised by their value for
the critical uniform string.  Three different resolutions, $60*25$,
$120*50$ and $240*100$, are used to generate this data, over the
ranges where the resolution still converges, and we take $m = 1$ and
$L = 2.4758$, the critical $S^1$ length for this $A, B, C = 0$
background. We see that for small deformation, where we can compare
all 3 resolutions, the different resolutions are very consistent, and
are compatible with second order scaling.  Even the accuracy of the
lowest resolution appears to be good.  As we have mentioned, the
reason to proceed to higher resolution is primarily to increase the
range of convergence, rather than to increase the accuracy of the
solution.  Also of interest is that, for a given resolution, the last
points which converge have the largest constraint violations, as
outlined earlier.  We do see small deviations appearing between the
failing lower resolution and the higher resolutions near the end of
converge of the lower resolution.  However the deviations are very
small, and thus we conclude that the physical error due to the
constraint violations is similarly very small. Thus even near the
point where the highest resolution breaks down, we believe that the
curves are accurate.

These plots are the key result of this paper. The plot of the mass
normalised by the critical string mass, $\mathcal{M} /
\mathcal{M}_{\lambda = 0}$, is the crucial one. Firstly we observe an
asymptotic behaviour in $\lambda$.  The mass appears to reach a fixed
value as the non-uniformity becomes very large, this asymptotic mass
being approximately twice that of the critical string. This indicates
that the non-uniform black string cannot be the end state of the
uniform string GL instability as all these non-uniform strings are more
massive than the unstable uniform ones, and thus such a decay is
classically forbidden. As for the uniform strings, as the mass of the
non-uniform solution increases, the entropy also increases
monotonically, and the temperature decreases. As with the mass, these
quantities stabilise to constant asymptotic values. We discuss the
implications at length in the later section \ref{sec:discussion}.

\begin{figure}[htb]
\centerline{\psfig{file=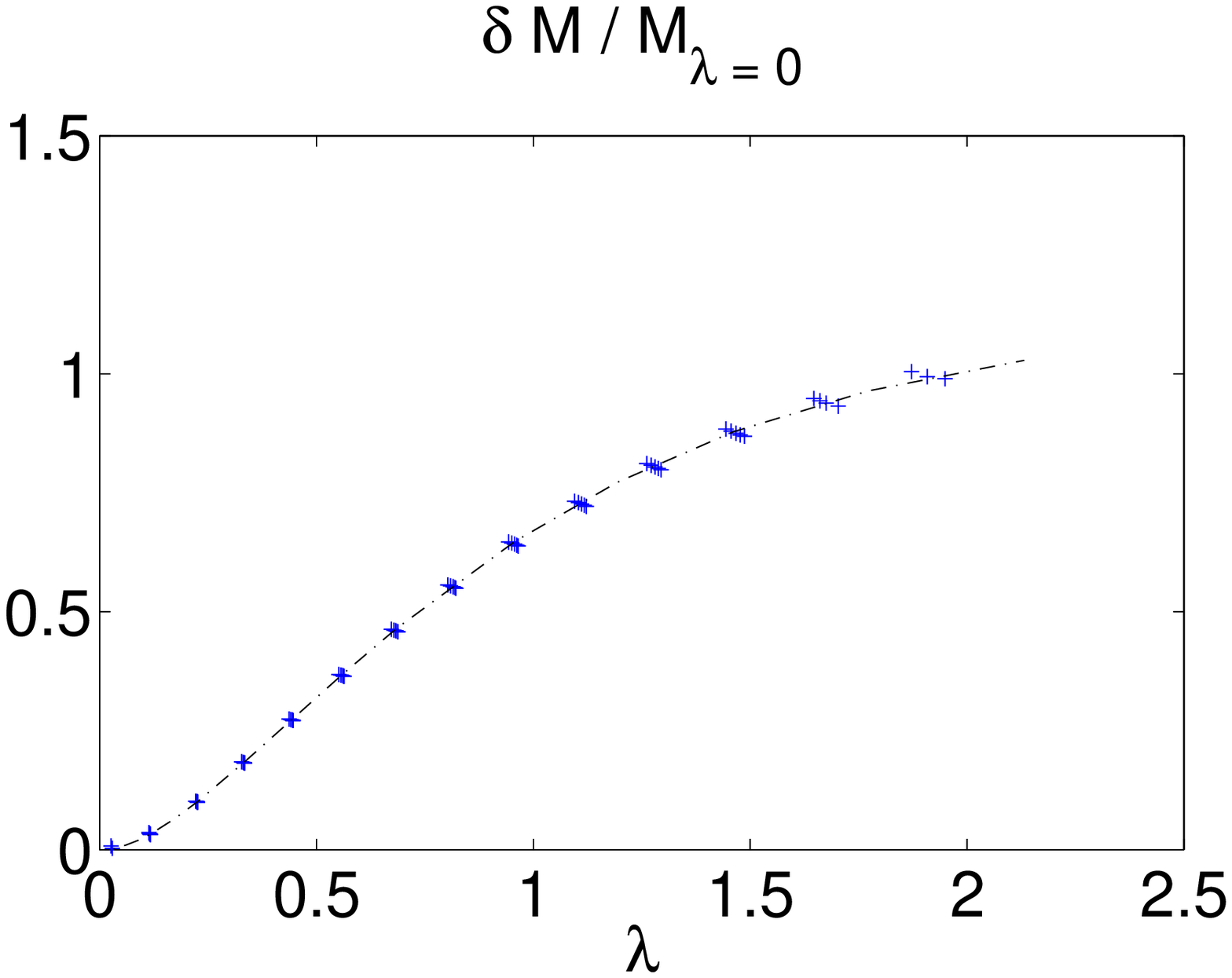,width=3in}\hspace{0.5cm}\psfig{file=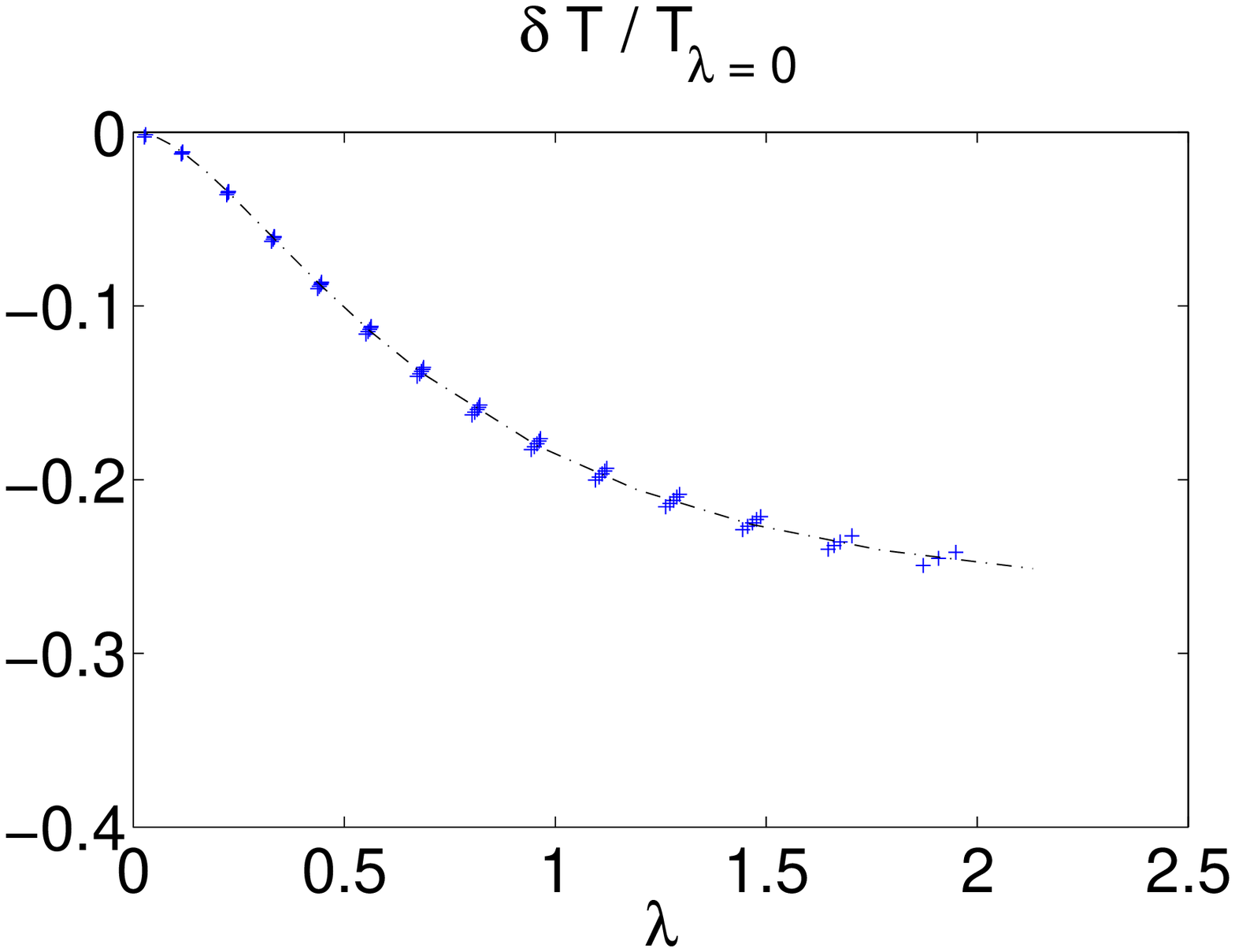,width=3in}}
\centerline{\psfig{file=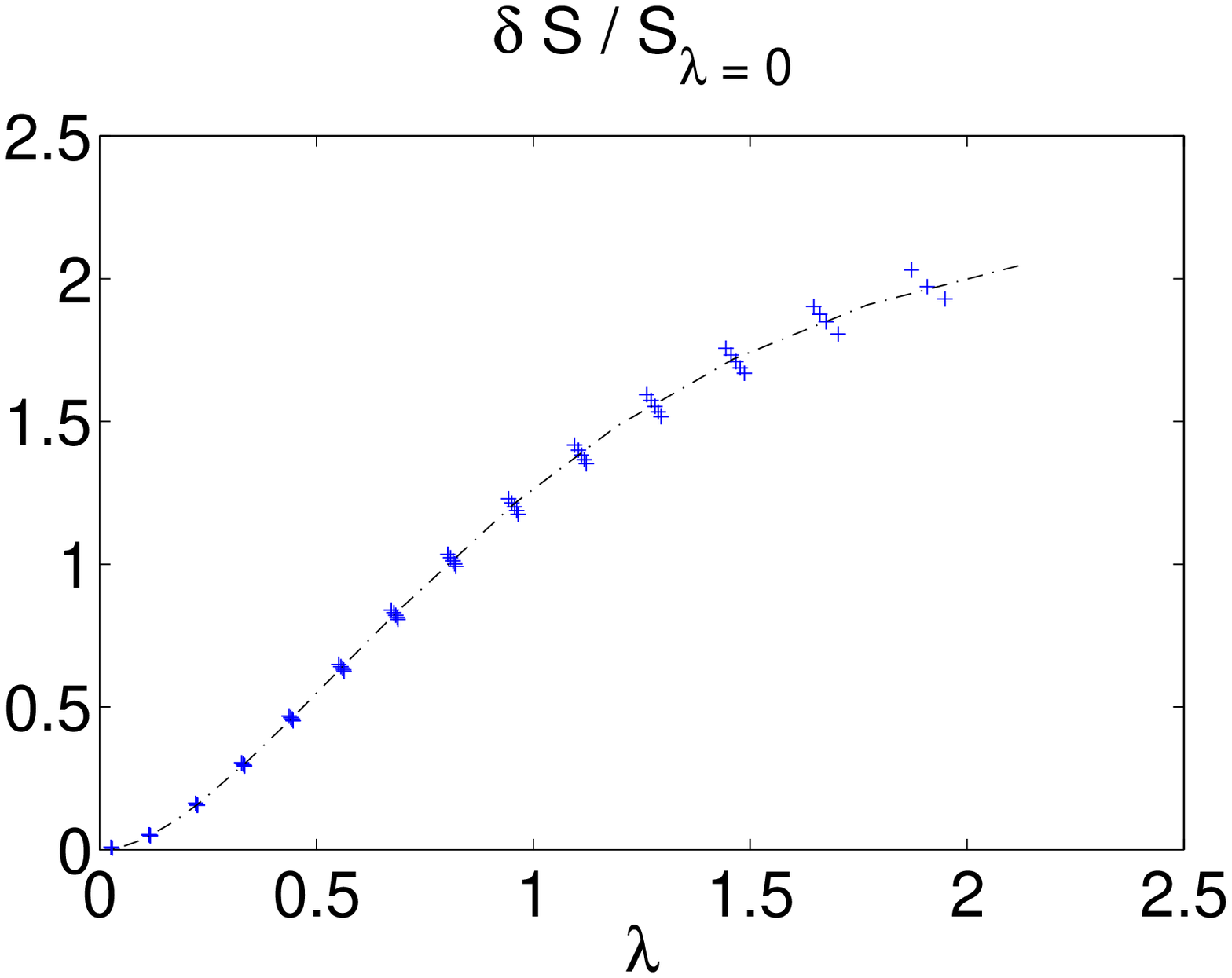,width=3in}}
\caption{ \figuremode
  Again we plot the thermodynamic quantities. For the middle
  resolution, $120*50$, we show in blue the critical $m = 1$, $L =
  2.4758$ scheme, but also two values of $L$ which are $20 \%$ higher
  and lower. The high resolution $240*100$ data for the critical $L$
  is shown in black. We see very good agreement to the few percent
  level for the scheme changing.
\label{fig:properties2} 
}
\end{figure}

We now check these results using simple consistency tests. Firstly, as
discussed earlier, we may change the `scheme' by choosing different
values of $L$. In figure \ref{fig:properties2}, we plot the same
quantities as in the previous figure, using the middle resolution,
$120*50$, choosing to plot the critical $L$ results, but in addition 2
values of $L$ offset by $\pm 20 \%$.  Overlaid on this is the high
resolution curve, $240*100$, for critical $L$, over the range where
the middle resolution converges. Again we see extremely good
consistency.  The values calculated with the critical $L$ agree for
the middle and high resolutions very well as we just observed for
figure \ref{fig:properties}. However the non-critical $L$ values also
agree very well, the errors being small, of order a few percent.

However, whilst $\lambda$ is large, we may still wonder how well the
perturbation expansion does. Indeed in section \ref{sec:compare_PT} we
showed agreement for a particular, and small, value of $\lambda$. We
will now show that the behaviour near the origin in $\lambda$ agrees
with the thermodynamic curves measured, but that the theory becomes
fully non-linear for $\bar{\lambda} \simeq 0.5$, where $\bar{\lambda}$
is the perturbation expansion parameter, as in Appendix
\ref{app:gubser_PT}.

The thermodynamic quantities $\mathcal{T, S}$ have leading order
contributions $ = \mathrm{const}\cdot\bar{\lambda}^2 +
O(\bar{\lambda}^4)$, so actually the perturbation theory is accurate
up to, and including, cubic order in $\bar{\lambda}$. Thus when
comparing the non-linear method with the perturbation theory, we wish
to determine the value of $\bar{\lambda}$ in the non-linear solutions
correct to quadratic order. We could just take $\lambda$, but this
only equals $\bar{\lambda}$ up to linear corrections. It is better to
extract the amplitude of the $\cos{K z}$ Fourier component of $C$, as
in \eqref{eq:fourier_decomp}, which is indeed equal to $\bar{\lambda}$
up to quadratic corrections. We term this amplitude $\tilde{\lambda}
\simeq \bar{\lambda} + O(\bar{\lambda}^3)$

\begin{figure}[phtb]
  \centerline{\psfig{file=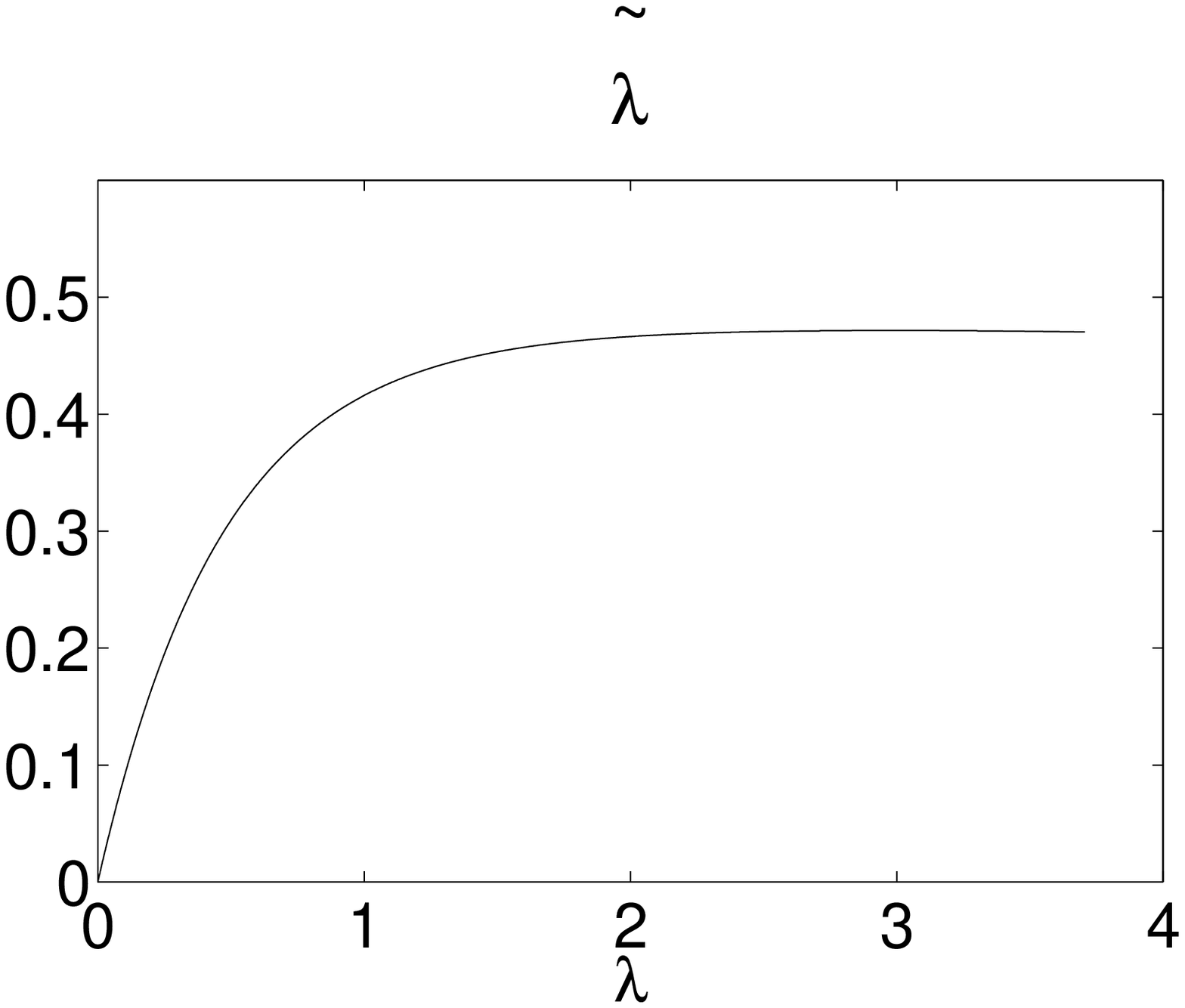,width=3in}\hspace{0.5cm}\psfig{file=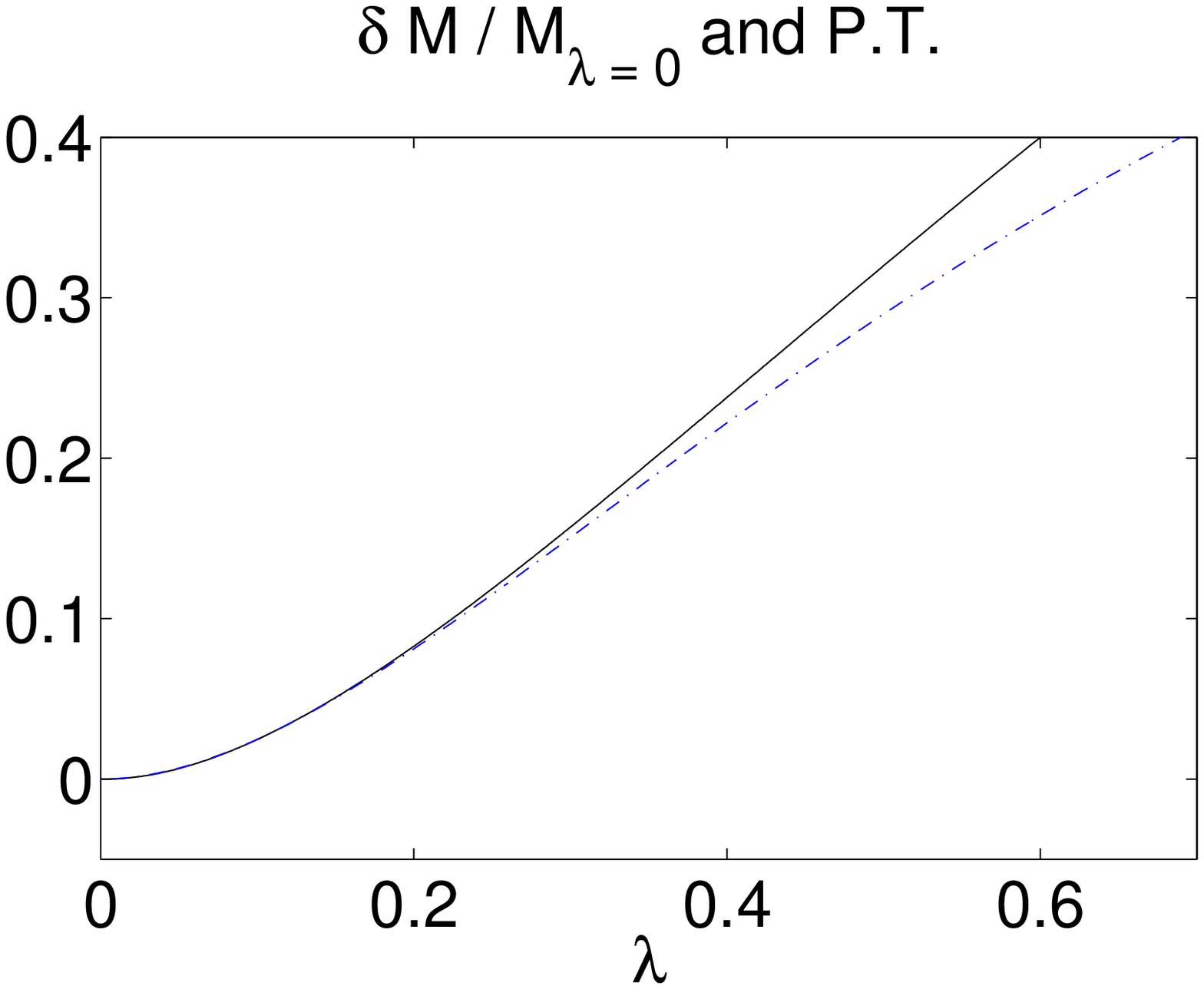,width=3in}}
  \centerline{\psfig{file=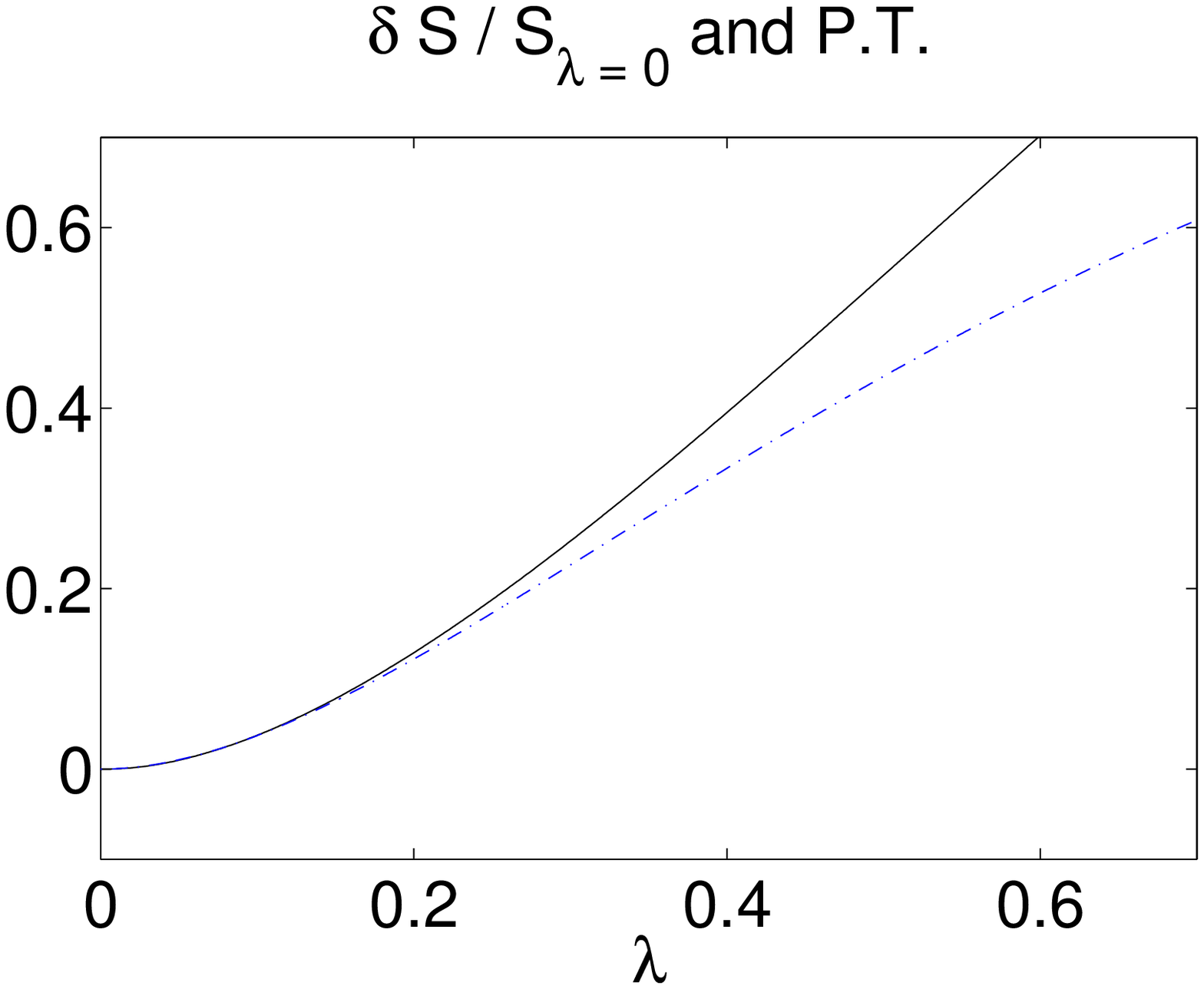,width=3in}\hspace{0.5cm}\psfig{file=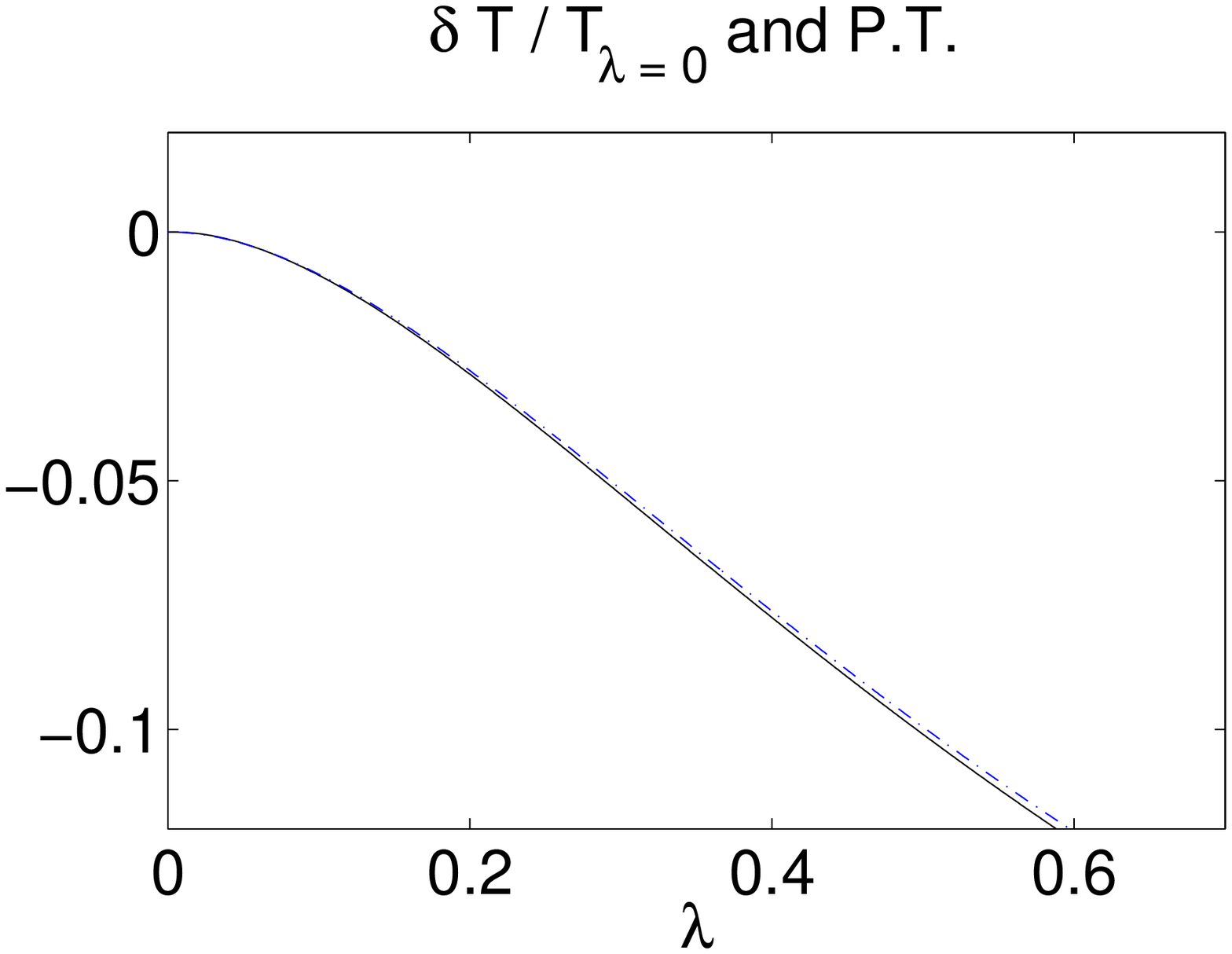,width=3in}}
\caption{ \figuremode
  The top left plot shows $\tilde{\lambda} = \bar{\lambda} +
  O(\bar{\lambda}^3)$, inferred by Fourier transforming the non-linear
  solutions at the horizon. We see that $\tilde{\lambda}$ asymptotes
  to approximately $0.5$, indicating that the perturbation theory
  breaks down for $\bar{\lambda}$ near to this value. Remaining frames
  show the actual thermodynamic quantities in solid black, together
  with the leading order perturbation theory results in dotted blue.
  These should agree to cubic order in $\bar{\lambda}$. We do indeed
  see excellent agreement near the critical point as expected.
\label{fig:PT_compare} 
}
\end{figure}

In figure \ref{fig:PT_compare} we firstly plot the inferred value of
$\tilde{\lambda}$ from the solutions, comparing this against the
actual $\lambda$. We find that $\tilde{\lambda}$ appears to asymptote
to around $0.5$. This shows that near this value of $\tilde{\lambda}$
($\simeq \bar{\lambda}$), the perturbation theory is no longer
accurate, and instead the full non-linear theory must be computed.
This occurs for $\lambda \gtrsim 1$ and indicates the regime where
corrections to the leading order results are larger than the leading
order results themselves, so all higher order corrections are
potentially relevant.  We might have expected that the perturbation
parameter $\bar{\lambda}$ would deviate from $\lambda$ by substantial
non-linear corrections before it reached one as some of the
coefficients in relevant quantities, such as the entropy, are quite
large. Also in the figure we plot the thermodynamic quantities
measured using the highest resolution against the perturbation theory
results, given in the Appendix \ref{app:gubser_PT}. As we extract
$\tilde{\lambda}$, and plot the perturbation theory results as
$\mathrm{const}\cdot\tilde{\lambda}^2 =
\mathrm{const}\cdot\bar{\lambda}^2 + O(\bar{\lambda}^4)$, these curves
should be accurate to cubic order in the perturbation expansion
parameter $\bar{\lambda}$, not just leading quadratic order. Indeed,
we see that excellent agreement is found near the critical point.

\begin{figure}[htb]
\centerline{\psfig{file=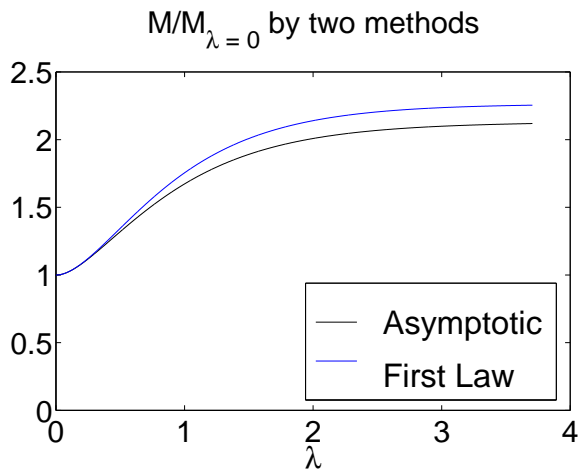,width=3in}}
\caption{ \figuremode
  Plot showing the mass calculated directly from examining the
  asymptotic behaviour, with that computed from the horizon geometry
  using the integrated form of the first law. We see very good
  agreement, confirming that this is indeed the correct behaviour, the
  mass tending to a constant asymptotically in $\lambda$, which is
  approximately twice that of the critical string.
\label{fig:first_law} 
}
\end{figure}

One nice test of the constraints we may perform is to compare the
asymptotic mass calculated directly from the solutions with the mass
integrated from the first law. Using the first law we may compute,
\begin{equation}
\frac{d \mathcal{M}}{d \lambda} = \mathcal{T} \frac{d \mathcal{S}}{d \lambda}
\end{equation}
where $\mathcal{S}$ is the horizon volume, or entropy, and
$\mathcal{T}$ the temperature. Both $\mathcal{T}$ and $\mathcal{S}$
are easily measured from the lattice at $r = 0$.  On the other hand,
the mass is determined by extrapolating the zero mode component of the
metric, as described in Appendix \ref{app:mass}. The errors inherent
in this mass computation not easily assessable.  Therefore this
provides another very useful, and non-trivial check of numerical
consistency of the whole scheme. In figure \ref{fig:first_law} we plot
the directly measured mass against the first law mass for the highest
resolution ($240*100$). We interpolate between data points of
$\mathcal{T}$ and $\mathcal{S}$ in order to calculate the necessary
derivatives, and then integrate to give $\mathcal{M}$. We see
satisfactory agreement, the asymptotic mass differing by about $6 \%$
for the two curves.

One slightly confusing point is that plotting the same curve for the
lower resolutions results in a similar difference between the two
curves. We would expect that the difference should improve with
resolution.  Furthermore, this seems to also be independent of
increasing the position of the large $r$ boundary.  This strongly
indicates that there is a systematic error that remains unidentified.
We expect that this error lies in the asymptotic mass determination
which is difficult to implement numerically. Thus we advocate using
the mass integrated from the first law, assuming the error to be in
the asymptotic mass, a point we return to shortly. However, whilst the
error does not appear to decrease as it should, we see that the
overall consistency of the two curves is good, and certainly the
qualitative form of the mass relation appears very robust. Coupled
with the good agreement for the perturbation theory results at small
$\lambda$, and the well satisfied constraints for all $\lambda$, we
believe that the systematic is likely to be a small effect, although
one worthy of further investigation in future work.

%
\subsection{`Difficult' Quantities: Entropy Difference and Specific Heat}
%

We now consider quantities that are `difficult' to determine
accurately, the entropy (or horizon volume) difference between
non-uniform and uniform strings, and the specific heat. The first is
difficult to determine as we must difference two almost equal
numerical quantities to compute it. The second is difficult to
determine at large $\lambda$ as we must take the ratio of two
quantities which both tend to zero. We will present results for the
entropy difference and specific heat, as they are of interest
thermodynamically. However, we wish to be clear that unlike the
thermodynamic curves for $\mathcal{T}, \mathcal{S}, \mathcal{M}$ in
the previous sub-section, which we believe to be robust, these
`difficult' quantities may contain some systematic errors, and
therefore we do not feel we can draw concrete inferences from them.

\begin{figure}[htb]
\centerline{\psfig{file=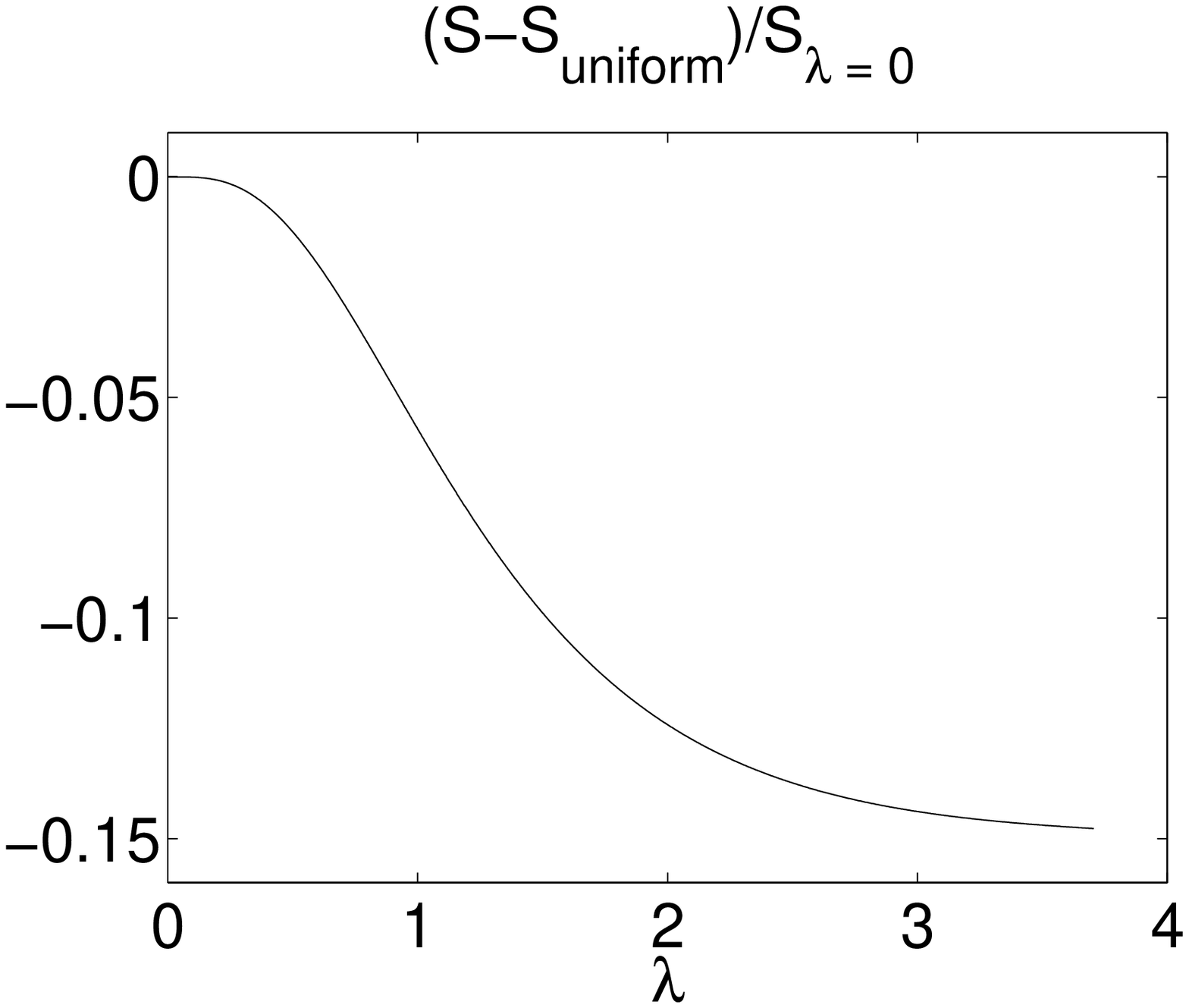,width=3in}\hspace{0.5cm}\psfig{file=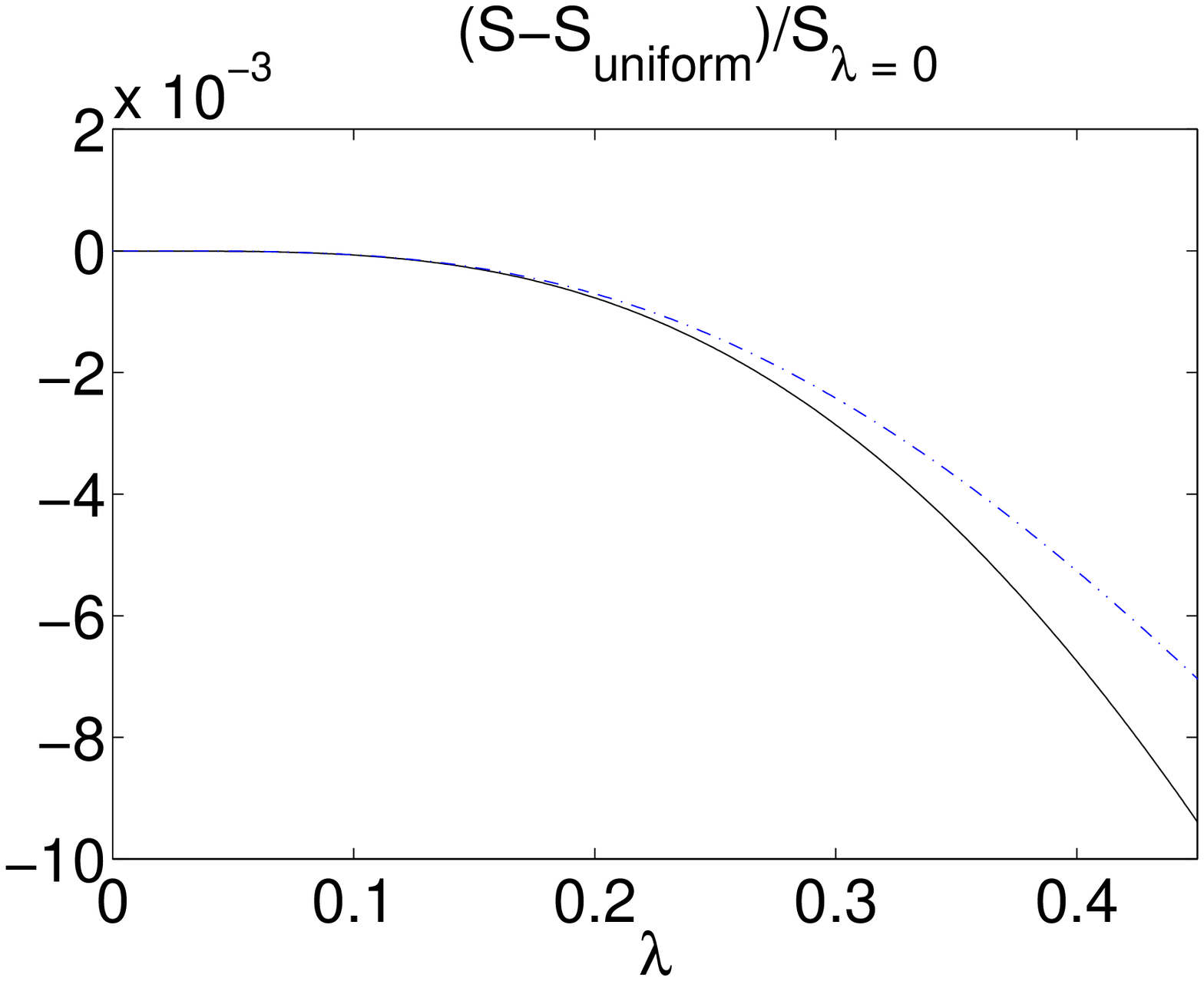,width=3in}}
\caption{ \figuremode
  The difference in entropy of the non-uniform string from a uniform
  string of the same mass. Computed using the mass calculated from the
  integrated first law.  The dotted curve is the leading order
  perturbation theory result, quartic in $\bar{\lambda}$, and gives
  excellent agreement near the critical point. For large $\lambda$ the
  entropy difference appears to asymptote to a small negative
  constant. If a non-uniform string is unstable, this implies that it
  could potentially classically decay to a stable uniform string of
  the same, or lower mass, provided the mass difference was relatively
  small.
\label{fig:entropy_diff} 
}
\end{figure}

Firstly consider the entropy of the non-uniform solution compared to a
uniform string with the same mass, as in \cite{Gubser}. If the
non-uniform solutions are found to be classically unstable, the
entropy difference will indicate whether a non-uniform string can
potentially classically decay to a stable uniform string of the same,
or lower mass.  However this involves differencing two almost equal
numerical quantities, that cancel to leading order $\bar{\lambda}^2$,
and therefore we expect it to be difficult to determine due to
sensitivity to systematic error.  The entropy of the uniform string
with mass $\mathcal{M}$ goes as, $\mathcal{S}_{uniform} \propto
\mathcal{M}^{3/2}$ and thus in order to calculate it we must know the
mass. We now have two ways to compute this mass, one directly from the
asymptotic behaviour of the metric functions, and the other from using
the horizon geometry and integrating the mass from the first law. In
figure \ref{fig:entropy_diff} we plot the entropy difference
calculated using the integrated mass from the first law, using the
highest resolution.  We also plot the perturbation theory to leading
order in $\bar{\lambda}$, which goes as $\sim \bar{\lambda}^4$.  Again
we see excellent agreement between the two for small $\lambda$, and
considering we are differencing two almost equal numerical quantities
this appears to accurately reproduce the quartic behaviour in
$\bar{\lambda}$. We see that the entropy difference between the
non-uniform string and a uniform one of the same mass also remains
monotonically decreasing and negative. Note that the entropy
difference is small compared to the actual entropy of the non-uniform
string.  If we calculate this entropy difference using the mass
computed directly from the asymptotic behaviour, we find poor
agreement with the perturbation theory result near the critical point,
giving more evidence to our previous claim that our direct mass
calculation contains some unidentified systematic error.  Whilst this
is evidently small, as discussed earlier, and seen in figure
\ref{fig:first_law}, it is large enough to upset the computation of
the very delicate entropy difference which involves differencing two
almost equal numerical quantities.

\begin{figure}[htb]
  \centerline{\psfig{file=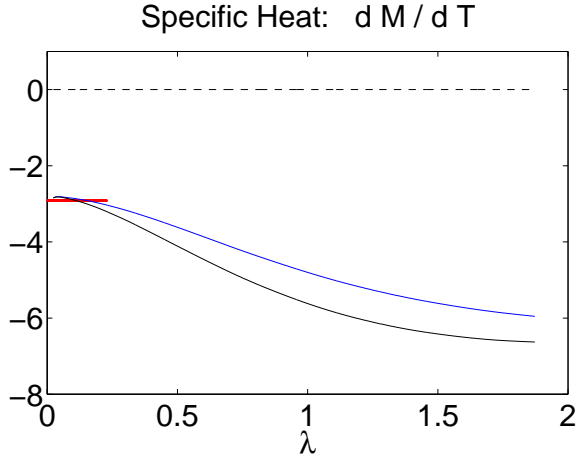,width=3in}}
\caption{ \figuremode
  Plot of the specific heat measured using the mass integrated from
  the first law (solid black) and the mass calculated directly from
  the asymptotic metric (solid blue). The red bar is the value
  predicted by perturbation theory for $\lambda = 0$, and we see good
  agreement there. The specific heat decreases from that of the
  critical string. However, asymptotically in $\lambda$ it is
  difficult to determine the behaviour, as both $d \mathcal{M} / d
  \lambda $ and $d \mathcal{T} / d \lambda$ become small and thus we
  expect amplified numerical errors in their ratio.  Above $\lambda
  \simeq 2$ we expect these curves to be unreliable.
\label{fig:spec_heat} 
}
\end{figure}

We now consider the specific heat of the non-uniform solutions, $d
\mathcal{M} / d \mathcal{T}$. It is immediately clear from the
`robust' thermodynamic curves \ref{fig:properties}, that the specific
heat is negative. We calculate it by taking the ratio of $d
\mathcal{M} / d \lambda$ and $d \mathcal{T} / d \lambda$. The reason
this quantity is `difficult' to determine is that at large $\lambda$
both these derivatives become small, and hence any systematic error is
likely to be vastly amplified in the resulting ratio, the specific
heat. In figure \ref{fig:spec_heat} we plot the curves yielded from
both estimators of the mass, the integrated first law mass and the
asymptotic metric mass. We see agreement for small $\lambda$, which we
expect as the form of the two masses agree with each other, as
discussed in the last sub-section, and shown in figure
\ref{fig:first_law}. However, we see from figure \ref{fig:properties}
that for $\lambda > 2$ the derivatives of $\mathcal{M, T}$ with
respect to $\lambda$ become very small, and therefore we do not have
confidence in these curves beyond $\lambda \simeq 2$. We can
confidently say that the specific heat initially decreases with
$\lambda$. It is tempting to say it appears to asymptote to a constant
at large $\lambda$, but this is highly speculative.  Again this
example serves to illustrate the requirement for future work that
improves the mass estimation and reduces systematic errors, so that
even these `difficult' quantities can be managed.

%
\section{Discussion}
\label{sec:discussion}
%

The main findings of this paper are found in figure
\ref{fig:properties}, namely that we can construct the compactified
non-uniform string in the full non-linear theory, and that the mass is
always larger than that of the critical uniform string.  This appears
to rule out the possibility that the Gregory-Laflamme instability can
have these non-uniform strings as an end state. We have also shown
that the non-uniform strings have lower entropy than uniform strings
of the same mass.  Unfortunately we cannot tell whether these
solutions are classically stable. There are several plausible options,
illustrated in the introduction in figure \ref{fig:results}.

Firstly the solutions are classically stable. This would result in a
very clear violation of black string uniqueness \cite{Kol3}. They are
then likely to play a crucial role in higher dimensional dynamics,
particularly in black hole formation in compactified extra dimensions
where the black hole mass is in the range where both the uniform and
non-uniform solutions exist. The behaviour would be similar to the
uniform strings above critical mass.  As with the uniform strings, the
non-uniform solutions have monotonically increasing entropy with mass,
and monotonically decreasing temperature.  Quantum mechanically, we
would therefore expect the compact non-uniform string to emit
radiation, and adiabatically move to a lower mass non-uniform string,
until the critical point is reached when the classical instability will occur.
Thus the radiation carries away mass and entropy from the string, and
in the process reduces its non-uniformity.

Secondly the strings are dynamically unstable, as advocated by Kol
\cite{Kol1}. If this is the case, it would then be improbable that the
non-uniform solutions would play much role in the dynamics of gravity
in higher dimensions, simply because it is unlikely any initial data
would evolve to a state near these static solutions.  However, it is
again interesting to consider the classical decay of such a solution.
The dynamics of the classical instability would then be governed by
the nature of the dominant unstable mode, and the amount of radiation
emitted in the process.  If little radiation is emitted, we could
conceive a non-uniform string classically decaying to a uniform stable
solution with slightly less mass. This is amusingly reminiscent of the
Horowitz-Maeda expectation, but now in reverse, the unstable
non-uniform strings decaying to the stable uniform ones. Note that as
we have shown the horizon volume, or entropy, of the non-uniform
solutions is less than that of the uniform solutions with the same
mass, this process is allowed by the second law. However, since this
horizon volume difference is small, it also means that the mass lost
to radiation must also be small. The volume difference allows a
transition to an equal mass uniform string, but the horizon volume of
less massive uniform strings decreases, and if too much mass is lost,
the process would then violate the second law. The alternative is that
the decay is to the same end state that the unstable uniform strings
reach, whatever that might be. Note that if much radiation is emitted
due to the classical instability, then as argued above, the latter
case is the only option.

The only arguments concerning classical stability are those made by
Kol in the interesting paper \cite{Kol1}, claiming the simplest
picture is for the strings to be unstable. This is largely based on
assuming that thermodynamic instability implies classical instability.
Whilst this is shown \cite{Gubser_Mitra1, Gubser_Mitra2, Reall} in the
uniform case, translational invariance is critical in the argument,
and there is no evidence that the relation should be more general.
Thus we feel that their classical stability is very much an open, and
interesting question.  It also remains an interesting question whether
this branch of non-uniform string solutions connects with the branch
of compactified black hole solutions, as suggested by Kol \cite{Kol1}.
In a future work, we will look at the geometry of our non-uniform
string solutions to infer whether it is plausible that these two
branches of solutions are linked.

The best way to understand the dynamics of the black strings is simply
to solve the full time dependent equations. Whilst this appears to be
under way, the end state of the full dynamical simulations is
apparently still inconclusive \cite{Horowitz, Choptuik}. From the
results presented here, we expect that the end state is not a
non-uniform string, at least in the branch of solutions connected to
the GL critical point.  This could considerably complicate the long
term dynamical evolution.  The best possibility from the numerical
point of view would have been an adiabatic motion to a non-uniform
string solution. Then all curvatures, and time derivatives would have
remained small during the simulation. Gubser showed that this was not
the case. The next best case would have been a brief non-adiabatic
period and then a slow motion to the non-uniform string. Presumably
the more dramatic the evolution, the harder it is to evolve for long
times. However, the present work indicates a different end state all
together, and so it could be that large curvatures and long dynamical
times may be involved in uncovering this end state, probably making
the numerical problem considerably more tricky.

Let us finally comment briefly on the interesting work of Harmark and
Obers \cite{Harmark_Obers}. As mentioned in the introduction they have
an ansatz that may solve the Einstein equations in terms of
only one unknown function.  This is shown to be a consistent ansatz to
second order in an asymptotic expansion, which is highly non-trivial
evidence supporting their claim. However the resulting equation for
the unknown is not pleasant, and would almost certainly require
numerical solution.  It is then very interesting whether we could
apply the methods of this paper to the solution of their equation, and
whether this would give a simpler method than the one using the
conformal gauge we use here. At first sight having only one function
to solve for would appear to be beneficial. However, really it is not
the number of equations to be solved that is critical, but rather the
stability of the equations under some relaxation algorithm.  The
powerful feature of our method is that because the conformal gauge
equations for the metric functions have Laplace second derivatives, at
least for small non-uniformity solutions can be easily found by a very
simple relaxation algorithm.  Thus the fact that we have more than one
equation to solve, due to our 3 metric functions does not really
complicate the method. In the Harmark and Obers ansatz, reducing the
metric to only one function involves substituting a metric function
that can be determined algebraically. This results in third order
derivative terms. It would be interesting to see whether standard
techniques could be used on this equation. We think it is likely to be
better to not eliminate the second function, and stick to reducing the
problem to two metric functions, and solving for these. Either way, it
appears that our methods might be applied. This could allow the
consistency of the ansatz to be tested. If correct, the ansatz may
provide a powerful way to implement these elliptic methods.

%
\section{Other Applications and Areas of Improvement}
\label{sec:improvement}
%

The guiding principles we have used in applying the elliptic method of
solution are the following:
\begin{itemize}
\item Perturb about a background solution, such that the deformation
  is parametrised by \emph{finite} valued metric functions. 
\item Choose a problem where at least some of the boundaries have data
  that is implemented locally, ie. not involving integration over a
  boundary, or worse, integration from one boundary to another. In the
  Randall-Sundrum star case \cite{Wiseman} the metric perturbation decayed to
  zero far from the brane and this satisfied the weighted constraints.
  On the brane one non-local integration was performed.  In the
  present example 2 boundaries, $z = 0, L$, are periodic, and the
  asymptotic boundary has simple local boundary conditions. Only the
  horizon boundary requires a non-local integration operation.
\item Most importantly, if one can solve the linear deformation
  problem elliptically in the conformal gauge, one should be able to
  use the non-linear methods employed here. Thus the method provides a
  way to extend static linear perturbation theory into the non-linear
  regime.
\end{itemize}
There are potentially numerous applications of this method. For
example, one can apply it to any axisymmetric problem in more than 4
dimensions, or indeed any situation where one has static
configurations with metric dependence on 2 variables. If possible,
extending these ideas to allow dependence on more than 2 variables
would allow insight into solutions to gravity where there is little
analytically known.

One exciting avenue for future research is to understand how to
include localised horizons in the scheme, which would allow access to
uncharged black hole solutions in Kaluza-Klein theory, and on branes,
if the latter solutions do indeed exist, (although the converse has
recently been conjectured by Tanaka \cite{Tanaka} and elaborated on in
\cite{Emparan_Kaloper}). In the Randall-Sundrum star case we tackled a
problem with a regular geometry and no horizon (other than the
asymptotic AdS horizon). In the current work, a continuous horizon was
considered. The challenge would be to understand how to merge these
two situations to allow localised horizons that do not extend along
the length of the symmetry axis.

Let us briefly consider the problem of a small neutral Kaluza-Klein
black hole.  Using a background metric that is flat space deformed by
$A, B, C$ in the conformal gauge, would not allow such solutions to be
found, as $A \rightarrow - \infty$ on the horizon.  Large deformations
of $A, B, C$ away from zero would most likely destroy the convergence
of the scheme. In this paper, we are able to tackle a horizon geometry
because we encoded the vanishing behaviour of the lapse at the horizon
in the background metric \eqref{eq:bs_metric}, essentially as we know
the translationally invariant black string solution. Then the
deformations of this required everywhere finite $A, B, C$. For a
localised black hole we similarly need to find a metric to perturb
around that encodes the horizon behaviour. The problem is that no
solutions are known. Thus, we must find a good guess, such that the
solutions will require non zero, but \emph{finite} $A, B, C$.  An
interesting guess we might try is the 5-dimensional Schwarzschild
metric in isotropic coordinates,
\begin{align}
  ds^2 & = - \left(\frac{m - \rho^2}{m + \rho^2}\right)^2 dt^2 +
  \left(1 + \frac{m}{\rho^2}\right)^2 \left( dr^2 + r^2 d\Omega_2^2 +
    dz^2 \right)
  \nonumber \\
  \rho^2 & = r^2 + z^2
\end{align}
which presumably closely approximates the horizon geometry for a very
small black hole. We see immediately that $B - C = 0$ for this metric,
which would be important due to the singular nature of the gauge at $r
= 0$ where the axis of symmetry is regular away from the horizon (as
discussed in \cite{Wiseman}). So we might try the ansatz,
\begin{equation}
ds^2 = - \left(\frac{m - \rho^2}{m + \rho^2}\right)^2
  e^{2 A} dt^2 + \left(1 +
  \frac{m}{\rho^2}\right)^2 \left( e^{2 B} (dr^2 + dz^2) + e^{2 C} r^2
  d\Omega^2 \right)
\end{equation}
We would hope that for a solution with small mass compared to the
compactification scale, $A, B, C$ would be small.  Note that due to
the conformal invariance in $r, z$ there is coordinate freedom to
choose the horizon to remain at the location $r^2 + z^2 = m$. On the
horizon itself, a constant temperature condition and regularity would
be imposed, just as for the black strings here. Whether such methods
can be employed, and whether the solutions exist remain exciting
avenues of research.

The numerical algorithms and computational power employed to produce
the results in this paper were very modest. The largest area of
potential improvement is in these numerical algorithms. This is
particularly important in progressing to higher values of $\lambda$
where we see that already the resolution $240*100$ is not sufficient
for $\lambda \gtrsim 3.9$. The time needed to relax still higher
resolutions using the elementary methods of this paper is prohibitive.

There are two potential remedies for this problem. Firstly using
higher order discretisation algorithms to increase the accuracy of the
solutions may well mean that lower resolutions are required to capture
a given $\lambda$. Alternatively, if we cannot remove the high
resolution requirement, using more sophisticated relaxation algorithms
might substantially reduce the time taken. For instance, we might
change the Gauss-Seidel relaxation discussed in the Appendix
\ref{app:details}, to a multi-grid algorithm. Whilst it is absolutely
clear that multi-grid vastly out-performs Gauss-Seidel as a Poisson
equation solver, it is not at all clear that it will perform so well
in our context, where the source terms and boundary conditions are
iteratively updated as functions of the metric. In any case, a better
understanding of the convergence properties of the method would
certainly be useful.

In addition to this, we have used a regular lattice to cover the $r,
z$ space. We have seen that the gauge gives rise to large gradients in
the $r = 0$, $z = L$ corner of the grid, which then require high
resolution.  However, using a variable grid spacing, or even an
adaptive mesh, we might hope to concentrate points in the areas with
strong gradients and thereby use the grid points more efficiently.
Furthermore, at large $r$, where the solutions tend to the $z$
independent asymptotics, we could possibly use less lattice points,
as the $z$ gradients would be small. Any reduction in the number of
lattice points is likely to make a huge difference to the relaxation
times.

%
\section{Conclusion}
\label{sec:conclusion}
%

We have developed elliptic numerical methods to find static
axisymmetric vacuum solutions in more than 4 dimensions, where no
analytic solutions are known.  The method is applied to the case of
the 6 dimensional compactified non-uniform black string, where the
implementation is straightforward compared to our previous work
\cite{Wiseman}. We show using perturbation theory that the 6
dimensional string has the same thermodynamic behaviour as Gubser
found for the 5 dimensional one near the critical point. The full
non-linear method produces solutions, the relaxation time required
increasing with increasing non-uniformity. We have configurations with
large non-uniformity, where $\lambda \simeq 3.9$, but this can no
doubt be improved by further investigation of the numerical algorithm,
and use of more sophisticated numerical methods. Such values of
$\lambda$ are well beyond the reach of a perturbative approach, the
corrections to the leading order being of order the quantity under
consideration, and thus all higher order corrections are likely to be
important.

The quality of the solutions is checked in several ways. The local
constraint violation is assessed, and shown to be small, and to
improve with increasing resolution. The solutions are checked against
the perturbation theory for small $\lambda$, and excellent agreement
is found. The ADM mass of the solution can be determined directly, and
via the horizon geometry using the first law, and these compare well.
Thus we claim, with confidence, that the mass of the non-uniform
strings is greater than that of the critical string for a given
asymptotic $S^1$ size. A further check is that solutions can be
relaxed with different asymptotic $S^1$ lengths, in an analogous
manner to the scheme changing used by Gubser in the perturbation
theory calculation.  Solutions with substantially different $L$ are
found to give thermodynamic curves which are highly consistent.
Overall, the method appears to perform extremely well, giving good
accuracy.

The mass of the non-uniform strings is robustly shown to increase with
$\lambda$, asymptoting to approximately twice the mass of the critical
string for very large $\lambda$. This intriguingly implies that these
stable non-uniform strings are not the end state of the dynamical
Gregory-Laflamme instability, and again raises the interesting
question of what this end state actually is. It may well be that full
dynamical numerical simulation is the only way to determine this,
although if the end state, or intermediate configurations, are highly
curved this may be an extremely difficult problem.  We also give
evidence that the entropy of the non-uniform solutions is always less
than that of the uniform solutions with the same mass, although this
is less firm, as we must difference two almost equal quantities to
calculate this.

Unfortunately we cannot assess the classical stability of the
solutions. If they are stable, our results indicate they will behave
in a similar way to the stable uniform solutions, and simply Hawking
radiate to the critical point. They would potentially play a crucial
role in dynamics, particularly for black hole formation in theories
with compactified extra dimensions when the black hole mass fell in
the range where both uniform and non-uniform solutions existed. If
unstable, they are unlikely to play a significant role in higher
dimensional dynamics. Starting with such a solution, it might either
classically decay to a stable uniform solution, or undergo the same
type of decay as the uniform unstable strings. The first possibility
is allowed by the second law, although the mass lost to radiation in
the decay must be small.

We have discussed possible improvements for the numerical algorithms
described here. We expect they can always be generalised to static
problems with symmetry such that the metric depends only on two
coordinates. The methods may be extendible to dependence on more than
just two coordinates. An interesting future application of the method
is to explicitly construct the Kaluza-Klein black hole in 5 or more
dimensions. We have discussed some technical subtleties involved, and
their possible solution. This open problem is complementary to
studying the black strings, and understanding the compactified black
hole solutions may also shed light on the dynamics involved in the
Gregory-Laflamme instability.

%
\section*{Acknowledgements}
%

I am grateful to Gary Gibbons for pointing out the elliptic nature of
the black string problem. I would like to thank Ruth Gregory, Steven
Gubser, Troels Harmark, Sean Hartnoll, Harvey Reall and Andrew Tolley
for useful discussions. In particular I would like to thank Harvey
Reall for comments on this manuscript. This work was supported by
Pembroke College, Cambridge, and computations were performed on COSMOS
at the National Cosmology Supercomputing Centre in Cambridge.

\newpage

\appendix

%
\section{Appendix: Perturbation Theory and Asymptotics}
\label{app:gubser_PT}
%

In this Appendix we review Gubser's perturbation theory method
\cite{Gubser}, and apply it to the 6-dimensional case. The Mathematica
code used to generate results stated in this section will be made
available \cite{Website}.

We Fourier decompose the metric as in equation
\eqref{eq:fourier_decomp}. As Gubser discusses, different values of
$K$ correspond to different `schemes'.  Geometrically, we might use
this freedom to choose the asymptotic size of the $S^1$ to be of fixed
proper length, as is relevant for comparison with our non-linear
method. Another scheme, the `standard scheme', allows $K$ to vary, and
instead fixes $C^{n (m)} = 0$ for $m \ge 2$.  This scheme reduces the
number of quantities that we must `shoot' for, although it has no
obvious geometrical interpretation.

Having fixed a scheme, the Einstein equations reduce to a series of
ordinary differential equations for $X^{n (m)}_i(r)$, the higher
orders in $\bar{\lambda}$ depending on the solutions obtained from the
lower orders. At each order some of the equations are shooting
problems, solved by demanding the metric functions tend to zero
asymptotically and have behaviour at the horizon to ensure regularity.
By Taylor expanding about the origin we obtain the condition that in
the conformal gauge, the $r$ derivative of $X^{n (m)}_i(r)$ must
vanish at $r = 0$.  

The quantities of interest can be determined with high accuracy,
typically with consistency better than $1 \%$ up to third order for
scheme changes of $C^{0 (1)} = \pm 1$. The same methods used to
generate the 6 dimensional results below were also used to recreate
the 5 dimensional results. Careful handling of the asymptotic
behaviour of the `zero modes' is crucial to obtain high precision and
we find a final `scheme error' in the entropy difference, $\sigma_2$
in Gubser's terminology, of only $1 \%$ for the 5-dimensional string.
In fact we find a slightly better result in the 6 dimensional case,
which may be attributable to the faster fall off of the zero modes
asymptotically.

Having employed Gubser's method in 6 dimensions, we now give the
`standard scheme' numbers for completeness, giving partial results for
third order;
\begin{align}
  A^{1 (0)}(0) & = -0.783  &  B^{1 (0)}(0) & = -0.783  &  C^{1 (0)}(0) & = 1  \nonumber \\
  k^{(0)} & = 1.269   \nonumber \\ \nonumber \\
  A^{0 (1)}(0) & = 0.63 &  B^{0 (1)}(0) & = 0.91 & C^{0 (1)}(0) & = 0 \nonumber \\
  A^{2 (0)}(0) & = 0.64 & B^{2
    (0)}(0) & = 0.64 &  C^{2 (0)}(0) & = -0.87 \nonumber \\ \nonumber \\
  A^{1 (1)}(0) & = -0.38  \nonumber \\
  k^{(1)} & = 1.02
\label{eq:standardscheme1}
\end{align}
The numerics verify the asymptotic behaviour we expect from the zero
mode equations.  These zero mode components, $X^{0 (1)}_i(r)$, do not
require shooting, going as,
\begin{align}
A^{0 (1)} & \rightarrow a_0 \frac{1}{r^2} & a_0 & = -0.93 \nonumber \\
B^{0 (1)} & \rightarrow b_0 \frac{1}{r^2} & b_0 & = 0.65 \nonumber \\
C^{0 (1)} & \rightarrow c_0 \frac{1}{r^2} + c_1 \frac{1}{r
} & c_0 & = -0.37, \qquad c_1 = 1.69 
\label{eq:standardscheme2}
\end{align}
where both $A^{0 (1)}$ and $B^{0 (1)}$ can be shifted by additive
constants, and are chosen to be zero asymptotically. The components
which depend on $z$ have exponentially growing and decaying solutions.
The method of shooting selects the decaying solution and thus the
asymptotic behaviour is dominated by the zero mode components above.
This is the same as in Kaluza-Klein theory where the leading
contribution to the reduced propagator is derived from the homogeneous
modes, and the inhomogeneous massive modes are Yukawa suppressed at
large distances.

Computation for the scheme where $k^{(1)} = 0$, and thus the
asymptotic size of the $S^1$ is held constant, yields,
\begin{align}
  A^{0 (1)}(0) & = -0.17 & C^{0 (1)}(0) & = 0.80 \nonumber \\
  A^{1 (1)}(0) & = -0.38  \nonumber \\ 
  k^{(1)} & = 0.00   
\end{align}
and,
\begin{align}
A^{0 (1)} & \rightarrow a_0 \frac{1}{r^2} & a_0 & = -1.73 \nonumber \\
C^{0 (1)} & \rightarrow c_0 \frac{1}{r^2} + c_1 \frac{1}{r} & c_0 &  = 0.43, \qquad c_1 = 1.69
\end{align}
where we have only reproduced the quantities in
\eqref{eq:standardscheme1} and \eqref{eq:standardscheme2} which are
scheme dependent. These numbers can then be directly compared with
those of the non-linear method which uses this scheme.

The ADM mass, $\mathcal{M}$, can be evaluated using the
Hawking-Horowitz expression \cite{Hawking_Horowitz}. To leading order
this gives the scheme independent quantity $\eta$,
\begin{align}
  \frac{\delta \eta}{\eta} & = \frac{\delta \mathcal{M}}{\mathcal{M}}
  + 2 \frac{\delta K}{K} \simeq \bar{\lambda}^2 \left( \frac{10}{3} b_0
    + 2 c_0 + 2 \frac{k^{(1)}}{k^{(0)}} \right)
  \nonumber \\ \nonumber \\
  & = 3.03 \bar{\lambda}^2
\end{align}
In our non-linear method, we fix the length of the $S^1$ for all
solutions, and thus $\delta K = 0$, and $\delta \eta / \eta = \delta
\mathcal{M} / \mathcal{M}$.  We may evaluate the horizon temperature
$\mathcal{T}$, as a scheme independent quantity, $\alpha$,
\begin{align}
  \frac{\delta \alpha}{\alpha} & = \frac{\delta
    \mathcal{T}}{\mathcal{T}} - \frac{\delta K}{K} \simeq
  \bar{\lambda}^2 \left( A^{0 (1)}(0) - B^{0 (1)}(0) -
    \frac{k^{(1)}}{k^{(0)}} \right)
  \nonumber \\ \nonumber \\
  & = -1.08 \bar{\lambda}^2
\end{align}
and similarly for the horizon volume, and thus entropy $\mathcal{S}$,
\begin{align}
  \frac{\delta \mathcal{S}}{\mathcal{S}} + 3 \frac{\delta K}{K} &
  \simeq \bar{\lambda}^2 \left( B^{0 (1)}(0) + 3 C^{0 (1)}(0) +
    \frac{1}{4} ( B^{1 (0)}(0) + 3 C^{1 (0)}(0) )^2 + 3 \frac{k^{(1)}}{k^{(0)}}
  \right)
  \nonumber \\ \nonumber \\
  & = 4.55 \bar{\lambda}^2
\end{align}
Using the first law as in \cite{Gubser}, this allows us to
calculate the difference in entropy between the deformed string and a
uniform string with the same mass, as is relevant in a dynamical
context. Indeed we find that,
\begin{align}
  \frac{\mathcal{S}_\mathrm{deformed} - \mathcal{S}_\mathrm{uniform}
  }{\mathcal{S}_\mathrm{uniform}} & = - \frac{3}{4} \left( \frac{d
      \alpha}{\alpha} \frac{d \eta}{\eta} + \frac{1}{2} \left( \frac{d
        \eta}{\eta} \right)^2 \right)
  \nonumber \\ \nonumber \\
  & = -0.98 \bar{\lambda}^4
\end{align}
and thus the deformed string has a lower entropy.  Therefore we find
the same qualitative features in 6 dimensions as Gubser found in 5
dimensions, namely that for fixed asymptotic $S^1$ radius, the mass
increases with $\bar{\lambda}$, and this entropy difference is
negative. We can compute systematic `errors' in the quantities
presented above, obtained by recalculating the quantities in a
different scheme such that $C^{1 (0)} = \pm 1$. This allows us some
indication of the accuracy of the method. The standard deviation of
the scheme independent quantities above for $C^{1 (0)} = {0, \pm 1}$
are all well below $1 \%$, the errors growing for higher order
quantities, the largest scheme error being for the entropy difference
giving $0.1 \%$.

%
\section{Appendix: Technical Details}
\label{app:details}
%

We now discuss the numerical scheme in a little more detail, and in
particular the algorithm for updating the boundary data. We begin with
the most time consuming step in the iteration, which is the relaxation
of the interior equations.

The 3 elliptic equations are solved by splitting the Laplacian term
from the source term as in \eqref{eq:general_equations}.  A relaxation
scheme is used to partially relax the Poisson equations holding the
sources constant. We use Gauss-Seidel combined with under-relaxation.
The Gauss-Seidel method is a local update procedure, and our general
philosophy in finding a stable method to solve these equations is to
keep operations as local as possible. Thus, whilst non-local solvers
such as multi-grid may give quicker solutions, we also expect them to
be less stable in generating a convergent solution to this complicated
non-linear problem.

We find that standard Gauss-Seidel works well for the $A, C$
equations. However we use under-relaxation for $B$, where the update
for $B$ is,
\begin{equation}
B_{update} = (1 - \epsilon) B_{old} + \epsilon B_{new} 
\label{eq:slow_update}
\end{equation}
where $B_{old}$ is the old value of $B$ at some point, $B_{new}$ is
the result of the Gauss-Seidel iteration and $B_{update}$ is the
updated value at the point. Thus for $A, C$, $\epsilon = 1$ and there
is no added `inertia'. For $B$ we find the method is unstable unless
`inertia' is added, and a value of $\epsilon = 0.01$ was used for the
results calculated in the paper. Understanding the nature of this
numerical instability in $B$ may be crucial for improving the
technique.

After one pass of Gauss-Seidel, the boundary conditions are updated,
as described below. We could update the Poisson source terms every
cycle, but it takes time to do so. The relaxation procedure is very
simple and quick, involving little floating-point arithmetic. However,
calculating the sources involves intensive arithmetic. Thus we have
found that updating the sources one in ten iterations is a good
compromise. 

The horizon boundary conditions are that $A_{,r} = C_{,r} = 0$ and
that $A_{,z} = B_{,z}$. We difference $A_{,r}$ and $C_{,r}$ to second
order and these allow the boundary value to be determined at $r = 0$.
The condition for $B$ is implemented by using a simple second order
integration from the $r=0$, $z=L$ lattice point, whose value is an
input parameter and specifies $B_{max}$, down to $r=0$, $z=0$.  As
with the interior relaxation, we cannot directly inject the integrated
value. We integrate along the boundary, storing the new values in a
buffer, and then inject them as in \eqref{eq:slow_update}. Again a
value of $\epsilon = 0.01$ was used to generate the results in this
paper. For larger $\epsilon$ the method becomes unstable.

The periodic boundaries at $z = 0, L$ are updated for $A, B, C$.
Differencing the zero normal gradient condition gives an update value
for the boundary points, but it is again important to introduce
`inertia' in the update. A value of $\epsilon = 0.02$ was used to
generate the results presented here. Without this inertia, the
solutions do not converge.

On the asymptotic boundary, we impose the conditions that $A, B \sim 1
/ r^2$ there. This is differenced to second order and the update
requires no `inertia'. The metric function $C$ is updated by
evaluating the linearised $G^{z}_{~z}$ constraint equation at the
midpoint of the boundary, $z = \frac{L}{2}$, and ignoring $z$ gradient
terms. Thus we assume that the metric functions are small and that
only the $z$ independent component remains in the metric
asymptotically.  Picking the mid point, $z = \frac{L}{2}$, reduces any
error from the exponentially suppressed $z$ dependent components in
the metric, as they go as $\cos{\pi z / L}$ to leading order.  This
linearised constraint then yields a boundary value for $C$.  We use
this to determine $C$ all along the $r = r_{max}$ boundary, again
assuming $z$ independence there. These $C$ values are injected with
`inertia', with a value of $\epsilon = 0.01$ giving good results.
Choosing no inertia again destroys convergence.

The above method is then iterated. After a hundred steps or so, the
error in the 3 elliptic equations is calculated, being taken as the
absolute of the difference of the left-hand side and right-hand side
of \eqref{eq:general_equations}.  This is then averaged over the
lattice for each metric function. Relaxation is stopped when these
quantities are less than $\sim 10^{-9}$, and the solutions have
converged.

As an initial guess for the relaxation procedure we simply take all
the metric functions to be zero, and put a cosine form for $B$ on the
horizon points with the correct amplitude. This appears to be quite
brutal but does indeed work for deformations with $B_{max} \le 0.3$.
After this value, larger $\lambda$ solutions are best calculated by
taking the relaxed solution for a lower $\lambda$, and then simply
updating the value of $B$ at $r = 0, z = L$ to the new $B_{max}$
value. This initial data will then relax provided that the jump in
deformation is not too great. Typically we found that a jump of
$\Delta B_{max} < 0.15$ gives good results for all the resolutions
used here. Note that perturbing the initial guess does not effect the
end solution, and thus there do not appear to be any additional static
perturbation modes present for the non-uniform strings.

The lattice used to cover the $rz$ plane is rectangular, with even
spacing $dr, dz$ in the $r, z$ directions.  We discretise the fields
over the lattice as $X_{(i,j)}$ where $i, j$ are the $r$ and $z$
positions respectively and run as $i=1,2,\ldots i_{\rm max}$,
$j=1,2,\ldots j_{\rm max}$. The interior equations are differenced
using standard second order templates, and the Laplace second
derivative terms imply they are naturally evaluated at $(i,j)$. When
evaluating the constraint equations, to check the consistency of the
solutions, the constraints are evaluated naturally at different
lattice positions due to their second derivative structure. The
constraint equation $\crz$ contains terms taking the generic form,
\begin{equation}
\partial_r \partial_z X + a \, \frac{1}{r} \partial_r Y + b \, \partial_r X
\partial_z Y + \ldots = 0
\end{equation}
where $X, Y$ are some metric functions. This is then evaluated at the
centre of a lattice cell $(i+\frac{1}{2},j+\frac{1}{2})$, compatible
with the second order mixed derivative operators. Similarly the
Einstein tensor components $G^{r}_{~r}$ and $G^{z}_{~z}$ reside
naturally at $z = z_{i+\frac{1}{2},j}$ and $z = z_{i,j+\frac{1}{2}}$
respectively, as indicated by their highest derivative operators, the
first having only first order $r$ derivatives and the second, only
first order $z$ derivatives. This is why we use $G^{z}_{~z}$ to
determine $C$ on the asymptotic boundary, rather than $\crrzz$.

%
\subsection{Mass Determination and Asymptotic Boundary Conditions}
\label{app:mass}
%

In this section we discuss the asymptotic boundary conditions and how
to compute the asymptotic mass from the metric at the large $r$
boundary.

In order to determine the mass of the solutions we assume that the
solution at $r = r_{max}$ has no dependence on $z$. The
remaining $z$ independent metric component may then be integrated out
to large $r$ to determine its asymptotic behaviour. At very large $r$,
we may linearise in the deformation, and then obtain the following
general asymptotic behaviour of the metric (similar to in Appendix
\ref{app:gubser_PT}),
\begin{eqnarray}
A \simeq a_0 + \frac{a_2}{r^2} + O(\frac{1}{r^3}) \\ \nonumber 
B \simeq b_0 + \frac{b_2}{r^2} + O(\frac{1}{r^3})\\ \nonumber 
C \simeq b_0 + \frac{c_1}{r} + \frac{c_2}{r^2} + O(\frac{1}{r^3})
\label{eq:asym_scaling}
\end{eqnarray}
This form of the asymptotic metric yields a mass per unit length,
$\rho$, of
\begin{equation}
\rho \propto \frac{3}{2} + 5 b_2 + 3 c_2   
\end{equation}
to linear order in the metric, and the total mass $\mathcal{M} =
\rho\, L$.  At $r = r_{max}$, where we take $r_{max}
\gtrsim 6$, we impose the asymptotic boundary conditions on the
lattice, namely we aim to have $a_0, b_0$ be zero, and to satisfy the
constraint $G^z_{~z}$.  However, whilst at $r_{max} = 6$ we
have largely eliminated the exponentially growing $z$ dependent
components of the metric, it is not large enough that the $z$
independent component satisfies the above asymptotic scaling very
accurately, mainly due to the slow $1 / r$ behaviour in $C$. Thus we
cannot extract $b_2, c_2$ at this boundary directly.  Really the
lattice would have to extend out to $r \sim O(1000)$ to be in the true
asymptotic regime. So, the metric functions and their $r$ derivatives
are averaged over the large $r$ boundary. These averaged values are
then integrated out to large $r \simeq 1000$, and the asymptotic
values of $b_2, c_2$ are determined by fitting functions of the
general form in \eqref{eq:asym_scaling}.  Then the mass is calculated.

Note that we have linearised the update equation $G^{z}_{~z}$ for
determining $C$ on the $r_{max}$ boundary, and use the scaling,
$A, B \simeq 1 / r^2$ which is based on the linear theory asymptotics.
Interestingly, we find best results when the equations used to evolve
the $z$ independent metric components out to very large $r$ are not
linearised in the metric perturbation. Although at the asymptotic
boundary the metric functions are small, even for large $\lambda$
solutions, non-linear corrections do appear to be important when
integrating the $z$ independent component out to very large $r$, and
do make a significant difference to the mass calculated from this
asymptotic metric.

As we have imposed boundary conditions at finite $r \simeq 6$, they
will not strictly enforce $a_0, b_0 = 0$ due to the approximate nature
of the boundary conditions, which really are only valid at very large
$r$. However, we find in practice only a very small $a_0, b_0$ which
does not concern us at all, and is simply a slight change of scheme.
The change of scheme is minute and can be neglected when we determine
the properties of the solutions.

%
\subsection{Finite Large $r$ Boundary Check}
\label{app:boundary}
%

In this section we test the sensitivity of the solutions to the finite
position of the asymptotic boundary, $r_{max}$.  In the table
below we compare two asymptotic quantities, the mass (measured
directly) and the peak value of the measure weighted constraint $\crz$
evaluated at $r = r_{max}$, and two horizon quantities, the horizon
volume or entropy $\mathcal{S}$, and the same constraint evaluated at
$r = 0$.  These are compared for solutions with $B_{max} = 0.1$
(so $\lambda \simeq 0.1$) for different asymptotic boundary positions,
$r_{max}$.  All solutions are at the same resolution as for $120*50$
with $r_{max} = 6$, namely $dr, dz \simeq 0.05$.
\begin{center}
\begin{tabular}{c|cc|cc}
  $r_{max}$ & $\mathcal{S}$ & weighted $\crz$ at $r = 0$ & $\eta$ & weighted $\crz$ at $r_{max}$ \\ \hline \\
  4.0     & 0.04330 & 0.003550 & 0.02848 & 0.048406 \\
  5.0    & 0.04850 & 0.000520 & 0.03190 & 0.020042 \\
  6.0    & 0.04905 & 0.000459 & 0.03225 & 0.007933 \\
  7.0    & 0.04908 & 0.000498 & 0.03226 & 0.002682 \\
  8.0    & 0.04907 & 0.000505 & 0.03224 & 0.000918 \\
  9.0    & 0.04905 & 0.000505 & 0.03223 & 0.000306 \\
  10.0   & 0.04904 & 0.000505 & 0.03221 & 0.000100
\end{tabular}
\end{center}
We see that increasing the position of the boundary does indeed yield
less asymptotic weighted constraint violations. The boundary
conditions imposed at $r = r_{max}$ are linearised and assume that the
metric is $z$ independent.  This assumption obviously proves to be
good enough in practice by $r_{max} = 6$.  The asymptotic mass,
measured directly, also tends to a constant as expected, and appears
to give a result close to the asymptotic value for $r_{max} \simeq 6$.
We see that the entropy evaluated at $r = 0$, and indeed other horizon
geometric quantities, and the weighted constraints evaluated at $r =
0$, are quite insensitive of the boundary position provided it is
above $r_{max} \gtrsim 6$.

We thus find the required insensitivity of the solutions to the value
of $r_{max}$ provided $r_{max} \ge 6$, and
$r_{max} \simeq 6$ does give good results, whilst not requiring
excessive relaxation times.

%
\subsection{Constraint Violation}
\label{app:constraints}
%

\begin{figure}[htb]
\centerline{\psfig{file=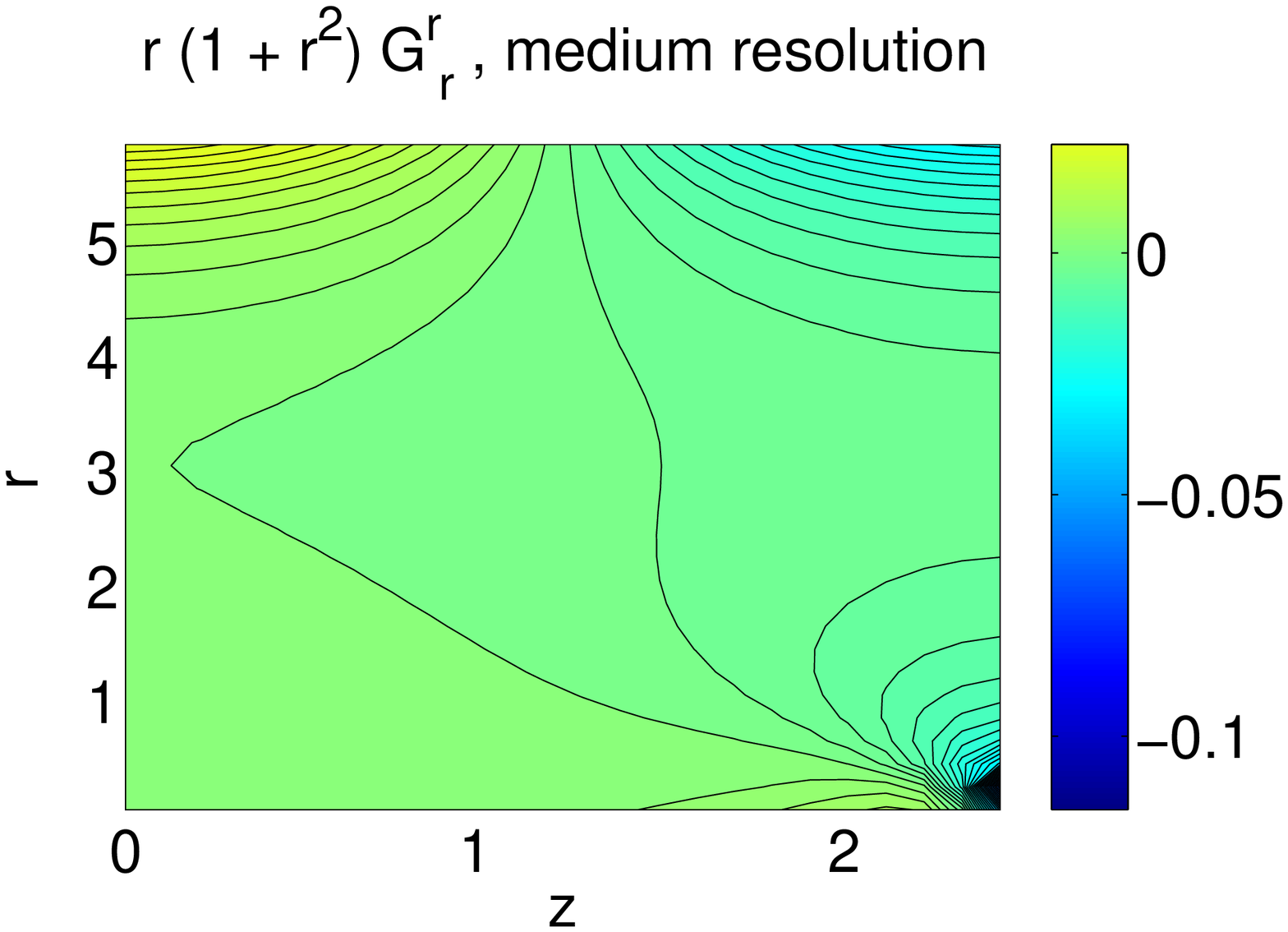,width=3in}
}
\centerline{\psfig{file=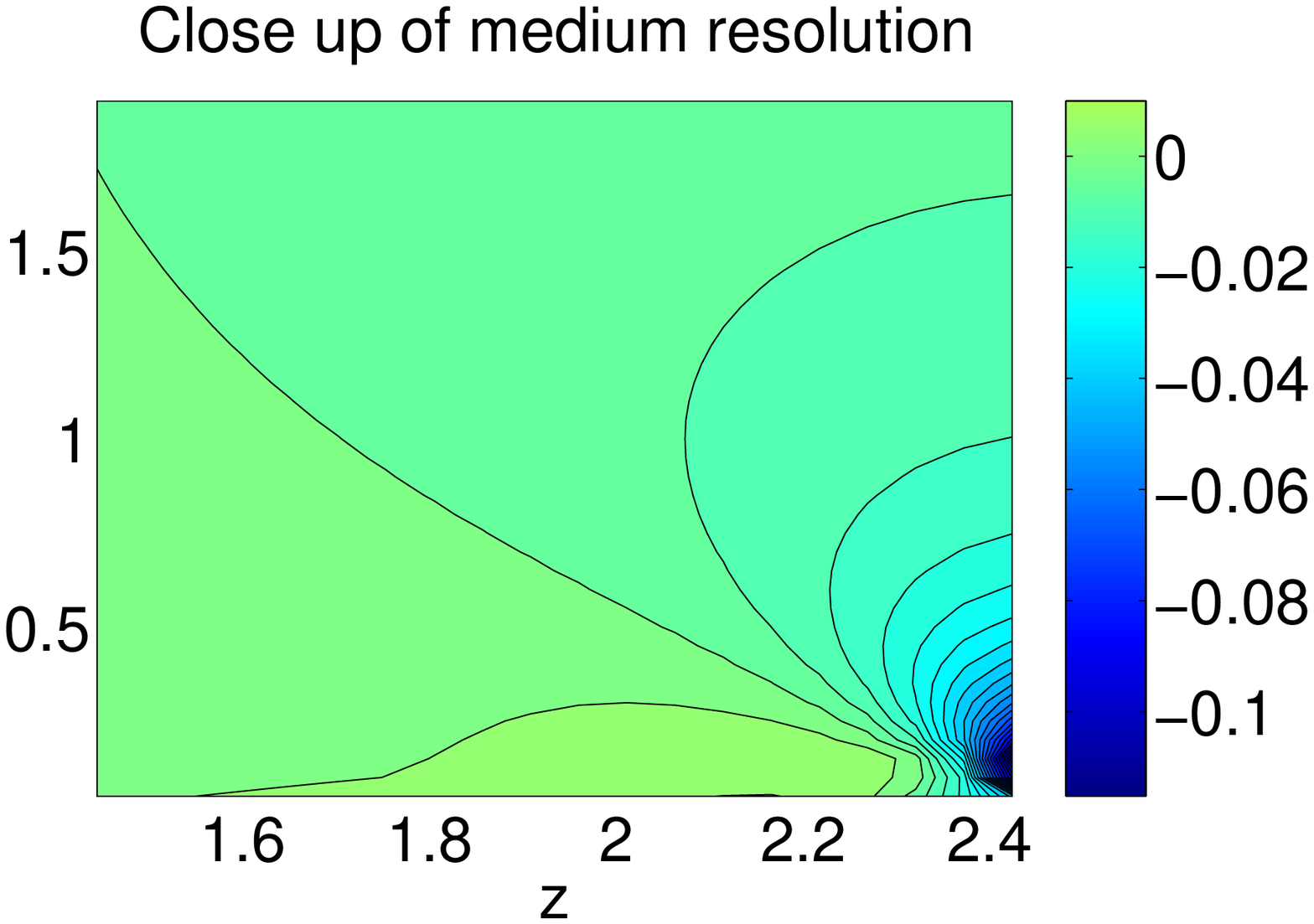,width=3in}\hspace{0.5cm}\psfig{file=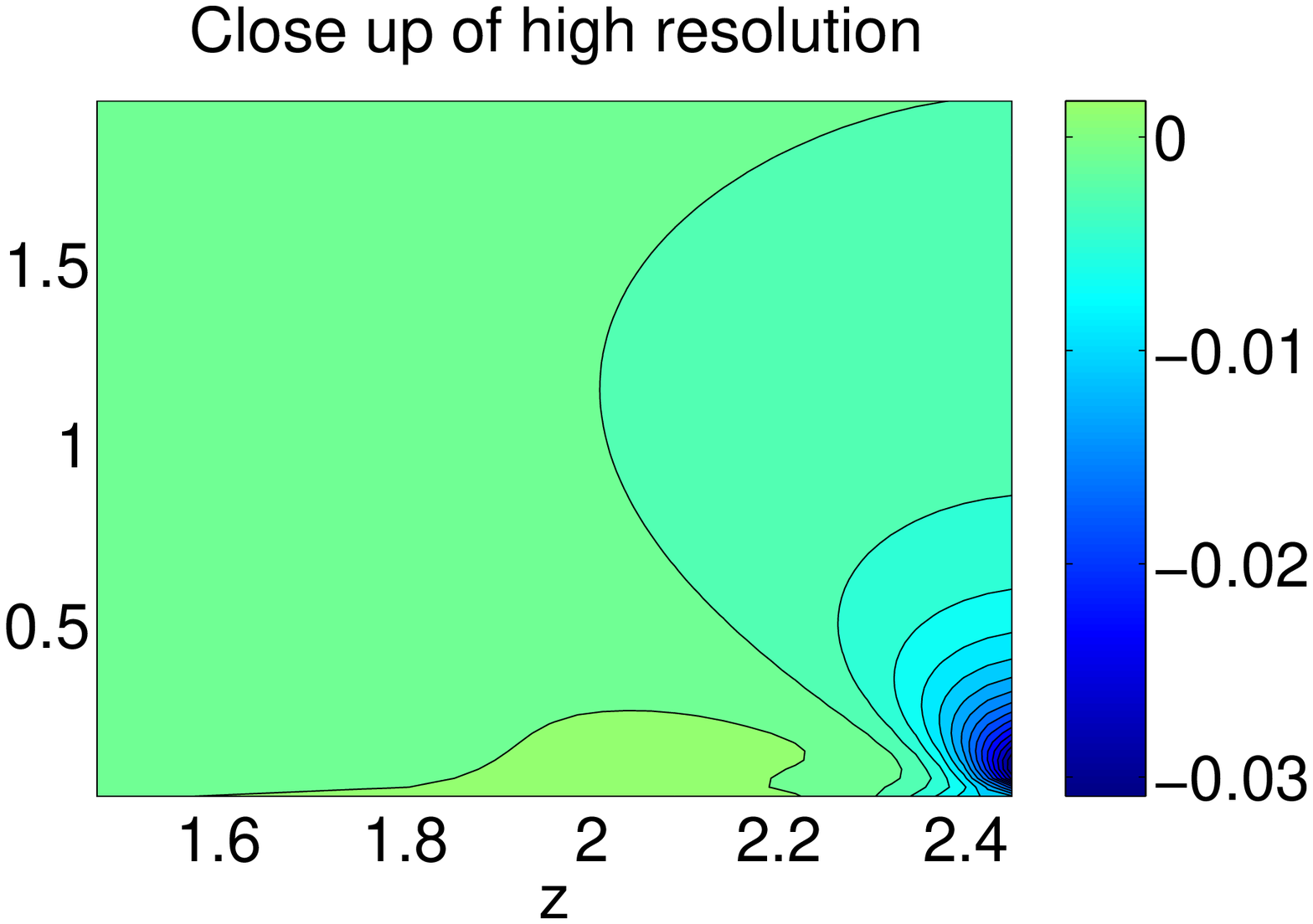,width=3in}}
\caption{ \figuremode
\label{fig:constraints1} 
The top plot shows the measure weighted constraint $\crz$ for the
medium resolution $120*50$ over the whole lattice for $B_{max} =
1.25$, so $\lambda \simeq 1.8$, near the point where this resolution
no longer converges. The bottom left plot shows a close up of the
lower right corner of the lattice, where the constraints become
increasingly violated as this convergence limit is reached. The bottom
right plot shows the same quantity, for the same $B_{max}$, but with
the high resolution $240*100$. We see that the constraints are
improved by a factor of approximately four, compatible with second
order scaling. The peak value of this function is plotted in the
following figure \ref{fig:constraints2} for a range of $\lambda$. }
\end{figure}

In this section we simply characterise the constraint violation of the
solutions for varying $\lambda$ and resolution, in order to
demonstrate the validity of the large $\lambda$ solutions. In figure
\ref{fig:constraints1}, we show the measure weighted constraint $\crz$
for 2 resolutions, medium ($120*50$), and high ($240*100$) with $B_{max}
= 1.25$ (so $\lambda \simeq 1.8$). Above $B_{max} \simeq 1.40$ the
$120*50$ resolution no longer converges, and at this $B_{max} = 1.25$
the constraint violation becomes large near $r = 0, z = L$.  We show
the constraint over the whole lattice for reference, but it is the
constraint in the lower right corner, $r = 0, z = L$ that is
substantially improved by going to higher resolution.  This lower
right corner is plotted for the two resolutions to show the
improvement.

\begin{figure}[htb]
\centerline{\psfig{file=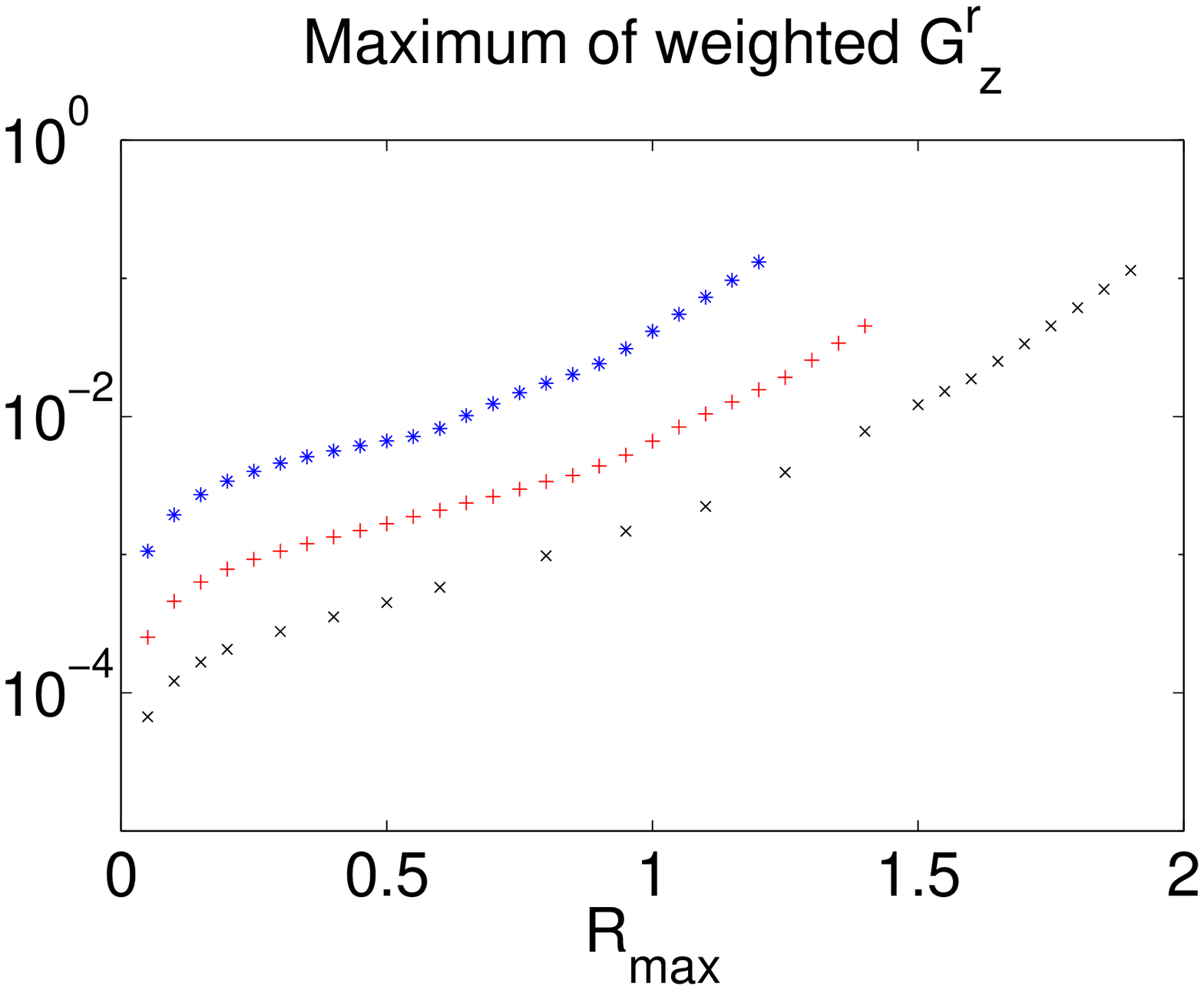,width=3in}\hspace{1.0cm}\psfig{file=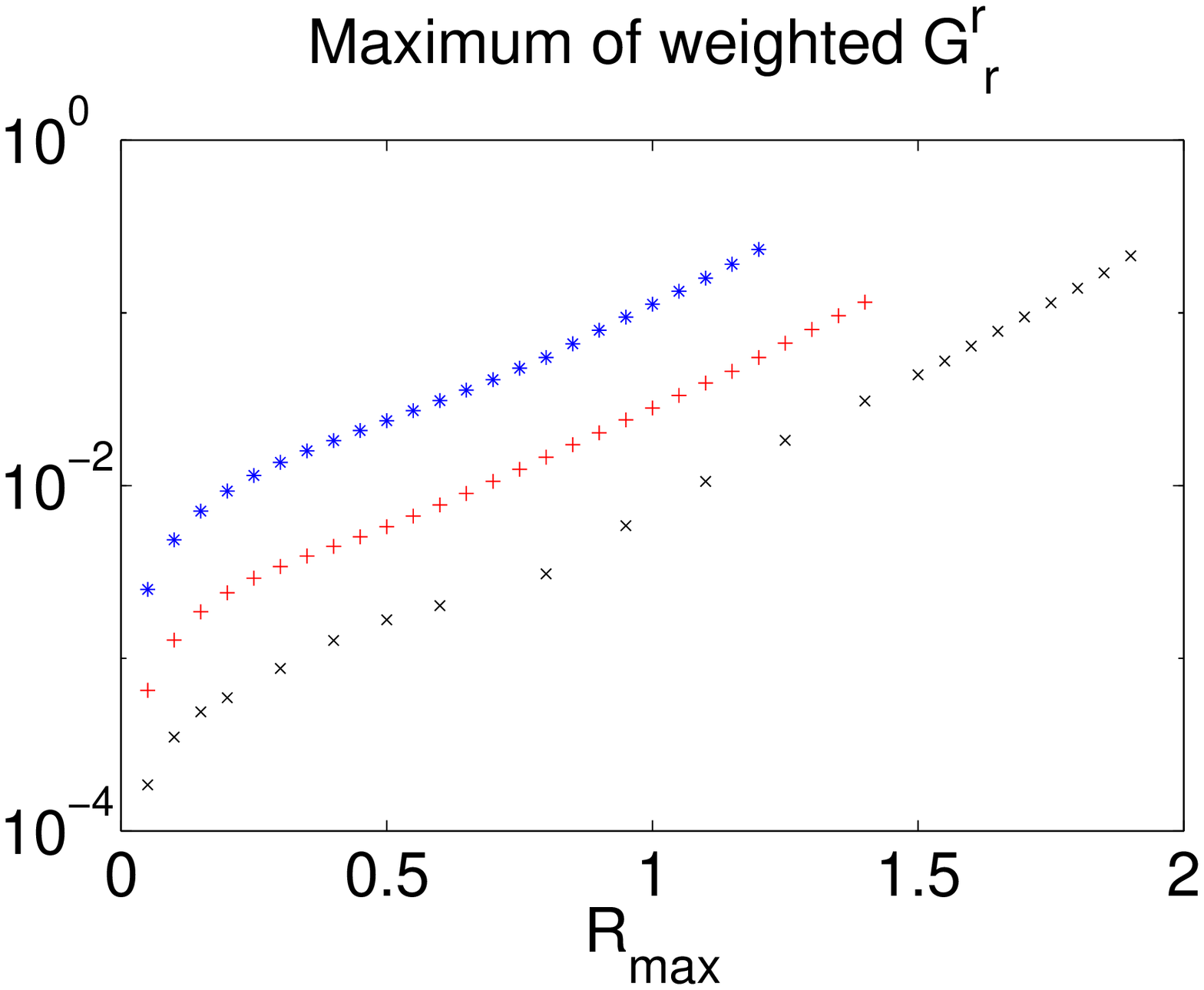,width=3in}}
\caption{ \figuremode
\label{fig:constraints2}
Plots showing the peak value of the two measure weighted constraints,
$\crz$ and $G^{r}_{~r}$, in the $r = 0$, $z = L$ corner of the
lattice, for three resolutions; $60*25$ (blue), $120*50$ (red) and
$240*100$ (black).  We see that increasing the resolution decreases
the constraint violation, compatible with second order scaling.
Furthermore, this shows that the large constraint violation that
occurs as the convergence limit of a resolution is reached is
unphysical, and is removed by simply going to a higher resolution. (Note that $B_{max} = 1.9$ corresponds to $\lambda \simeq 3.9$)}
\end{figure}

This constraint violation can be characterised by taking the absolute
peak value of the measure weighted constraints in the lower half of the
lattice, thus picking out the trouble spot near $r = 0$, $z = L$.  In
figure \ref{fig:constraints2} we plot this peak violation value for
varying $\lambda$ and for 3 grid resolutions.  Note that the
constraint violation in the upper half of the lattice is due to the
position of the large $r$ boundary and is uneffected by changing
resolution.

Firstly we see that increasing the resolution does indeed reduce these
values, indicating that the constraints improve, as we would hope for.
Indeed for each doubling of resolution, the constraints improve by
approximately a factor of 4, indicating second order scaling,
consistent with our second order differencing. Furthermore, we can
clearly see the effect discussed in section \ref{sec:large_lambda},
where at some large value of $\lambda$ the constraints become strongly
violated at a fixed resolution before convergence is lost, and we must
proceed to a higher resolution to continue to larger $\lambda$.  It is
always difficult to assess how physically bad a constraint violation
is.  The absolute values of these measure weighted constraints can be
compared to the typical values of the similarly weighted curvatures.
These typically take values of order one, as shown in the earlier
figure \ref{fig:example2}, and thus the violation in the constraints,
even for $B_{max}$ quite near to the loss of convergence point, is
much smaller than the curvatures.  However a more physical error
assessment is to compare the thermodynamic properties of a solution at
low resolution with large constraint violation, with a higher
resolution where the constraints are much better satisfied. For
example, comparing the temperature, mass and entropy for the low and
high resolutions for $B_{max} \simeq 1.25$ ($\lambda \simeq 1.8$)
where the low resolution only just converges, but the high resolution
performs very well. This comparison can be made in figure
\ref{fig:properties} and shows that although the weighted constraints
become violated just before convergence is lost, this hardly effects
the physical properties of the solution.

%
%

%
\newpage
%

\newcommand{\href}[1]{}%
\newcommand{\dhref}[1]{}%
\newenvironment{hpabstract}{%
  \renewcommand{\baselinestretch}{0.2}
  \begin{footnotesize}%
}{\end{footnotesize}}%
\newcommand{\hpeprint}[1]{%
  \href{http://arXiv.org/abs/#1}{\texttt{#1}}}%
\newcommand{\hpspires}[1]{%
  \dhref{http://www.slac.stanford.edu/spires/find/hep/www?#1}{\ (spires)}}%

%

\begin{thebibliography}{999}
\bibitem{Weyl}
\textsc{H.~Weyl}: \textsl{Ann. Phys. (Leipzig)} \textbf{54} (1917) 117.

\bibitem{Emparan_Reall1}
\textsc{R.~Emparan {\upshape and} H.~Reall}: Generalized Weyl solutions.
\newblock \textsl{Phys. Rev.} \textbf{D65} (2002) 084025,
  \hpeprint{hep-th/0110258}.

\bibitem{Emparan_Reall2}
\textsc{R.~Emparan {\upshape and} H.~Reall}: A rotating black ring in five
  dimensions.
\newblock \textsl{Phys. Rev. Lett.} \textbf{88} (2002) 101101,
  \hpeprint{hep-th/0110260}.

\bibitem{Gregory_Laflamme1}
\textsc{R.~Gregory {\upshape and} R.~Laflamme}: Black strings and p-branes are
  unstable.
\newblock \textsl{Phys. Rev. Lett.} \textbf{70} (1993) 2837--2840,
  \hpeprint{hep-th/9301052}.

\bibitem{Gregory_Laflamme2}
\textsc{R.~Gregory {\upshape and} R.~Laflamme}: Hypercylindrical black holes.
\newblock \textsl{Phys. Rev.} \textbf{D37} (1988) 305.

\bibitem{Gregory_Laflamme3}
\textsc{R.~Gregory {\upshape and} R.~Laflamme}: The Instability of charged
  black strings and p-branes.
\newblock \textsl{Nucl. Phys.} \textbf{B428} (1994) 399--434,
  \hpeprint{hep-th/9404071}.

\bibitem{Kol1}
\textsc{B.~Kol}: Topology change in general relativity and the black-hole
  black-string transition (2002).
\newblock Preprint \hpeprint{hep-th/0206220}.

\bibitem{Kol2}
\textsc{B.~Kol}: Explosive black hole fission and fusion in large extra
  dimensions (2002).
\newblock Preprint \hpeprint{hep-ph/0207037}.

\bibitem{Reall2}
\textsc{A.~Chamblin, S.~Hawking {\upshape and} H.~Reall}: Brane-world black
  holes.
\newblock \textsl{Phys. Rev.} \textbf{D61} (2000) 065007,
  \hpeprint{hep-th/9909205}.

\bibitem{Gregory}
\textsc{R.~Gregory}: Black string instabilities in anti-de Sitter space.
\newblock \textsl{Class. Quant. Grav.} \textbf{17} (2000) L125--132,
  \hpeprint{hep-th/0004101}.

\bibitem{Kang}
\textsc{T.~Hirayama {\upshape and} G.~Kang}: Stable black strings in anti-de
  Sitter space.
\newblock \textsl{Phys. Rev.} \textbf{D64} (2001) 064010,
  \hpeprint{hep-th/0104213}.

\bibitem{Gibbons_Hartnoll1}
\textsc{G.~Gibbons {\upshape and} S.~Hartnoll}: A gravitational instability in
  higher dimensions (2002).
\newblock Preprint \hpeprint{hep-th/0206202}.

\bibitem{Gubser_Mitra1}
\textsc{S.~Gubser {\upshape and} I.~Mitra}: Instability of charged black holes
  in anti-de Sitter space (2000).
\newblock Preprint \hpeprint{hep-th/0009126}.

\bibitem{Gubser_Mitra2}
\textsc{S.~Gubser {\upshape and} I.~Mitra}: The evolution of unstable black
  holes in anti-de Sitter space.
\newblock \textsl{JHEP} \textbf{08} (2001) 018, \hpeprint{hep-th/0011127}.

\bibitem{Hubeny_Rangamani}
\textsc{V.~Hubeny {\upshape and} M.~Rangamani}: Unstable horizons.
\newblock \textsl{JHEP} \textbf{05} (2002) 027, \hpeprint{hep-th/0202189}.

\bibitem{Reall}
\textsc{H.~Reall}: Classical and thermodynamic stability of black branes.
\newblock \textsl{Phys. Rev.} \textbf{D64} (2001) 044005,
  \hpeprint{hep-th/0104071}.

\bibitem{Ross}
\textsc{J.~Gregory {\upshape and} S.~Ross}: Stability and the negative mode for
  Schwarzschild in a finite cavity.
\newblock \textsl{Phys. Rev.} \textbf{D64} (2001) 124006,
  \hpeprint{hep-th/0106220}.

\bibitem{Horowitz_Maeda1}
\textsc{G.~Horowitz {\upshape and} K.~Maeda}: Fate of the black string
  instability.
\newblock \textsl{Phys. Rev. Lett.} \textbf{87} (2001) 131301,
  \hpeprint{hep-th/0105111}.

\bibitem{Gubser}
\textsc{S.~Gubser}: On non-uniform black branes (2001).
\newblock Preprint \hpeprint{hep-th/0110193}.

\bibitem{Horowitz}
\textsc{G.~Horowitz}: Playing with black strings (2002).
\newblock Preprint \hpeprint{hep-th/0205069}.

\bibitem{Choptuik}
\textsc{M.~Choptuik, L.~Lehner, I.~Olabarrieta, R.~Petryk, F.~Pretorius
  {\upshape and} H.~Villegas}: To appear.

\bibitem{Lehner}
\textsc{L.~Lehner}: Numerical relativity: A review.
\newblock \textsl{Class. Quant. Grav.} \textbf{18} (2001) R25--R86,
  \hpeprint{gr-qc/0106072}.

\bibitem{Harmark_Obers}
\textsc{T.~Harmark {\upshape and} N.~Obers}: Black holes on cylinders.
\newblock \textsl{JHEP} \textbf{05} (2002) 032, \hpeprint{hep-th/0204047}.

\bibitem{Horowitz_Maeda2}
\textsc{G.~Horowitz {\upshape and} K.~Maeda}: Inhomogeneous near-extremal black
  branes.
\newblock \textsl{Phys. Rev.} \textbf{D65} (2002) 104028,
  \hpeprint{hep-th/0201241}.

\bibitem{DeSmet}
\textsc{P.~De Smet}: Black holes on cylinders are not algebraically special
  (2002).
\newblock Preprint \hpeprint{hep-th/0206106}.

\bibitem{Wiseman}
\textsc{T.~Wiseman}: Relativistic stars in Randall-Sundrum gravity.
\newblock \textsl{Phys. Rev.} \textbf{D65} (2002) 124007,
  \hpeprint{hep-th/0111057}.

\bibitem{Randall_Sundrum1}
\textsc{L.~Randall {\upshape and} R.~Sundrum}: A large mass hierarchy from a
  small extra dimension.
\newblock \textsl{Phys. Rev. Lett.} \textbf{83} (1999) 3370--3373,
  \hpeprint{hep-ph/9905221}.

\bibitem{Randall_Sundrum2}
\textsc{L.~Randall {\upshape and} R.~Sundrum}: An alternative to
  compactification.
\newblock \textsl{Phys. Rev. Lett.} \textbf{83} (1999) 4690--4693,
  \hpeprint{hep-th/9906064}.

\bibitem{Website}
\textsc{T.~Wiseman}: http://www.damtp.cam.ac.uk/user/tajw2.

\bibitem{Myers}
\textsc{R.~Myers}: Higher dimensional black holes in compactified space-times.
\newblock \textsl{Phys. Rev.} \textbf{D35} (1987) 455.

\bibitem{Gregory2}
\textsc{R.~Gregory {\upshape and} A.~Padilla}: Nested braneworlds and strong
  brane gravity.
\newblock \textsl{Phys. Rev.} \textbf{D65} (2002) 084013,
  \hpeprint{hep-th/0104262}.

\bibitem{Horowitz2}
\textsc{R.~Emparan, G.~Horowitz {\upshape and} R.~Myers}: Exact description of
  black holes on branes.
\newblock \textsl{JHEP} \textbf{01} (2000) 007, \hpeprint{hep-th/9911043}.

\bibitem{Shinkai}
\textsc{A.~Chamblin, H.~Reall, H.~Shinkai {\upshape and} T.~Shiromizu}: Charged
  brane-world black holes.
\newblock \textsl{Phys. Rev.} \textbf{D63} (2001) 064015,
  \hpeprint{hep-th/0008177}.

\bibitem{Casadio}
\textsc{R.~Casadio {\upshape and} L.~Mazzacurati}: Bulk shape of brane-world
  black holes (2002).
\newblock Preprint \hpeprint{hep-th/0205129}.

\bibitem{Wiseman2}
\textsc{T.~Wiseman}: Strong brane gravity and the radion at low energies.
\newblock \textsl{Class. Quant. Grav.} \textbf{19} (2002) 3083--3106,
  \hpeprint{hep-th/0201127}.

\bibitem{Soda}
\textsc{S.~Kanno {\upshape and} J.~Soda}: Radion and holographic brane gravity
  (2002).
\newblock Preprint \hpeprint{hep-th/0207029}.

\bibitem{ADM}
\textsc{R.~Arnowitt, S.~Deser {\upshape and} C.~Misner}: The dynamics of
  general relativity.
\newblock \textsl{Gravitation: An introduction to current research, edited by
  L. Witten (John Wiley, New York)}  (1962).

\bibitem{numrecp}
\textsc{W.~Press, S.~Teukolsky, W.~Vetterling {\upshape and} B.~Flannery}:
  Numerical recipes.
\newblock Cambridge University Press.

\bibitem{Kol3}
\textsc{B.~Kol}: Speculative generalization of black hole uniqueness to higher
  dimensions (2002).
\newblock Preprint \hpeprint{hep-th/0208056}.

\bibitem{Tanaka}
\textsc{T.~Tanaka}: Classical black hole evaporation in Randall-Sundrum infinite braneworld (2002).
\newblock Preprint \hpeprint{hep-th/0203082}.

\bibitem{Emparan_Kaloper}
\textsc{R.~Emparan, A.~Fabbri {\upshape and} N.~Kaloper}: Quantum black holes
  as holograms in AdS braneworlds (2002).
\newblock Preprint \hpeprint{hep-th/0206155}.

\bibitem{Hawking_Horowitz}
\textsc{S.~Hawking {\upshape and} G.~Horowitz}: The gravitational Hamiltonian,
  action, entropy and surface terms.
\newblock \textsl{Class. Quant. Grav.} \textbf{13} (1996) 1487--1498,
  \hpeprint{gr-qc/9501014}.

\end{thebibliography}
\end{document}